\documentclass[11pt,twoside]{article}

\usepackage[utf8]{inputenc}
\usepackage[colorlinks=true
]{hyperref}
\hypersetup{urlcolor=blue, citecolor=red, linkcolor=blue}
\usepackage{mleftright}
\usepackage{stmaryrd}
\SetSymbolFont{stmry}{bold}{U}{stmry}{m}{n}
\usepackage{microtype}

\usepackage{bbm}
\usepackage{colortbl}

\usepackage{graphicx}
\usepackage[active]{srcltx}

\usepackage{amsmath,amssymb}

\numberwithin{equation}{section}

\usepackage{physics}
\usepackage{bm}
\usepackage{enumerate}

\usepackage{faktor}
\usepackage{amsthm}
\usepackage[all,2cell]{xy}
\usepackage{mathrsfs}
\usepackage{mathtools}
\mathtoolsset{showonlyrefs}

\usepackage{tikz-cd}
\usetikzlibrary{cd}

\usepackage{xcolor}
\DeclareUnicodeCharacter{0301}{\'{}}

\usepackage[a4paper, left=25mm, right=25mm, top=30mm, bottom=25mm]{geometry}

\usepackage[symbol]{footmisc}

\usepackage{marginnote}

\usepackage{imakeidx}
\makeindex[intoc]

\usepackage[nottoc]{tocbibind}
\usepackage[colorinlistoftodos]{todonotes} 

\theoremstyle{plain}
\newtheorem{theorem}{Theorem}[section]
\newtheorem{proposition}[theorem]{Proposition}
\newtheorem{corollary}[theorem]{Corollary}

\theoremstyle{definition}
\newtheorem{definition}[theorem]{Definition}
\newtheorem{remark}[theorem]{Remark}

\newcommand\restr[2]{{
  \left.\kern-\nulldelimiterspace #1 \right|_{#2} 
}}

\newcommand*{\longhookrightarrow}{\ensuremath{\lhook\joinrel\relbar\joinrel\rightarrow}}
\newcommand{\R}{\mathbb{R}}

\renewcommand{\d}{\mathrm{d}}

\newcommand{\df}{\Omega}
\newcommand{\Cinfty}{\mathscr{C}^\infty}
\newcommand{\T}{\mathrm{T}}
\newcommand{\Tan}{\mathrm{T}}
\newcommand{\cT}{\mathrm{T}^\ast}

\newcommand*{\innp}[1]{\iota\left(#1\right)}

\newcommand{\Lie}{\mathscr{L}}
\newcommand{\vf}{\mathfrak{X}}

\newcommand{\F}{\mathcal{F}}

\renewcommand{\P}{\mathcal{P}}

\newcommand*{\bd}{\overline{\mathrm{d}}}

\newcommand{\bfX}{\mathbf{X}}
\newcommand{\bfY}{\mathbf{Y}}

\def\derpar#1#2{\frac{\partial{#1}}{\partial{#2}}}
\newcommand{\parder}[2]{\frac{\partial #1}{\partial #2}}
\newcommand{\dparder}[2]{\dfrac{\partial #1}{\partial #2}}
\newcommand{\tparder}[2]{\partial #1/\partial #2}
\newcommand{\parderr}[3]{\frac{\partial^2 #1}{\partial #2\partial #3}}

\DeclareMathOperator{\Ima}{Im}

\DeclareMathOperator{\Hom}{Hom}

\DeclareMathAlphabet{\mathpzc}{OT1}{pzc}{m}{it}
\def\d{\mathrm{d}}

\let\ds\displaystyle

\newcommand{\dfn}[1]{{\sl\textbf{#1}}}

\usepackage{xcolor}
\usepackage{fancyhdr}

\begin{document}

\parskip=4pt

{\Huge\sffamily\raggedright
Extended Hamiltonian formalism\\[10pt]for multicontact field theories}

\vspace{1em}

{\Large\raggedright\sffamily
    Manuel de León
}\vspace{1mm}\newline
{\raggedright
    Instituto de Ciencias Matemáticas (ICMAT--CSIC) $\&$
    Real Academia de Ciencias Exactas, Físicas y Naturales de España,
    Madrid, Spain.\\[0pt]    
    e-mail: \href{mailto:mdeleon@icmat.es}{mdeleon@icmat.es} --- orcid: \href{https://orcid.org/0000-0002-8028-2348}{0000-0002-8028-2348}
}

\vspace{0.5em}

{\Large\raggedright\sffamily
    Rub\'en Izquierdo-López
}\vspace{1mm}\newline
{\raggedright
    Instituto de Ciencias Matem\'aticas (ICMAT--CSIC),
    Madrid, Spain.\\[0pt]
    e-mail: \href{mailto:ruben.izquierdo@icmat.es}{ruben.izquierdo@icmat.es} --- orcid: \href{https://orcid.org/0009-0007-8747-344X}{0009-0007-8747-344X}
}

\vspace{0.5em}

{\Large\raggedright\sffamily
    Xavier Rivas
}\vspace{1mm}\newline
{\raggedright
    Department of Computer Engineering and Mathematics, Universitat Rovira i Virgili,
    Tarragona, Spain.\\[0pt]
    e-mail: \href{mailto:xavier.rivas@urv.cat}{xavier.rivas@urv.cat} --- {\sc orcid}: \href{https://orcid.org/0000-0002-4175-5157}{0000-0002-4175-5157}
}

\vspace{0.5em}

{\Large\raggedright\sffamily
    Narciso Román-Roy
}\vspace{1mm}\newline
{\raggedright
    Department of Mathematics $\&$
Institute of Mathematics (IMTech), Universitat Politècnica de Catalunya,
Barcelona, Spain.\\[0pt]
e-mail: 
\href{mailto:narciso.roman@upc.edu}{narciso.roman@upc.edu} --- {\sc orcid}: 
    \href{https://orcid.org/0000-0003-3663-9861}{0000-0003-3663-9861}
}

\vspace{0.5em}

{\Large\raggedright\sffamily
    Annamaria Villanova
}\vspace{1mm}\newline
{\raggedright
    Escuela Superior de Ingenier\'ia y Tecnolog\'ia, Univ. Internacional de La Rioja. La Rioja, Spain
    $\&$ Department of Mathematics, Universitat Politècnica de Catalunya,
    Barcelona, Spain.\\[0pt]   
    e-mail: \href{mailto:annamaria.villanova@unir.net}{annamaria.villanova@unir.net } --- orcid: \href{https://orcid.org/0009-0002-2350-8109}{0009-0002-2350-8109}
}

\vspace{1em}

{\large\bf\raggedright
    Abstract
}\vspace{1mm}\newline
{\parindent 0pt
We develop an extended Hamiltonian formalism for action-dependent (Herglotz-type) classical field theories within the framework of multicontact geometry. While the standard Hamiltonian description of these theories is formulated on the restricted multimomentum bundle, we work instead on the extended multimomentum bundle, which carries an additional energy-type coordinate together with a canonical multicontact structure. In this setting, an extended Hamiltonian multicontact system is governed by a Hamiltonian density and its associated Hamiltonian one-form, and the submanifold they determine recovers the restricted formalism. We characterise the admissible one-forms and prove that the associated Hamilton--de Donder--Weyl field equations, stated equivalently for sections, multivector fields, and Ehresmann connections, are mutually equivalent. We then establish the precise relation between the extended and restricted formalisms in both the regular (multicontact) and the almost-regular (premulticontact) settings, the latter requiring a constraint algorithm. The formalism is illustrated with two examples: the action-dependent Klein--Gordon equation, which includes the telegrapher equation as a particular case, and the heat equation, whose contactification yields Burgers' equation.
}
\vspace{1em}

\noindent\textbf{Keywords:}
Lagrangian and Hamiltonian classical field theories,
Action-dependent theories,
Extended Hamiltonian formalism,
Multicontact geometry,
Contact structures.

\vspace{1em}

\noindent\textbf{MSC\,2020 codes:} \
{\sl Primary:}
70S05, 
70S10,
53D10, 
35R01,
37J55, 
\\
\indent\indent\indent\indent\indent\indent
{\sl Secondary:}
35Q99, 
53C15, 
53Z05, 
58A10, 
70G45. 





\newpage

{\setcounter{tocdepth}{2}
\def\baselinestretch{0.9}
\def\addvspace#1{\vskip 1pt}
\parskip 0pt plus 0.1mm
\tableofcontents
}

\pagestyle{fancy}

\fancyhead[LE]{Extended Hamiltonian formalism for multicontact field theories} 
\fancyhead[RE]{}
\fancyhead[RO]{M. de León, R. Izquierdo-López, X. Rivas, N. Román-Roy, and A. Villanova} 
\fancyhead[LO]{}    

\fancyfoot[L]{}     
\fancyfoot[C]{\thepage}                  
\fancyfoot[R]{}            

\setlength{\headheight}{17pt}

\renewcommand{\headrulewidth}{0.1pt}  
\renewcommand{\footrulewidth}{0pt}    

\renewcommand{\headrule}{%
    \vspace{3pt}                
    \hrule width\headwidth height 0.4pt 
    \vspace{0pt}                
}

\setlength{\headsep}{30pt}  

\section{Introduction}
\label{sec:intro}
Dissipation, damping and, more generally, non-conservative behaviour are
ubiquitous in physics, but they fall outside the scope of the classical
variational principles, whose Euler--Lagrange equations are
conservative in the autonomous case. A very fruitful way of incorporating these phenomena into a
geometric and variational framework is provided by the {\sl Herglotz
variational principle}~\cite{Herglotz}, in which the Lagrangian depends not
only on the positions and velocities but also on the action itself. In
mechanics, this principle is the variational counterpart of {\sl contact
Hamiltonian dynamics}, which has been intensively developed in recent
years~\cite{BCT_17,LL_19,GGMRR_20a} and has found applications in thermodynamics,
dissipative mechanics, and many other areas \cite{Bra_18,CCM_18,CBG_21}.

The extension of these ideas to classical field theories leads to the so-called
{\sl action-dependent} (or Herglotz-type) field theories \cite{GGB_03}, in which the
Lagrangian density depends on the fields, their derivatives, and an
$(m-1)$-form, the {\sl action density}, whose components $s^\mu$ play the role
of the contact variable of mechanics
\cite{LPAF2018}. A first geometric description of these
theories was given in terms of $k$-contact structures~\cite{GGMRR_20,GGMRR_21}, and a
covariant one was later obtained with the introduction of {\sl multicontact
geometry}~\cite{LGMRR_23}, which is for action-dependent field theories what
multisymplectic geometry is for conservative ones~\cite{CCI_91,GIMM_98,Rom_09}.
A systematic presentation of the Lagrangian and Hamiltonian multicontact
formalisms, together with a wide collection of examples (damped vibrating
strings and membranes, the telegrapher equation, Burgers' equation, Maxwell's
equations with dissipation, \dots) is found in~\cite{LGMRR_25}, and the
algebraic structures associated with multicontact manifolds (brackets and
multisymplectisation) have been studied in~\cite{LIR_25}.

In all these works, the Hamiltonian formalism of action-dependent field theories is constructed starting from a configuration bundle $\pi\colon E\to M$.
If $\bigwedge^m_2\mathrm{T}^*E$ and
$\bigwedge^m_1\mathrm{T}^*E$
denote the bundles of $2$-semibasic $m$-forms and $1$-semibasic $m$-forms in $E$, respectively;
the Hamiltonian formalism is stated on the {\sl restricted multimomentum phase bundle}
$$
\mathcal{P}^* =J^{1*}\pi\times_M\bigwedge\nolimits^{m-1}\Tan^*M \,,
$$
with natural coordinates $(x^\mu,y^A,p^\mu_A,s^\mu)$.
Here, $J^{1*}\pi$ is the quotient $\bigwedge^m_2\mathrm{T}^*E/\bigwedge^m_1\mathrm{T}^*E$,
where $\bigwedge^m_2\mathrm{T}^*E\equiv\mathcal{M}\pi$ is called the {\sl extended multimomentum bundle}
and $J^{1*}\pi$ is then called the {\sl restricted multimomentum bundle}.
The dynamics is
introduced by means of a {\sl Hamiltonian section}
${\bf h}\colon\mathcal{P}^*\to\widetilde{\mathcal{P}}$, locally given by a Hamiltonian function
$h(x^\mu,y^A,p^\mu_A,s^\mu)$, which is used to pull back the canonical
structure with which $\widetilde{\mathcal{P}}$ is endowed, and obtain the Hamiltonian multicontact form $\Theta_{\bf h}$ on $\mathcal{P}^*$.

This is the natural counterpart for action-dependent field theories of the restricted Hamilton--de Donder--Weyl
formalism of multisymplectic field theories. However, as is well known in the
multisymplectic setting~\cite{CCI_91,GIMM_98,Rom_09}, the bundle
$J^{1*}\pi$ does not carry any canonical multisymplectic structure,
but $\mathcal{M}\pi\equiv\bigwedge^m_2\mathrm{T}^*E$ is endowed with the
canonical (Liouville) $m$-form $$
\vartheta=p^\mu_A\,\d y^A\wedge\d^{m-1}x_\mu+p\,\d^mx \ \,
$$
and the role of Hamiltonian sections is precisely to bring this canonical
structure down to $J^{1*}\pi$. The extra coordinate $p$ is an energy-type variable,
the covariant analogue of the momentum conjugate to time in the extended phase space of time-dependent mechanics. 
Therefore, the {\sl extended multisymplectic Hamiltonian formalism} \cite{LEMR_2007,PR_2002},
in which one works directly on $\mathcal{M}\pi$, 
is the most intrinsic and economical way of stating the Hamiltonian field equations.
This formalism is a generalisation to field theories of the {\sl extended Hamiltonian formalism} developed by R. Kuwabara 
\cite{Ku-84} for time-dependent mechanical systems.
There, the Hamiltonian section is replaced
by a constraint $\mathrm{H}=p+{\rm h}=0$ in the spirit of P.A.M. Dirac and P.G. Bergmann \cite{AB_51,Dir_50} and
M.J. Gotay {\it et al}~\cite{GNH_78}, 
and no choice of section is required a priori: the
Hamiltonian is a genuine function or, more generally, a closed $1$-form on a
manifold with a canonical structure.

The aim of this paper is to develop this extended Hamiltonian formalism for
action-dependent field theories, and to show that it is equivalent to the
restricted one. Our phase space is the {\sl extended multimomentum phase
bundle}
$$
\widetilde{\cal P}={\cal M}\pi\times_M\bigwedge\nolimits^{m-1}\Tan^* M \,,
$$
with natural coordinates $(x^\mu,y^A,p^\mu_A,p,s^\mu)$, endowed with the
canonical $m$-form
\[
  \widetilde\Theta = -p^\mu_A\,\d y^A\wedge\d^{m-1}x_\mu - p\,\d^mx
  + \d s^\mu\wedge\d^{m-1}x_\mu\,,
\]
built from the canonical Liouville forms of $\mathcal{M}\pi$ and $\bigwedge\nolimits^{m-1}\Tan^* M$.

It should be stressed that this is not a mere rewriting of the restricted
formalism on a larger manifold. Some features make the extended setting different and more appropriate depending on the scenario. 
For instance, this framework is the natural one to study the
algebraic structures of action-dependent field theories. 
Brackets of Hamiltonian forms on multicontact manifolds are constructed via
the multisymplectisation \cite{LIR_25}; the canonical structure of $\widetilde{\mathcal{P}}$ suggests that the extended formalism
provides the appropriate setting to relate these brackets with the (graded) Poisson brackets of multisymplectic field theories. 
This study is postponed to a forthcoming work.

In particular, the main contributions of this text are the following:
\begin{itemize}\itemsep=-1pt
\item 
A first problem is that the form $\widetilde\Theta$ on the extended multimomentum phase bundle $\widetilde{ \mathcal{P}}$, 
although canonical, does {not} define a special multicontact structure in the sense of~\cite{LGMRR_23}. 
As a consequence, $(\widetilde{\mathcal{P}},\widetilde\Theta)$ is not covered by the notions of multicontact manifold
 available in the literature. The objective of Section \ref{mmb} is to adopt a more general
definition, inspired by the correspondence between contact manifolds and
homogeneous symplectic manifolds~\cite{GG_22}, and based on \cite{LIR_25}: an $m$-form is
{\sl multicontact} when its canonical multisymplectisation is
multisymplectic. This definition extends the previous ones
(see~\cite{LGMRR_25,LIR_25}) and includes both the extended and restricted structures. 
In the extended multimomentum phase bundle, it is canonical and recovers the known notion; 
but, in the restricted multimomentum phase bundle, it is not canonical and requires the choice of a Hamiltonian section.
A definition of premulticontact form is also included, in order to deal with singular theories in the sequel. 
We focus on generalising the notion of dissipation $1$-form (an essential object to define the dynamics) to include every case at the same time (extended, restricted, and singular). 

\item In Section \ref{ehs}, after recalling the theory of restricted multicontact systems, we focus on introducing extended Hamiltonian multicontact system $(\widetilde{\mathcal{P}},\widetilde\Theta;\mathcal{H})$. 
Instead of by a Hamiltonian section, these are determined by giving a {\sl Hamiltonian density} $\mathcal{H}$, which allows us to define an appropriate
exact $1$-form $\alpha\in\Omega^1(\widetilde{\mathcal{P}})$ which is the differential of the Hamiltonian function
${\rm H}=p+{\rm h}$ associated with the Hamiltonian density.
In addition, extended Hamiltonian systems defined by non-closed forms are also studied,
which provide a natural geometric description of external sources.
\item The local field equations on $\widetilde{\mathcal{P}}$ are given in Theorem \ref{holsecreg3}. These, besides recovering the Herglotz--Hamilton--de Donder--Weyl equations, include an equation 
which gives account of a geometric property: the tangency of the solutions to the level sets of the Hamiltonian function ${\rm H}$.
We would like to remark that this last equation has no counterpart in the restricted formalism, but this property allows us to establish the relationship between the restricted and extended Hamiltonian formalisms.
\item In Section \ref{arhs} we adapt the theory to include almost-regular Lagrangians. Now, the
restricted formalism must be stated in a submanifold $\mathcal{P}^*_\circ\subset\mathcal{P}^*$, which is usually given by the image of the Legendre transformation (see \cite{LGMRR_23}).
Under certain conditions, which are analysed here, a premulticontact structure is available, and a constraint
algorithm is needed. In the extended setting, the primary constraints coming
from the Legendre map and the Hamiltonian constraint $p=-h_\circ$ are treated
on the same footing, on a submanifold $\widetilde{\mathcal{P}}_\circ\subset\widetilde{\mathcal{P}}$.
\end{itemize}

The paper concludes with Sections \ref{exs} and \ref{conclusions}, where we give some examples illustrating both the regular and singular cases, and draw some conclusions.

Throughout the paper, all manifolds and maps are smooth, and the
summation convention over repeated indices is used. 
If $M$ is an $m$-dimensional
orientable manifold with a fixed volume form $\omega_{M}\equiv\d^mx$,
making a slight abuse of notation, the pull-backs of $\omega_{_M}$ to all
bundles over $M$ are denoted by $\omega$. 
We also denote $\ds\d^{m-1}x_\mu=\iota\Big(\parder{}{x^\mu}\Big)\d^mx$,
and $\d^{m-2}x_{\mu\nu}$ is defined analogously.

 \section{Preliminaries}
 \protect\label{mmb}

We start by stating and reviewing the foundations of multicontact geometry on which the geometric formulation of action-dependent field theories is based.

\subsection{Multicontact structures}
\protect\label{msms}

In this section, we define multicontact manifolds and study the  conditions under which, when in the presence of a
fibration, a dissipation form can be defined. Our approach is inspired by the equivalence between contact and homogeneous symplectic manifolds \cite{BGG_17,GG_22}.
The comparison with other notions present in the literature \cite{LGMRR_23,LIR_25,GRR_26,RRV_26,Vit_15} is made in Remark \ref{remark:comparison_geometries}.

\begin{definition}
\label{genmulticontact}
Let $\mathcal{M}$ be a differentiable manifold. 
An $m$-form $\Omega\in\df^m(\mathcal{M})$, with $m\leq\dim\mathcal{M}$, is a \dfn{premultisymplectic form} if it is closed, that is,
$\d\Omega = 0$.
In this case, $(\mathcal{M},\Omega)$ is a \dfn{premultisymplectic manifold}. 
In addition, if the induced map
\begin{align}
\flat_\Omega\colon\mathrm{T}\mathcal{M} & \longrightarrow \bigwedge\nolimits^{m-1} \mathrm{T}^*\mathcal{M} \\
v & \longmapsto \ \innp{v}\Omega
\end{align}
is a monomorphism, then $\Omega$ is a \dfn{multisymplectic form}, and $(\mathcal{M},\Omega)$ is a \dfn{multisymplectic manifold}.
(This is equivalent to $\Omega$ be $1$-nondegenerate).
\end{definition}

Since contact forms can be
symplectised, we introduce the following definition, that encompasses previous notions in the literature.

\begin{definition}
\label{MCasMSYMP}
Let $\mathcal{M}$ be a differentiable manifold and an $m$-form $\Theta \in \Omega^m(\mathcal{M})$. 
The canonical \dfn{premultisymplectisation} of $(\mathcal{M}, \Theta)$ is a pair $(\widetilde{\mathcal{M}}, \Omega)$ such that:
\begin{enumerate}[{\rm (i)}]\itemsep=-1pt
\item 
$\widetilde{\mathcal{M}} =\mathcal{M}\times\mathbb{R}_\times$,
where $\mathbb{R}_\times=\mathbb{R}\setminus\{ 0\}$,
\item
$\Omega = -\mathrm{d}\Gamma$,
where $\Gamma = z \cdot \Theta$, 
and $z$ is the global coordinate in $\mathbb{R}_\times$,
\item
$(\widetilde{\mathcal{M}},\Omega)$
is a premultisymplectic manifold.
\end{enumerate}
\end{definition}

\begin{definition}
\label{multicontact}
An $m$-form $\Theta \in \Omega^m(\mathcal{M})$ is a \dfn{multicontact form} if its canonical premultisymplectisation is a multisymplectisation;
that is, $(\widetilde{\mathcal{M}},\Omega)$
is a multisymplectic manifold.
Then, $(\mathcal{M},\Theta)$ is said to be a \dfn{multicontact manifold}.
\end{definition}

\begin{remark}
\label{remark:comparison_geometries}
We would like to remark that there are several notions of multicontact manifold in the literature, and it is still an evolving subject. In this text we adopt the above definition as it incorporates both extended and restricted multicontact field theories. It is important to note that this definition is strictly more
general:
\begin{enumerate}[\rm (i)]\itemsep=-1pt
\item 
The definition given in \cite{LIR_25,GRR_26,RRV_26}, where a multicontact form is a differential form $\Theta \in \Omega^m (\mathcal{M})$ such that $\ker \Theta \cap \ker \d \Theta = \{0\}$ and $\ker \d \Theta \neq \{0\}$. 
Notice that this implies that the premultisymplectisation $(\mathcal{M}\times\mathbb{R}_\times, - \d \Gamma)$ is, in fact, a multisymplectic manifold in the sense of Definition \ref{multicontact}.
\item 
The definition introduced in \cite{LGMRR_23}, which falls under the definition given here (see Section \ref{msms2}).
\item 
The higher-dimensional version of contact distributions proposed in \cite{Vit_15},
where a distribution $D\subset\Tan\mathcal{M}$ of codimension $m$ is called {\sl multicontact} if it is maximally non-integrable. 
If $D = \ker \Theta$, where $\Theta$ is a nowhere vanishing locally decomposable $m$-form, the distribution $D$ is maximally non-integrable if, and only if, $\Theta$ is multicontact, in the sense of Definition \ref{multicontact}.
(See \cite{GRR_26} for the relation with other definitions).
\end{enumerate}
In particular, as we will see below, the conditions of Definition \ref{MCasMSYMP} allow us to introduce the structures necessary to set the equations of action-dependent field theories in a generic way.
\end{remark}

First, bearing in mind the notions and notation about multivector fields and their operations given in the Appendix \ref{mvfc}, if $\alpha\in\df^k(\mathcal{M})$, we denote
$$
\ker_m\alpha\coloneqq\{ {\bf X}\in\vf^m(\mathcal{M})\, \vert\, \innp{\bf X}\alpha=0;\ m\leq k\} \, ,
$$
and, as usual, we write simply $\ker\alpha$ for $\ker_1\alpha$.

\begin{definition} 
\label{genpremulticontact}
A pair $(\mathcal{M},\Theta)$, where $\mathcal{M}$ is a smooth manifold and $\Theta \in \Omega^m(\mathcal{M})$ is called a \dfn{premulticontact manifold} if:
\begin{enumerate}[{\rm (i)}]\itemsep=-1pt
\item \label{item:genpremulticontact:1}
The form $\Theta$ is nowhere vanishing. 
\item \label{item:genpremulticontact:2}
The distribution $\ker\Theta\cap\ker\d\Theta$ has constant rank.
\item \label{item:genpremulticontact:3}
Either $\Theta \in \mathrm{Im}\,{\flat_{\d \Theta}}$ at every point, or $\Theta \not \in \mathrm{Im}\,{\flat_{\d \Theta}}$ at every point.
\end{enumerate}
\end{definition}

\begin{remark}
\label{remark:multilinear_algebra}
Notice that condition (\ref{item:genpremulticontact:3}) can be written in another equivalent way. In fact, recall that the following statements are equivalent for a couple of forms $\beta \in \bigwedge^m V^\ast$ and $\gamma \in \bigwedge^{m +1} V^\ast$ over a real vector space $V$:
\begin{enumerate}[\rm (i)]\itemsep=-1pt
\item There exists a vector $v \in V$ satisfying $\innp{v} \gamma = \beta$.
\item There exists a vector $v \in \ker_1 \beta$ satisfying $\innp{v} \gamma = \beta$
\item For every ${\bm U} \in \bigwedge^m V$ such that $\innp{\bm U} \gamma = 0$, we obtain $\innp{\bm U} \beta = 0$. 
\item We have $\ker_m \gamma \subseteq \ker_m \beta$.
\end{enumerate}
In particular, applied to Definition \ref{genpremulticontact} we have that condition (\ref{item:genpremulticontact:3}) is equivalent to either $\ker_m \d\Theta \subseteq \ker_m \Theta$ at every point, or $\ker_m \d \Theta \not \subset \ker_m \Theta$ at every point. We will use these conditions indistinctively. 
\end{remark}

We have the following characterisation:

\begin{proposition}
\label{equivalenceMC}
Let $\Theta \in \Omega^m(\mathcal{M})$ be an $m$-form. Then, the following assertions are equivalent:
\begin{enumerate}[{\rm (i)}]\itemsep=-1pt
\item The form $\Theta$ is a multicontact form.
\item 
The form $\Theta\notin\flat_{\mathrm{d}\Theta}(\ker\Theta)$ and
$\ker\Theta\cap\ker\d\Theta=\{ 0\}$.
\item The following pairing is nondegenerate:
\begin{align}
\ker \Theta \otimes \ker_m \Theta & \longrightarrow \bigwedge\nolimits^{m} \mathrm{T}\mathcal{M}\big /\ker_m \Theta \\
v \otimes{\bf U} & \longmapsto [v,{\bf U}] + \ker_m \Theta
\end{align}
\end{enumerate}
\end{proposition}
\begin{proof}
The multisymplectic form $\Omega = - \d (z \cdot \Theta)$ reads locally as 
$$
\Omega = - \d z \wedge \Theta - z\ \d \Theta \,.
$$
Let us first check that, if $\Omega$ is nondegenerate, then {$\flat_{\d\Theta}\vert_{\ker\Theta}$} is injective, and $\Theta \not \in \mathrm{Im}\, (\flat_{\d \Theta})$. 
First, suppose that there are $v \in \ker\Theta$ such that $\innp{v}\d\Theta = 0$. 
Identifying $\T(\mathcal{M} \times \mathbb{R}_\times) = \T\mathcal{M} \oplus \mathbb{R}$ 
we have $v \in \ker \Omega$, so that, since it is nondegenerate, we have $v = 0$. 
Now, by contradiction, assume that $\Theta = \innp{v} \d \Theta$, for some $v \in \ker \Theta$. 
Then, the vector $\ds z\pdv{z} - v \in \ker \Omega$, which we know is trivial. Hence, $\Theta \not \in \mathrm{Im}\, (\flat_{\d \Theta})$.

Now, let us check that if $\flat_{\d \Theta}$ is injective and $\Theta \not \in \mathrm{Im}\, (\flat_{\d \Theta})$, then the pairing
$$
\ker \Theta \otimes \ker_m \Theta \rightarrow \bigwedge\nolimits^m \T\mathcal{M} \big / \ker_m \Theta
$$
is nondegenerate. Suppose that there are $X \in \mathfrak{X}(\mathcal{M})$ {with $\innp{X} \Theta = 0$} and such that $\innp{[X,{\bf U}]} \Theta = 0$, for every ${\bf U} \in \ker_m \Theta$. Then, using Cartan calculus, we get:
\[
0 = \innp{[X,{\bf U}]} \Theta = \Lie_{X}(\innp{{\bf U}} \Theta) - \innp{{\bf U}} \Lie_X \Theta =  - \innp{{\bf U}} \innp{X} \d \Theta\,,
\]
for every ${\bf U} \in \ker_m \Theta$. In particular, $\innp{X} \d \Theta$ must be a multiple of $\Theta$. The hypotheses imply that we necessarily have $\innp{X} \d \Theta = 0$, so that $X = 0$.

Finally, let us check that if the above pairing is nondegenerate, then $\Theta$ is a multicontact form. 
Indeed, Suppose that $\ds v + a \pdv{z}$ satisfies 
\[
\innp{v+a\pdv{z}}\Omega = \d z \wedge\innp{v}\Theta-z\,\innp{v}\d \Theta-a\,\Theta = 0\,.
\]
This splits into two conditions: $\innp{v}\Theta=0$ and $\ds\innp{v} \d\Theta=-\frac{a}{z}\,\Theta$. 
In particular, $v$ satisfies $[v,{\bf U}]\in \ker_m \Theta$, for every ${\bf U} \in \ker_m \Theta$, 
which implies $v = 0$ and therefore $a = 0$.
\end{proof}

Similarly, for the premulticontact case, we have:

\begin{proposition} 
A form $\Theta \in \Omega^m (\mathcal{M})$ is premulticontact if, and only if, its premultisymplectisation $(\mathcal{M} \times \mathbb{R}_\times, \Omega = - \d \Gamma)$ satisfies that $\ker\Omega\cap\ker\Gamma$ has constant rank.
\end{proposition}
\begin{proof}
From the above observation, it follows  that 
\[
\ker\Omega\cap\ker\Gamma = \left\{ \left(a\pdv{z}+v\right)\ \Big\vert\ \innp{v}\d\Theta = -\frac{a}{z}\, \Theta \ 
{\text{and} \ v \in \ker \Theta}\right\}\,,
\]
and the conclusion holds.
\end{proof}

\begin{remark} 
Note that asking for either $\ker_m \d \Theta \subset \ker_m \Theta$ or $\ker_m \d \Theta \not \subset \ker_m \Theta$ at every point in Definition \ref{genpremulticontact} is necessary.
Indeed if we were to weaken the conditions not requiring it, the differential form $\Theta = x_1 \d x_2 \wedge \d x_3 + (1 + x_1^2) \d x_4\wedge \d x_5$ provides a counterexample of the above result. 
\end{remark}

\subsection{The dissipation $1$-form}
In order to write equations for field theories, let us consider the following particular situation:

\begin{definition}
\label{fibered}
Let $(\mathcal{M},\Theta)$ be a (pre)multicontact manifold such that $\varkappa\colon \mathcal{M}\to M$ is a surjective submersion onto an $m$-dimensional manifold $M$ (usually an orientable manifold with volume form $\omega_{_M}\in\df^m(M)$,
and $\omega=\varkappa^\ast \omega_{_M}$),
and then $\dim\mathcal{M}=m+N$.
In this case, we say that $(\mathcal{M},\Theta)$ is a \dfn{fibred (pre)multicontact manifold}.
\end{definition}

\begin{remark}
\label{remequiv}
It is important to point out that a fibred (pre)multicontact manifold is equivalent to a (pre)multicontact manifold $(\mathcal{M},\Theta)$,
endowed with another differential form $\omega\in\df^m(\mathcal{M})$; so $\dim\mathcal{M}=m+N$, $N\geq m\geq 1$.
In fact, assuming that the quotient $M=\mathcal{M}\big / \ker\omega$ is a differentiable manifold, 
we have a natural surjective submersion $\varkappa\colon \mathcal{M}\to M$ and,
if $\omega$ is closed, it is $\varkappa$-projectable to a form $\omega_{_M}\in\Omega^m(M)$, which defines a volume form in $M$.
Furthermore, denoting by $V(\varkappa)$ the $\varkappa$-vertical distribution and by $\mathfrak{X}^{V(\varkappa)}(\mathcal{M})$ the module of $\varkappa$-vertical vector fields, we have
$\ker\omega=\mathfrak{X}^{V(\varkappa)}(\mathcal{M})$.
\end{remark}

\begin{definition}
\label{Reebdist}
The \dfn{Reeb distribution} associated with $\Theta$ is defined as
\begin{equation}
\label{multiReeb}
\mathfrak{R} \coloneqq \{v \in  \ker \omega \mid \innp{v} \d\Theta\ \text{ is a $\varkappa$-semibasic form}\} \,. 
\end{equation}
The set of sections of the Reeb distribution is also denoted by $\mathfrak{R}$,
and its elements $R\in\mathfrak{R}$ are called \dfn{Reeb vector fields}.
\end{definition}

From this distribution, we define the associated Reeb tensor field.
First, notice that the map
\begin{align}
\flat_\Theta\colon\mathfrak{R} & \longrightarrow \bigwedge\nolimits^{m-1} \mathrm{T}^*\mathcal{M} \\
R & \longmapsto \innp{R} \Theta
\end{align}
is an epimorphism onto its image, $\mathcal{Z} \coloneqq \mathrm{Im}\, \flat_\Theta \subset \bigwedge^{m-1} \mathrm{T}^*\mathcal{M}$. Hence, by the first isomorphism theorem, the map $\widetilde\flat_\Theta \colon \mathfrak{R}\big /\ker \flat_\Theta \rightarrow  \mathcal{Z}$ is an isomorphism. Let us consider its inverse 
\[
\widetilde\flat_{\Theta}^{-1} \colon \mathcal{Z} \rightarrow\mathfrak{R}\big / \ker \flat_\Theta \,. 
\]
Then, $\widetilde\flat_\Theta^{-1}$ can be regarded as an element
\[
\widetilde\flat_{\Theta}^{-1} \in \Hom \left( \mathcal{Z},\mathfrak{R}\big / \ker \flat_\Theta \right) \cong  \left(\mathfrak{R}\big / \ker \flat_\Theta \right) \otimes \mathcal{Z}^\ast\,.
\]
Finally, let us notice that for any subspace of forms $\mathcal{Z} \subset \bigwedge^{m-1} \cT \mathcal{M}$, the dual $\mathcal{Z}^\ast$ can be canonically identified with the quotient $\bigwedge^{m-1} \T \mathcal{M} \big / \mathcal{Z}^\circ$, where 
\[
\mathcal{Z}^\circ = \Big\{ U \in \bigwedge\nolimits^{m-1} \T \mathcal{M} \colon \innp{U} \alpha = 0 \,, \text{for all }\alpha \in \mathcal{Z}\Big\}\,.
\]
In particular, we may consider $\widetilde\flat_\Theta^{-1} \in \left(\mathfrak{R}\big /\ker \flat_\Theta\right) \otimes \left( \bigwedge^{m-1} \T \mathcal{M} \big / \mathcal{Z}^\circ \right)$.
\begin{remark}
Notice that if $\flat_\Theta \colon \mathfrak{R}\rightarrow {\mathcal{Z}}$ is an isomorphism, then $\ker \flat_\Theta = \{0\}$, so that $\widetilde\flat_\Theta^{-1} \in \mathfrak{R}\otimes \left(\bigwedge^{m-1} \T\mathcal{M} \big / \mathcal{Z}^\circ\right)$ by the above construction.
\end{remark}

\begin{definition}
\label{Reebmvf}
The \dfn{Reeb tensor field} is defined as 
$$
\mathcal{R} \coloneqq \widetilde\flat_\Theta^{-1} \in  \left(\mathfrak{R}\big /\ker \flat_\Theta\right) \otimes \left( \bigwedge\nolimits^{m-1} \T \mathcal{M} \big / \mathcal{Z}^\circ \right)\,. 
$$
\end{definition}

The contraction of $\mathcal{R}$ with differential forms is well-defined only when it is independent of the representative. 
When this contraction makes sense, we denote it as usual by $\innp{\mathcal{R}}$.

Finally, consider the case where $\varkappa\colon \mathcal{M} \to M$ is a surjective submersion onto an orientable $m$-dimensional manifold $M$;
let $\omega_{_M}\in\df^m(M)$ be the volume form on $M$, and $\omega=\varkappa^\ast \omega_{_M}$. Following \cite{LGMRR_23},
for every $k\in\mathbb{N}$, we define
the set 
\[
\mathcal{A}^k(\mathcal{D})\coloneqq\big\{ \alpha\in\df^k(\mathcal{M})\mid
\innp{Z}\alpha=0 \,,\ \text{ for every}\ Z\in\Gamma({\cal D})\big\} \,.
\]
Then we have:

\begin{proposition}
\label{prop:existence_sigma}
 Suppose that $\varkappa \colon \mathcal{M}\rightarrow M$ is a surjective submersion, and that $\mathcal{A}^{m-1}(\ker\omega) \subseteq \Ima \flat_{\Theta}$.
Let $\alpha \in \Omega^1(\mathcal{M})$, such that for every $R \in \ker \flat_\Theta$ we have 
\begin{equation}
\label{condi}
\innp{R} \d \Theta = -\innp{R} (\alpha \wedge \omega) \, .    
\end{equation}
Then, considering the Reeb tensor field $\mathcal{R}$, the $1$-form
\[
\sigma_\alpha \coloneqq {(-1)^{m-1}} \innp{\mathcal{R}} (\d \Theta + \alpha \wedge \omega)\,,
\]
is well-defined and is called a \dfn{dissipation form}.
\end{proposition}
\begin{proof} It is enough to observe that the condition in the hypotheses implies that $\d \Theta + \alpha \wedge \omega$ is annihilated by $\ker \flat_ \Theta \otimes \bigwedge^{m-1} \T \mathcal{M}$ and that for every Reeb vector field $R \in \Gamma(\mathfrak{R})$ and every ${\bm U} \in \Gamma(\mathcal{Z}^\circ)$, we have 
\[
\innp{R \otimes {\bm U}} (\d \Theta + \alpha \wedge \omega) = \innp{\bm U} \left( \innp{R} \d \Theta + \alpha(R) \omega \right) = 0\,,
\]
as the form on the right-hand side is semibasic and is annihilated by ${\bm U}$ (by definition of $\mathcal{Z}^\circ$).
\end{proof}

\begin{remark} 
Note that, when satisfying the above conditions, the choice $\alpha = 0$ allows us to build a canonical dissipation form which is denoted $\sigma_\Theta = \innp{\mathcal{R}}\d \Theta$. 
For instance, this is the case where $\flat_\Theta$ defines a linear isomorphism, since in this particular case the Reeb ``multivector field'' is truly a multivector field. However, notice that there are scenarios where the choice of $\alpha$ is necessary (see Section \ref{subsec:extended-Hamiltonian}).
\end{remark}

\begin{remark} Notice that in the case where $\ker \flat_\Theta = \{0\}$, any $\alpha \in \Omega^1(\mathcal{M})$ works; in particular, $\alpha = 0$. 
In this last case, we denote the associated $1$-form by 
$$\sigma_\Theta  = \innp{\mathcal{R}} \d \Theta \, .$$
\end{remark}

Finally, associated with a dissipation form $\sigma_\alpha$, we define:

\begin{definition}\label{def:bard}
\label{bard}
Let $\sigma_{\alpha}\in\df^1(\mathcal{M})$ be a dissipation form. 
The \dfn{Lichnerowicz differential operator} associated with $\sigma_\alpha$ is defined by
\begin{align*}
\bd\colon\df^k(\mathcal{M})&\longrightarrow\df^{k+1}(\mathcal{M})
\\
\beta&\longmapsto\bd\beta=\d \beta+\sigma_\alpha\wedge\beta\,.
\end{align*}
\end{definition}

\begin{remark}
The operator $\bd$ satisfies $\bd^2=\bd\circ\bd=0$ if, and only if, $\d\sigma_\alpha=0$.
In this case, the operator $\bd$ defines a {\sl Lichnerowicz--Jacobi cohomology} \cite{LLMP_03}, and $(\Theta,\omega)$ is called a \dfn{closed multicontact structure}. 
\end{remark}

\subsection{Special multicontact structures}
\protect\label{msms2}

The definition of multicontact structure given in the above section is very general.
For instance, the existence of adapted or Darboux coordinates as well as other useful properties is not guaranteed
for these multicontact forms, unless additional conditions are imposed
(as is also the case for multisymplectic structures \cite{LMS_03,Mar_88,Ryv_26}).
Furthermore, multicontact forms appearing in the standard Lagrangian and Hamiltonian formalisms of action-dependent field theories are a special kind of multicontact structures.
Taking all this into account leads to the definition of multicontact structure introduced in \cite{LGMRR_23,LGMRR_25},
which is then recalled as a special case of the more general definition previously established in Definitions \ref{genmulticontact} and \ref{genpremulticontact}.

In the cases of interest to us, we will be in a particular situation in which
$(\mathcal{M},\Theta)$ is a fibred (pre)multicontact manifold, with 
$\varkappa\colon \mathcal{M}\to M$,
and there is some compatibility between the multicontact form $\Theta$ and $\varkappa$.
This compatibility is studied in this section. 

Thus, as a more general framework, let $\mathcal{M}$ be a differentiable manifold and $\Theta,\omega\in\df^m(\mathcal{M})$,
so $\dim\mathcal{M}=m+N$, $N\geq m\geq 1$.
For every regular distribution $\mathcal{D}\subset\Tan \mathcal{M}$,
we denote by $\Gamma(\mathcal{D})$ the $\Cinfty(\mathcal{M})$-module of sections of $\mathcal{D}$.
Now, the Reeb distribution associated with
$(\Theta,\omega)$ is the distribution
$\mathfrak{R}\subset\Tan \mathcal{M}$ may be written as
$$
\mathfrak{R}\coloneqq\big\{ R\in\Gamma(\ker\omega)\mid \innp{R}\d\Theta\in \mathcal{A}^m(\ker\omega)\big\} \,.
$$
Note that $\ker\omega\cap\ker\d\Theta\subset\mathfrak{R}$.
Furthermore, if $\omega\in\df^m(\mathcal{M})$ has constant rank and is a closed form, then $\mathfrak{R}$ is involutive \cite{LGMRR_23},
and this is assumed from now.

\begin{remark}
Observe that, in the case where $\varkappa\colon \mathcal{M} \to M$ is a surjective submersion onto an orientable $m$-dimensional manifold $M$, with volume form $\omega_{_M}\in\df^m(M)$ and $\omega=\varkappa^\ast \omega_{_M}$,
then ${\cal A}^k(\ker\omega)$ coincides with the bundle of $\varkappa$-semibasic $k$-forms, and we recover the Definition \ref{Reebdist}.
In this case, vector fields $X\in\ker\omega$ are also called \dfn{$\omega$-vertical}.
\end{remark}

Therefore, from the involutive distributions $\ker\omega$, $\mathfrak{R}$, and $\mathcal{C}=\ker\omega\cap\ker\Theta\cap\ker\d\Theta$, 
we obtain charts of local coordinates in $\mathcal{M}$, adapted to the geometric structure, as follows \cite{LGMRR_23}:

\begin{theorem}
\label{thm:coordenadas}
Let $\mathcal{M}$ be a differentiable manifold and $\Theta,\omega\in\df^m(\mathcal{M})$,
where $\omega$ is a closed form, such that the following conditions hold
\begin{enumerate}[{\rm (i)}]\itemsep=-1pt
\item \label{item:thm:coordenadas:1} {$\rank\ker\omega = N$},
where $0 \leq k \leq N - m$ is fixed.
\item \label{item:thm:coordenadas:2} $\rank\mathfrak{R} = m + k$,
\item \label{item:thm:coordenadas:3} $\rank(\ker\omega\cap\ker\Theta\cap\ker\d\Theta)=k$,
\item \label{item:thm:coordenadas:4} ${\cal A}^{m-1}(\ker\omega)=\{\innp{R}\Theta\mid R\in\mathfrak{R}\}$,
\end{enumerate}
Then, around every point ${\rm p}\in\mathcal{M}$, there exist a local chart of adapted coordinates
$(U;x^\mu,u^I,s^\mu,w^r)$; $\mu=1,\dotsc,m$; $I=1,\dotsc,N-m-k$; $r=1,\dotsc,k$;
such that
$$
\ker\omega\vert_U=
\Big\langle\frac{\partial}{\partial u^I},
\frac{\partial}{\partial s^\mu},
\frac{\partial}{\partial w^r}\Big\rangle \,,\quad
\mathfrak{R}\vert_U=
\Big\langle\frac{\partial}{\partial s^\mu},
\frac{\partial}{\partial w^r}\Big\rangle \,,\quad
\left(\ker\omega\cap\ker\Theta\cap\ker\d\Theta\right)\vert_U=
\Big\langle\frac{\partial}{\partial w^r}\Big\rangle\,,
$$
satisfying
\[
\omega\vert_U=\d x^1\wedge\dots\wedge\d x^m\equiv\d^m x\, , \quad \innp{\pdv{s^\mu}} \Theta = \d^{m - 1} x_\mu \,.
\]
(For $k=0$, $\mathcal{C}=\{0\}$, and no coordinates $w^r$ appear.)
\end{theorem}

In addition, if the conditions in Theorem \ref{thm:coordenadas} hold,
we can consider the quotient spaces defined by the involutive distributions $\ker\omega$ and $\mathfrak{R}\subset\ker\omega$, namely
$M \equiv \mathcal{M}/\ker\omega$ and $\mathcal{Q} \equiv \mathcal{M}/\mathfrak{R}$,
which are assumed to be differentiable manifolds. Then, we have the natural projections
\[
\begin{array}{ccccc}
\varkappa\colon & \mathcal{M} &\longrightarrow & M \\
& (x^\mu,u^I,s^\mu,w^r) &\longmapsto & (x^\mu) \,, \\[0.3cm]
\varsigma\colon & \mathcal{M} &\longrightarrow & \mathcal{Q} \\
& (x^\mu,u^I,s^\mu,w^r) &\longmapsto & (x^\mu,u^I) \,, \\[0.3cm]
\varepsilon\colon & \mathcal{Q} &\longrightarrow & M \\
& (x^\mu,u^I) &\longmapsto & (x^\mu) \, .
\end{array}
\]
Moreover, since $\omega$ is closed, it is $\varkappa$-projectable to a form $\omega_{_M}\in\Omega^m(M)$, which defines a volume form in $M$. Hence, we conclude that:

\begin{proposition}
Let $(\mathcal{M},\Theta,\omega)$, where $\omega$ is a closed form, be such that the conditions of Theorem \ref{thm:coordenadas} hold.
Then, there are two manifolds $\mathcal{Q}$ and $M$ such that:
\begin{enumerate}[{\rm (i)}]\itemsep=-1pt
\item 
$\mathcal{M}$ is fibred over $\mathcal{Q}$, $\varsigma\colon \mathcal{M}\to\mathcal{Q}$, and 
$\mathcal{Q}$ is fibred over $M$, $\varepsilon\colon \mathcal{Q}\to M$.
\item
$M$ is an orientable manifold with volume form $\omega_{_M}$ and $\omega=\varkappa^*\omega_{_M}$,
where $\varkappa=\varepsilon\circ\varsigma$.
\end{enumerate}
\end{proposition}

\begin{definition}
Let $(\mathcal{M},\Theta)$ be a fibred (pre)multicontact manifold and let $\omega \in\df^m(\mathcal{M})$
be a closed locally decomposable form, with $m \geq 2$.

\noindent The tuple $(\mathcal{M},\Theta,\omega)$ is a \dfn{special premulticontact manifold} if the map
$\ds\flat_\Theta \colon \mathfrak{R} \to\mathcal{A}^{m-1}(\ker\omega)$
is a surjective submersion satisfying that:
\begin{enumerate}[{\rm (i)}]\itemsep=-1pt
\item
The following distributions coincide:
$\ker \flat_\Theta = \mathfrak{R} \cap \ker\Theta \cap \ker\d \Theta$.
\item
The distribution $\ker\Theta \cap \ker\d \Theta$ has constant rank.
\end{enumerate}

\noindent In particular, $(\mathcal{M},\Theta,\omega)$ is said to be a \dfn{special multicontact manifold} when $\flat_\Theta$
is a vector bundle isomorphism.

\noindent Then $\Theta\in\Omega^m(\mathcal{M})$ is called a \dfn{special (pre)multicontact form} on $\mathcal{M}$. 
\end{definition}

\begin{remark}
In action-dependent classical field theories, the relevant manifolds are of the form
$\mathcal{M}=\mathcal{Q}\times_M\bigwedge^{m-1}(\T^*M)$,
where $\mathcal{Q}\to M$ is a (pre)multisymplectic bundle, typically a jet bundle or a bundle of forms. 
Moreover, as we have the identification $\ker\omega = \mathfrak{X}^{V(\varkappa)}(\mathcal{M})$,
condition (\ref{item:thm:coordenadas:1}) in Theorem~\ref{thm:coordenadas} is automatically satisfied. 
In this case, $(\Theta,\omega)$ is said to be a \dfn{(pre)multicontact bundle structure}, 
and $(\mathcal{M},\Theta,\omega)$ is called a \dfn{(pre)multicontact bundle}.
\end{remark}

The above definition recovers the notion of special (pre)multicontact manifold
given in \cite{LGMRR_23,LGMRR_25},
which is characterised by the conditions stated in Theorem \ref{thm:coordenadas}, since:

\begin{theorem}
\label{multicontactdef}
Let $(\mathcal{M},\Theta)$ be a (pre)multicontact manifold, where $\omega \in\df^m(\mathcal{M})$ 
is a closed form,
$\dim\,M=m$, and $\dim\,\mathcal{M}=m+N$,
with $m \geq 2$.
Then $(\mathcal{M},\Theta,\omega)$ is a special (pre)multicontact manifold if, and only if, the conditions in Theorem  \ref{thm:coordenadas} hold.

In this case, the distribution $\mathcal{C} = \ker\omega\cap\ker\Theta\cap\ker\d\Theta$
is called the \dfn{characteristic distribution} of the special (pre)multicontact manifold $(\mathcal{M},\Theta,\omega)$.
\end{theorem}
\begin{proof}\,
[$\Longleftarrow$]\quad 
Condition (\ref{item:genpremulticontact:1}) in Definition \ref{genpremulticontact} follows from Theorem \ref{thm:coordenadas}, as the local property $\ds\iota\Big(\pdv{s^\mu}\Big) \Theta =  \d^{m - 1} x_\mu$ implies that $\Theta$ is nowhere vanishing. Condition (\ref{item:genpremulticontact:2}) in Definition \ref{genpremulticontact} follows from the equality 
$\ker \omega \cap \ker \Theta \cap \ker \d \Theta = \ker \Theta \cap \ker \d \Theta$,
as $\ker \Theta \cap \ker \d \Theta \subseteq \ker \omega$, which follows by the hypothesis on $m \geq 2$ and, again, the equality $\ds\iota\Big(\pdv{s^\mu}\Big)\Theta =  \d^{m - 1} x_\mu$. 

Finally, we show that $\Theta \not \in \mathrm{Im}\, \flat_{\d \Theta}$ everywhere. 
Using the coordinates from Theorem \ref{thm:coordenadas}, we show that if $\innp{X} \d \Theta = a \Theta$, for $\ds X \in \mathfrak{X}(\mathcal{M})$ and $a \in \mathbb{R}$, then we necessarily have $a = 0$. 
Indeed, simply counting terms involving $\d s^\mu$ and $dx^\mu$, it is easy to show that if $a \neq 0$, we would have $\ds X \in \left\langle \pdv{u^I}\right\rangle$. 
With another change of coordinates, we may assume $\ds X = \pdv{u^1}$. 
Contracting again by $X$, we obtain $\ds\innp{\pdv{u^1}} \Theta = 0$, 
and we have that $\Theta$ is semibasic with respect to the projection $(x^\mu, s^\mu, u^1, u^I) \mapsto (s^\mu, u^1)$. 
In particular, $\d\Theta$ is at most $1$-vertical, so $\ds\innp{\pdv{u^1}} \d\Theta$ cannot involve terms with $\d s^\mu$, arriving at a contradiction, and finishing this proof.

[$\Longrightarrow$]\quad 
Set $\dim \mathcal{M} = N + m$ and $k=\rank \left(\ker \Theta \cap \ker \d \Theta \right)$. Then, condition (\ref{item:thm:coordenadas:1}) of Theorem \ref{thm:coordenadas} follows from the requirement on $\omega$ being locally decomposable. For condition (\ref{item:thm:coordenadas:2}), by the rank-nullity Theorem, it is enough to prove \[
m = \rank \mathrm{Im}\, \flat_\Theta \quad \text{and} \quad \rank \ker \flat_{\Theta} =  k\,.
\]
The first equality follows from $\flat_{\Theta} \colon \mathfrak{R} \rightarrow \mathcal{A}^{m - 1}(\ker\omega)$ being surjective, as $\rank \mathcal{A}^{m - 1}(\ker\omega) = m$. The second follows from the requirement $\ker \flat_\Theta = \mathfrak{R} \cap \ker\Theta \cap \ker\d \Theta$. Finally conditions (\ref{item:thm:coordenadas:3}) and (\ref{item:thm:coordenadas:4}) simply follows by the definition of $k$ and by the requirement on $\flat_\Theta$ to be surjective.
\end{proof}

\begin{remark} The condition $m\geq 2$ is compulsory. Indeed, when $m=1$, multicontact structures are just contact structures; but a particular instance of special multicontact structure with $m = 1$ recovers \emph{co-contact structures} \cite{LGGMR_23} and they cannot be isomorphic
(simply by dimensional requirement).
\end{remark}

\begin{proposition}
\label{prop:dissipation_form_special}
Let $(\mathcal{M},\Theta,\omega)$ be a special (pre)multicontact manifold, then
there exists a unique $1$-form
$\sigma_{\Theta}\in\df^1(\mathcal{M})$, called the \dfn{dissipation form}, satisfying
\begin{equation}
\label{sigma}
\sigma_{\Theta}\wedge\innp{R}\Theta=\innp{R}\d\Theta \,,\quad \text{for every }\,R\in\mathfrak{R} \,.
\end{equation}
\end{proposition}

\begin{proof} See \cite{LGMRR_23}.
\end{proof}

\begin{remark}
\label{alpha_especial_and_alpha_pre}
It should be noted that the dissipation form $\sigma_\Theta$ of Proposition \ref{prop:dissipation_form_special}, 
coincides with the form $\sigma_\alpha$ of Proposition \ref{prop:existence_sigma}, with $\alpha = 0$ (see \cite[Lemma 5.25]{LIR_25}).
\end{remark}

Then, as in Definition \ref{bard}, we can define the \dfn{Lichnerowicz differential operator} associated with $\sigma_\Theta$.
We will commit an abuse of notation denoting with $\bd$ both
the Lichnerowicz differential operators associated with $\sigma_\alpha$ and $\sigma_\Theta$.

The (pre)multicontact structures corresponding to action-dependent field theories arising from the Herglotz variational principle (see \cite{GLMR_24}) satisfy the following additional requirement:

\begin{definition}
\label{def:variational}
Let \((\mathcal{M},\Theta,\omega)\) be a special (pre)multicontact manifold satisfying
\begin{equation}
\label{varcond}
\innp{X}\innp{Y}\Theta = 0 \,, \quad
\text{for every } X,Y\in\ker\omega \,.
\end{equation}
Then $(\mathcal{M},\Theta,\omega)$ is a \dfn{variational (pre)multicontact manifold} and $(\Theta,\omega)$ is said to be a \dfn{variational (pre)multicontact structure}.
\end{definition}

This terminology reflects the fact that condition \eqref{varcond} is the analogue of the variational requirement for (pre)multisymplectic structures to ensure that the system derives from a Lagrangian. In that case, $\ker\omega$ coincides with the vertical distribution of the corresponding bundles.

Using Theorem~\ref{thm:coordenadas}, we get a Darboux-type result for variational (pre)multicontact manifolds.

\begin{theorem}
\label{prop-adapted-coord}
Let $(\mathcal{M},\Theta,\omega)$ be a variational (pre)multicontact manifold. 
Then, there exist adapted local coordinates $(x^\mu,u^I,s^\mu,w^r)$, with $1\leq\mu\leq m$, $1\leq I\leq N-m-k$, and $1\leq r\leq k$, such that
\begin{equation}
\label{thetacoor}
\Theta\big|_U=
H(x^\nu,u^I,s^\nu,w^r)\,\mathrm{d}^m x
+f_I^\mu(x^\nu,u^J)\,\mathrm{d}u^I\wedge \mathrm{d}^{m-1}x_\mu
+\mathrm{d}s^\mu \wedge \mathrm{d}^{m-1}x_\mu \, ,
\end{equation}
and the dissipation form reads
$\ds\sigma_\Theta\big|_U=\frac{\partial H}{\partial s^\mu}\,\mathrm{d}x^\mu$.
\end{theorem}
\begin{proof} See \cite{LGMRR_23}.
\end{proof}

\section{Hamiltonian multicontact systems}
\protect\label{ehs}

We introduce {\sl extended Hamiltonian multicontact systems}
and study their relationship with the Hamiltonian systems of the standard Hamiltonian formalism of action-dependent field theories.

\subsection{Multimomentum bundles and their geometric structures}

Consider the bundle $\pi\colon E\to M$,
where $M$ is an $m$-dimensional oriented manifold with a fixed volume form $\omega_{_M}\in\df^m(M)$.
Let ${\cal M}\pi\equiv\bigwedge\nolimits_2^m\Tan^*E$ denote the bundle of $m$-forms on
$E$ vanishing by contraction with two $\pi$-vertical vector fields. We refer to it as the {\sl extended multimomentum bundle}.
The natural coordinates adapted to the bundle structure ${\cal M}\pi\to E\to M$ are $(x^\mu,y^A,p^\mu_A,p)$, representing the differential form 
$p\, \d^m x + p^\mu_A \d y^A \wedge \d^{m-1}x_\mu\in\bigwedge\nolimits_2^m\Tan^*E$,
with $\mu=1,\ldots,m$ and $A=1,\ldots,n$.
Notice that $\dim\, {\cal M}\pi=nm+n+m+1$.

Let $\bigwedge\nolimits^m_1\Tan^*E$ be the bundle of $\pi$-semibasic $m$-forms on $E$.
Then, we define the {\sl restricted multimomentum bundle} as the quotient
$J^{1*}\pi={\cal M}\pi/\pi^*\bigwedge\nolimits^m\Tan^*M$.
From the coordinates defined above we can define coordinates $(x^\mu,y^A,p_A^\mu)$. Now we have $\dim\,J^{1*}\pi=nm+n+m$.
As usual, we also denote by $\omega$ the pull-back of $\omega_{_M}$ to all bundles over $M$.

For the Hamiltonian formalism of action-dependent field theories, the {\sl configuration bundle} is ${\mathscr{E}=E\times_M\bigwedge\nolimits^{m-1}\Tan^*M}$,
with coordinates $(x^\nu,y^A,s^\mu)$.
Then, we can consider the {\sl extended} and {\sl restricted multimomentum phase bundles}, respectively
$$
\widetilde{\cal P}={\cal M}\pi\times_M\bigwedge\nolimits^{m-1}\Tan^* M
\, , \quad
\mathcal{P}^* =J^{1*}\pi\times_M\bigwedge\nolimits^{m-1}\Tan^*M \,,
$$
with natural coordinates $(x^\mu, y^A,p_A^\mu,p,s^\mu)$ and $(x^\mu,y^A,p_A^\mu,s^\mu)$, respectively. 
The relation of all bundles is represented in the following diagram
\begin{equation}
\label{firstdiag}
\begin{tikzcd}
&& {\widetilde{\mathcal{P}} = \mathcal{M}\pi \times_M\bigwedge\nolimits^{m-1}\Tan^*M} && \\
\\
{\mathcal{M}\pi} && {\mathcal{P}^*=J^{1*}\pi\times_M\bigwedge\nolimits^{m-1}\Tan^*M} \\
\\
{J^{1*}\pi} && {\mathscr{E}=E\times_M\bigwedge\nolimits^{m-1}\Tan^*M} && {\bigwedge\nolimits^{m-1}\Tan^*M} \\
\\
E \\
&& M
\arrow["{\widetilde{\rho}_1}"', shift right=1, bend right=10, from=1-3, to=3-1]
\arrow["{\widetilde{\mathfrak{p}}}"', shift right, from=1-3, to=3-3]
\arrow["{\widetilde{\rho}}"', shift right, bend right = 10, from=1-3, to=5-1]
\arrow["{\widetilde{\tau}_1}", bend left = 20, from=1-3, to=5-5]
\arrow["{\mathfrak{p}}"', swap, from=3-1, to=5-1]
\arrow["{\widetilde{\kappa}^1}"', bend right = 35, from=3-1, to=7-1]
\arrow["{\mathbf{h}}"', shift right=1, bend right=10, from=3-3, to=1-3]
\arrow["\varrho"', from=3-3, to=5-1]
\arrow["\tau"', from=3-3, to=5-3]
\arrow["{\overline{\tau}_1}", shift left=1, bend left = 5, from=3-3, to=5-5]
\arrow["{\varrho_E}"', from=3-3, to=7-1]
\arrow["{\overline{\tau}}"', swap, bend left = 75, from=3-3, to=8-3]
\arrow["{\kappa^1}"', swap, from=5-1, to=7-1]
\arrow["{\tau_E}"', swap, shift left = 1, from=5-3, to=7-1]
\arrow["{\tau_M}"', from=5-3, to=8-3]
\arrow["{\tau_\Lambda}", shift left=3, bend left = 10, from=5-5, to=8-3]
\arrow["\pi", from=7-1, to=8-3]
\end{tikzcd}
\end{equation}
where $\widetilde{\mathfrak{p}}=(\mathfrak{p},{\rm Id}_{\bigwedge\nolimits^{m-1}\Tan^*M})$.
We also denote $\widetilde\tau= \overline\tau\circ\widetilde{\mathfrak{p}}\colon\widetilde{\mathcal{P}}\to M$ and $\widetilde\kappa= \tau\circ\widetilde{\mathfrak{p}}\colon\widetilde{\mathcal{P}}\to\mathscr{E}$.

The bundle $\widetilde{\mathcal{P}}$ is endowed with a canonical multicontact structure, induced from the canonical forms $\vartheta\in\df^{m}(\mathcal{M}\pi)$ (sometimes called the {\sl Liouville form\/}) 
and $\theta\in\df^{m-1}(\bigwedge\nolimits^{m-1}\Tan^*M)$, whose local expressions are
\begin{equation}
\label{canformmulticot}
\vartheta=p_A^\mu\d y^A\wedge\d^{m-1}x_\mu+p\,\d^m x
\,,\quad
\theta=s^\mu\,\d^{m-1}x_\mu \,.
\end{equation}
Indeed, let us define the form
\begin{equation}
\label{wideTheta}
\widetilde\Theta := -\widetilde\rho_1^{\,*}\vartheta+\d(\widetilde\tau_1^{\,*}\theta)\in\df^m(\widetilde{\mathcal{P}})\,  .    
\end{equation}
From the coordinate expressions above we get that locally
\begin{equation}
\label{canoncontact}
\widetilde\Theta =-p_A^\mu \d y^A\wedge\d^{m-1}x_\mu-p\,\d^m x+\d s^\mu\wedge \d^{m-1}x_\mu\, ,
\end{equation}
Then, $(\widetilde{\mathcal{P}}, \widetilde{\Theta})$ is a multicontact manifold. Indeed,
from \eqref{canoncontact} we obtain
\begin{align}
\ker\widetilde\Theta=\left\langle
\frac{\partial}{\partial y^A}+
p_A^\mu\frac{\partial}{\partial s^\mu},
\frac{\partial}{\partial p_A^\mu},
\frac{\partial}{\partial p}
\right\rangle\, ,& \quad
\ker\d\widetilde\Theta=\left\langle
\frac{\partial}{\partial s^\mu}
\right\rangle\, ,
\\
\ker\omega=\left\langle
\frac{\partial}{\partial y^A},
\frac{\partial}{\partial p_A^\mu},
\frac{\partial}{\partial p},
\frac{\partial}{\partial s^\mu}
\right\rangle\, , \quad
\mathfrak{R}=
\left\langle
\frac{\partial}{\partial p},
\frac{\partial}{\partial s^\mu}
\right\rangle\, , & \quad\ker\flat_{\widetilde\Theta}=\Big\langle\derpar{}{p}\Big\rangle=\vf^{{\rm V}(\widetilde{\mathfrak{p}})}(\widetilde{\mathcal{P}}) \, .  
\end{align}
Furthermore, it is easy to prove that
$\widetilde\Theta\not\in\flat_{\d\widetilde\Theta}(\ker\widetilde\Theta)$,
(since $\d\widetilde\Theta$ does not contain terms $\d s^\mu$).
Hence, according to Definitions \ref{multicontact} and \ref{fibered}, and Proposition \ref{equivalenceMC}, $\widetilde\Theta$ is a multicontact form on $\widetilde{\cal P}$
and $(\widetilde{\cal P},\widetilde\Theta)$ is a fibred multicontact manifold.
However, in general, $(\widetilde\Theta,\omega)$ is not a special multicontact (nor special premulticontact) structure,
since $\ker\flat_{\widetilde\Theta} \neq \mathfrak{R}\cap \ker{\widetilde\Theta}\cap \ker{\d\widetilde\Theta} = \{0\}$; but $\rank \mathfrak{R} = m +1$.

As $\ds\ker \flat_{\widetilde\Theta} = \left\langle \pdv{p}\right\rangle$,
then $\ds \mathfrak{R}\big /\ker \flat_{\widetilde\Theta} = \left\langle \pdv{s^\mu}\right\rangle$
and, following Definition \ref{Reebmvf}, the Reeb tensor field for this fibred multicontact structure is  
\begin{equation}
\label{Rmvf}
\mathcal{R} =\pdv{s^\mu} \otimes \pdv{^{m-1}}{^{m-1} x_\mu}\,,
\end{equation}
where the notation $\ds\pdv{^{m-1}}{^{m-1} x_\mu}$ stands for 
\[
\ds\pdv{^{m-1}}{^{m-1} x_\mu}\equiv\innp{\d x^\mu} \pdv{^m}{^m x} = \innp{\d x^\mu} \left( \pdv{x^1} \wedge \cdots \wedge \pdv{x^m}\right)=(-1)^{\mu-1} \pdv{x^1} \wedge \cdots \widehat{\pdv{x^\mu}} \cdots \wedge \pdv{x^m}\,.
\]
Notice that the choice $\alpha = 0$ does not satisfy the conditions in Proposition \ref{prop:existence_sigma}, since $\ds\pdv{p} \in \ker \flat_{\widetilde\Theta}$ and $\ds\innp{\pdv{p}}\d\widetilde\Theta=-\d^m x$.

\subsection{Restricted Hamiltonian multicontact systems in \texorpdfstring{${\cal P}^* = J^{1*}\pi\times_M\bigwedge\nolimits^{m-1}\Tan^*M$}{}}
\protect\label{rhs}

First, we recall the standard definition of
Hamiltonian systems in action-dependent field theories,
which is stated using the restricted action-dependent multimomentum bundle
${\cal P}^* = J^{1*}\pi\times_M\bigwedge\nolimits^{m-1}\Tan^*M$.
(See \cite{LGMRR_23} for details).

Let
${\bf h}\colon{\cal P}^*\to\widetilde{\cal P}$
be a section of $\widetilde{\mathfrak{p}}$
(see Diagram \eqref{firstdiag}).
Locally, it is determined by a {\sl local Hamiltonian function}
$h\in\Cinfty(U)$, $U\subset{\cal P}^*$
such that
\[
{\bf h}(x^\mu,y^A,p_A^\mu,s^\mu)=(x^\mu,y^A,p_A^\mu,p=-h(x^\nu,y^B,p_B^\nu,s^\nu),s^\mu)\, .
\]
Denoting ${\rm h}\equiv\widetilde{\mathfrak{p}}^*h$, this means that the submanifold $\mathcal{S}={\rm Im}({\bf h})$ 
is locally defined in $\widetilde{\mathcal{P}}$ by the constraint 
$$
{\rm H}=p+{\rm h}=p+{\rm h}(x^\nu,y^B,p_B^\nu,s^\nu)=0\ .
$$

\begin{definition}
The section ${\bf h}$ and the function $h$ are called a
\dfn{Hamiltonian section} and its associated
\dfn{local Hamiltonian function}, respectively.

\noindent The \dfn{Hamilton--Cartan form} associated with ${\bf h}$ is
\begin{equation}
\label{thetaH}
\Theta_{\bf h} \coloneqq {\bf h}^*\widetilde\Theta={\bf h}^*(-\widetilde\rho_1^*\vartheta+\widetilde\tau_1^{\,*}\d\theta )\in\df^m({\cal P}^*)\,.
\end{equation}
\end{definition}

It follows that $(\Theta_{\bf h},\omega)$ defines a variational
multicontact structure on ${\cal P}^*$
(see Definition \ref{def:variational})
and natural coordinates in $\mathcal{P}^*$ are adapted coordinates for this structure because
\begin{equation}
\label{thetaHcoor}
\Theta_{\bf h}=-p_A^\mu\,\d y^A\wedge\d^{m-1}x_\mu+h\,\d^m x+\d s^\mu\wedge\d^{m-1}x_\mu\, .
\end{equation}
The associated Reeb vector fields and the dissipation form are
\begin{equation}
\label{Reebsigma}
(R_{\bf h})_\mu=\frac{\partial}{\partial s^\mu}\, , \quad 
\sigma_{\Theta_{\bf h}}=\frac{\partial h}{\partial s^\mu}\,\d x^\mu\, .
\end{equation}

\begin{definition}
The triple $({\cal P}^*,\Theta_{\bf h},\omega)$
is called a \dfn{restricted Hamiltonian multicontact system}.
\end{definition}

\begin{definition}
\label{mconteqH}
The field equations for the restricted Hamiltonian multicontact system $({\cal P}^*,\Theta_{\bf h},\omega)$ can be
written in the following equivalent ways:
\begin{enumerate}[{\rm(i)}]\itemsep=-1pt
\item The \dfn{multicontact Herglotz--Hamilton--de Donder--Weyl equations for sections} ${\bm\psi}_{\bf h}\colon M\to{\cal P}^*$ are
\begin{equation}
\label{sect1Hr}
{\bm\psi}_{\bf h}^*\Theta_{\bf h}=0 \, ,\quad
{\bm\psi}_{\bf h}^*\innp{Y}\bd\Theta_{\bf h}=0 \,,
\ \text{for every $Y\in\mathfrak{X}({\cal P}^*)$} \,,
\end{equation}
or equivalently
\begin{equation}
\label{sect2Hr}
\innp{{\bm\psi}_{\bf h}^{(m)}}(\Theta_{\bf h}\circ{\bm\psi}_{\bf h})=0 \,,\quad
\innp{{\bm\psi}_{\bf h}^{(m)}}(\bd\Theta_{\bf h}\circ{\bm\psi}_{\bf h})=0 \,.
\end{equation}
\item The \dfn{multicontact Herglotz--Hamilton--de Donder--Weyl equations for $\overline\tau$-transverse locally decomposable multivector fields} ${\bf X}_{\bf h}\in\mathfrak{X}^m({\cal P}^*)$ are
\begin{equation}
\label{vfHr}
\innp{{\bf X}_{\bf h}}\Theta_{\bf h}=0 ,\quad
\innp{{\bf X}_{\bf h}}\bd\Theta_{\bf h}=0 \,,
\end{equation}
with $\overline\tau$-transversality imposed by $\innp{{\bf X}_{\bf h}}\omega=1$.
\item The \dfn{multicontact Herglotz--Hamilton--de Donder--Weyl equations for Ehresmann connections} $\nabla_{\bf h}$ on ${\cal P}^*\to M$ are
\begin{equation}
\label{EcHr}
\innp{\nabla_{\bf h}}\Theta_{\bf h}=(m-1)\Theta_{\bf h}\,,\quad
\innp{\nabla_{\bf h}}\bd\Theta_{\bf h}=(m-1)\bd\Theta_{\bf h}\,.
\end{equation}
\end{enumerate}
\end{definition}

\begin{remark}
As $(\Theta_{\bf h},\omega)$ is a  multicontact structure, the above equations are compatible in ${\cal P}^*$.   
\end{remark}

The integral sections ${\bm\psi}_{\bf h}(x) =(x^\mu,y^A(x^\nu),p^\mu_A(x^\nu),s^\mu(x^\nu))$ of integrable multivector fields and connections, solutions to equations
\eqref{vfHr} and \eqref{EcHr}, respectively, are the solutions to equations \eqref{sect1Hr} or \eqref{sect2Hr}.
In coordinates, they read as follows:
\begin{equation}
\frac{\partial s^\mu}{\partial x^\mu}=\Big(p_A^\mu\frac{\partial h}{\partial p_A^\mu}-h\Big)\circ\widetilde{\bm\psi}_{\bf h}\,, \quad
\derpar{y^A}{x^\mu}=\derpar{h}{p^\mu_A}\circ\widetilde{\bm\psi}_{\bf h} \, ,\quad
\derpar{p_A^\mu}{x^\mu}=-\Big(\derpar{h}{y^A}+\derpar{h}{s^\nu}p^\nu_A\Big)\circ\widetilde{\bm\psi}_{\bf h} \,.
\end{equation}
and are called \dfn{restricted Herglotz--Hamilton--de Donder--Weyl equations}

\subsection{Extended Hamiltonian multicontact systems in \texorpdfstring{$\widetilde{\cal {P}} = {\cal M}\pi\times_M\bigwedge\nolimits^{m-1}\Tan^*M$}{}}
\label{subsec:extended-Hamiltonian}

\begin{definition}
\label{hameqbis}
Consider the multicontact manifold $(\widetilde{\mathcal{P}},\widetilde\Theta)$
and let $\mathcal{H}\in\df^m(\widetilde{\mathcal{P}})$ be a $\widetilde\tau$-semibasic form,
which is called a \dfn{Hamiltonian density}. 
Then, $\mathcal{H} \equiv{\mathrm{H}}\,\omega$, where ${\mathrm{H}}\in\Cinfty(\widetilde{\mathcal{P}})$ is the associated \dfn{extended Hamiltonian function}.

\noindent We say that $(\widetilde{\mathcal{P}},\widetilde\Theta;\mathcal{H})$ is
an \dfn{extended Hamiltonian multicontact system} if
the exact form ${\alpha}:=\d{\rm H}\in\df^1(\widetilde{\mathcal{P}})$ satisfies the properties of Proposition \ref{prop:existence_sigma}.
It is called the \dfn{Hamiltonian form} associated with $\mathcal{H}$
and, obviously, $\d\mathcal{H}=\alpha\wedge\omega$.

\noindent The form $\alpha$ defines a dissipation form $\sigma_\alpha$ which is used to construct the corresponding Lichnerowicz differential $\bd$
(Def. \ref{bard}).
Then, the field equations of the system are
\begin{equation}
\label{hameq2}
\innp{{\bf X}_{\alpha}}(\widetilde\Theta+ \mathcal{H})= 0\, , \quad
\innp{{\bf X}_{\alpha}}\overline{\mathrm{d}}\widetilde\Theta=(-1)^{m-1}{\alpha}  \, ;
\end{equation}
where ${\bf X}_{\alpha}\in \vf^m(\widetilde{\mathcal{P}})$ are $\widetilde\tau$-transverse, locally decomposable multivector fields,
and $\widetilde\tau$-transversality can be imposed by demanding $\innp{{\bf X}_{\alpha}}\omega=1$.
\end{definition}

\begin{remark}
Multivector fields solutions to  \eqref{hameq2}
do not necessarily have to be integrable; nevertheless, this condition must also be imposed for solutions to field equations in field theories.
\end{remark}

An important observation is that, as a consequence of the conditions on $\alpha$ in Definition \ref{hameqbis}, 
not every $\widetilde\tau$-semibasic form $\mathcal{H}$ and hence, not every exact $1$-form ${\alpha}\in\df^1(\widetilde{\mathcal{P}})$,
define an extended Hamiltonian multicontact system. In fact, as the next proposition shows, they must take the following form:
\[
\mathcal{H} = (p + \mathrm{h}) \, \omega \,, \quad \alpha = \d p + \d\mathrm{h}\,.
\]

\begin{proposition}
\label{Hprop}
Let $(\widetilde{\mathcal{P}}, \widetilde\Theta;\mathcal{H})$ be an extended Hamiltonian multicontact system
and $\alpha$ the associated Hamiltonian form. Then,
for every
$\widetilde{\mathfrak{p}}$-vertical vector field $Y\in\vf^{{\rm V}(\widetilde{\mathfrak{p}})}(\widetilde{\cal P})$, we have 
$$
\innp{Y}\d\widetilde\Theta=-Y({\rm H})\,\omega \, .
$$
As a consequence, in a chart of natural coordinates
$(U;x^\mu,y^A,p_A^\mu,p,s^\mu)$ in $\widetilde{\mathcal{P}}$, 
adapted to the bundle structure (with $\omega=\d^mx$), we find
$\ds\derpar{{\rm H}}{p}=1$ and hence, there exists
${\rm h}=\widetilde{\mathfrak{p}}^*h\in\Cinfty(U)$, with $h\in\Cinfty(\widetilde{\mathfrak{p}}(U))$, such that
\begin{equation}
{\rm H}\underset{U}{=}p+{\rm h}\equiv p+{\rm h}(x^\mu,y^A,p^\mu_A, s^\mu) \, .
\label{Hcoor}
\end{equation}
\end{proposition}
\begin{proof}
Recall that $\ds\ker\flat_{\widetilde\Theta}=\vf^{{\rm V}(\widetilde{\mathfrak{p}})}(\widetilde{\mathcal{P}})$.
Since $\d\mathcal{H}=\alpha\wedge\omega=\d{\rm H}\wedge\omega$,
condition \eqref{condi} in Proposition \ref{prop:existence_sigma} leads to 
$$
\innp{Y}\d\widetilde\Theta=-\innp{Y}({\alpha}\wedge\omega)=
-\innp{Y}(\d{\rm H}\wedge\omega)=
-(\Lie_Y{\rm H})\,\omega \, .
$$
In the local chart $U\subset\widetilde{\mathcal{P}}$ we have $\ds\ker\flat_{\widetilde\Theta}=\left\langle\pdv{p}\right\rangle$. 
Then, if $\ds Y=f\pdv{p}$ denotes an arbitrary vector field in $\ker\flat_{\widetilde \Theta}$,
bearing in mind the local expression \eqref{canoncontact} of $\widetilde\Theta$, we obtain:
$$
\innp{Y}\d\widetilde\Theta=-f\,\d^mx \, ,\quad
-(\Lie_Y{\rm H})\omega=-f\,\derpar{{\rm H}}{p}\,\d^mx \, ,
$$
and taking $f=1$, we have $\ds\derpar{{\rm H}}{p}=1$.
Finally, \eqref{Hcoor} is a direct consequence of the last result.
\end{proof}

As a straightforward consequence of this Proposition \ref{Hprop}, we have:

\begin{corollary}
\label{orthog}
For the Hamiltonian form $\alpha$ the following relation holds,
for every
$\widetilde{\mathfrak{p}}$-vertical vector field $Y\in\vf^{{\rm V}(\widetilde{\mathfrak{p}})}(\widetilde{\cal P})$,
\begin{equation}
\label{ortogonal}
{\alpha}(Y) \omega = -\innp{Y}\d\widetilde\Theta \, ;
\end{equation}
hence, in a chart of natural coordinates, we obtain \
$\ds\innp{\derpar{}{p}}{\alpha}=1$.
Therefore, there exists an exact and $\widetilde{\mathfrak{p}}$-basic local $1$-form $\beta\in\df^1(U)$
such that
\begin{equation}
{\alpha}\underset{U}{=}\d p+\beta
\equiv\d p+\d{\rm h} \, .
\label{alphacoor}
\end{equation}    
\end{corollary}

As the functions $h\in\Cinfty(\mathcal{P}^*)$ and ${\rm h}=\widetilde{\mathfrak{p}}^*h\in\Cinfty(\widetilde{\mathcal{P}})$
have the same coordinate expression;
taking into account
the local expression \eqref{Rmvf} for the Reeb tensor field and the definition of the dissipation form given in Proposition \ref{prop:existence_sigma} $\widetilde\Theta$ and $\alpha$,
we obtain
\begin{equation}
\label{sigmalpha}
\sigma_\alpha=\innp{\mathcal{R}} (\d\widetilde\Theta + \alpha \wedge \omega)=\derpar{{\rm h}}{s^\mu}\d x^\mu \, .    
\end{equation}
Therefore, from Definition \ref{bard} and using the local expression \eqref{canoncontact} we obtain the following expression
\begin{equation}
\label{bardTheta}
\bd\widetilde\Theta=-\d p_A^\mu\wedge\d y^A\wedge\d^{m-1}x_\mu-\d p\wedge\d^m x+\derpar{{\rm h}}{s^\mu}(p_A^\mu\d y^A-\d s^\mu)\wedge\d^mx \, .
\end{equation}

\begin{theorem}
\label{holsecreg3}
Let $(\widetilde{\mathcal{P}}, \widetilde\Theta;\mathcal{H})$ be an extended Hamiltonian multicontact system.
Then:
\begin{enumerate}[\rm(i)]\itemsep=-1pt
\item 
There are locally decomposable multivector fields
${\bf X}_{\alpha}\in\vf^m(\widetilde{\cal P})$
satisfying equations \eqref{hameq2}.

In a local system of natural coordinates they depend on $n(m^2-1)+m^2-1$ arbitrary functions.
\item
If ${\bf X}_{\alpha}$ are integrable, then their integral sections $\widetilde{\bm\psi}(x) =(x^\mu,y^A(x^\mu),p^\nu_A(x^\mu),p(x^\mu), s^\nu(x^\mu))$
are the solutions to the \dfn{extended Herglotz--Hamilton--de Donder--Weyl equations} 
\begin{equation}
\label{HHDWeq}
\frac{\partial s^\mu}{\partial x^\mu}=\Big(p_A^\mu\frac{\partial {\rm h}}{\partial p_A^\mu}-{\rm h}\Big)\circ\widetilde{\bm\psi}\,, \quad
\derpar{y^A}{x^\mu}=\derpar{{\rm h}}{p^\mu_A}\circ\widetilde{\bm\psi} \, ,\quad
\derpar{p_A^\mu}{x^\mu}=-\Big(\derpar{{\rm h}}{y^A}+\derpar{{\rm h}}{s^\nu}p^\nu_A\Big)\circ\widetilde{\bm\psi} \,,
\end{equation}
\begin{equation}
\label{balanceq}
\pdv{p}{x^\mu}=-\derpar{({\rm h}\circ\widetilde{\bm\psi})}{x^\mu} \,.
\end{equation}
\end{enumerate}
\end{theorem}
\begin{proof} \ (i) In natural coordinates of $\widetilde{\mathcal{P}}$, the most general expression for a locally decomposable multivector field 
${\bf X}_{\alpha} \in\vf^m(\widetilde{\mathcal{P}})$ satisfying the $\widetilde\tau$-transversality condition $\innp {{\bf X}_{\alpha}}\omega=1$ is
\begin{equation}
\label{locmvf3}
{\bf X}_{\alpha}=\bigwedge_{\nu=1}^m \Big(\derpar{}{x^\nu}+\widetilde F_\nu^A\derpar{}{y^A}+\widetilde G^\mu_{A\nu}\derpar{}{p^\mu_A}+\widetilde G_\nu\derpar{}{p}+\widetilde g^\mu_\nu\derpar{}{s^\mu}\Big)
\equiv\bigwedge_{\nu=1}^m X_\nu \,.
\end{equation}
Using \eqref{wideTheta} and that 
\begin{equation}
\label{ThetaH}
\widetilde \Theta+\mathcal{H}=-p^\mu_A\d y^A \wedge \d^{m-1}x_\mu+{\rm h}\,\d^m x+\d s^\mu\wedge\d^{m-1}x_\mu \, ,
\end{equation}
from the first equation in \eqref{hameq2} we obtain
\begin{equation}
\label{eqsG0}
\widetilde g^\mu_\mu=p_A^\mu\widetilde{F}^A_\mu-{\rm h}=
p_A^\mu\frac{\partial{\rm h}}{\partial p_A^\mu}-{\rm h}\, ,  
\end{equation}
and using \eqref{bardTheta},
the second equation in \eqref{hameq2} leads to
\begin{align}
\label{eqsG1}
\widetilde F^A_\mu &=
\derpar{{\rm h}}{p_A^\mu} \, ,  \\
 \label{eqsG2}
 \widetilde G^\mu_{A\mu} &=
 -\derpar{{\rm h}}{y^A}-\parder{{\rm h}}{s^\mu} \, p^\mu_A \, , \\
\widetilde G_\mu &=
-\derpar{{\rm h}}{x^\mu}-\derpar{{\rm h}}{y^A}\derpar{{\rm h}}{p^\mu_A}-
\derpar{{\rm h}}{p^\nu_A}\widetilde G^\nu_{A\mu}-\parder{{\rm h}}{s^\nu}\widetilde g^\nu_\mu \\
& =-\derpar{{\rm h}}{x^\mu}-\derpar{{\rm h}}{y^A}\widetilde F^A_\mu-
\derpar{{\rm h}}{p^\nu_A}\widetilde G^\nu_{A\mu}-\parder{{\rm h}}{s^\nu}\widetilde g^\nu_\mu \, , 
\label{eqconst} 
\end{align}
where equation \eqref{eqsG1} is used in the last equality of equation \eqref{eqconst},
and there are two more equations that are identities.
Thus, the coefficients $\widetilde G^\nu_{A\mu}$ are related by 
\eqref{eqsG2};
\eqref{eqsG1} and \eqref{eqconst} uniquely determine all functions $\widetilde F^A_\mu$ and components $\widetilde G_\mu$
as functions of the other components $\widetilde G^\nu_{A\mu}$, respectively,
and  \eqref{eqsG2} are a compatible system of
$n$ linear equations for the functions $\widetilde G^\mu_{A\mu}$.
These results ensure the local existence of multivector fields ${\bf X}_{\alpha}$, depending on $n(m^2-1)+m^2-1$ arbitrary functions.

Global solutions are obtained using a partition of unity subordinated to a covering of $\widetilde{\cal P}$ made of natural charts.

\noindent (ii) \
If ${\bf X}_{\alpha}$ is integrable and $\widetilde{\bm\psi}(x) =(x^\mu,y^A(x^\nu),p^\mu_A(x^\nu),p(x^\nu), s^\mu(x^\nu))$
is an integral section of ${\bf X}_{\alpha}$, then
\begin{equation}
\label{eqsG4}
\derpar{s^\mu}{x^\nu}=\widetilde g^\mu_\nu\circ\widetilde{\bm\psi} \, , \quad
\derpar{y^A}{x^\mu}=\widetilde F^A_\mu\circ\widetilde{\bm\psi}  \, , \quad
\derpar{p^\mu_A}{x^\nu}=\widetilde G^\mu_{A\nu}\circ\widetilde{\bm\psi} \, , \quad 
\derpar{p}{x^\mu}=\widetilde G_\mu\circ\widetilde{\bm\psi} \, ,
\end{equation}
and from  \eqref{eqsG0}, \eqref{eqsG1}, and \eqref{eqsG2}
we recover the Herglotz--Hamilton--de Donder--Weyl equations \eqref{HHDWeq};
while \eqref{eqconst} become
$$
\frac{\partial p}{\partial x^\mu}=
-\left(\derpar{\rm h}{x^\mu}\circ\widetilde{\bm\psi}+\Big(\derpar{{\rm h}}{y^A}\circ\widetilde{\bm\psi}\Big)\derpar{y^A}{x^\mu}+\Big(\derpar{{\rm h}}{p_A^\nu}\circ\widetilde{\bm\psi}\Big)\derpar{p_A^\nu}{x^\mu}+
\Big(\derpar{{\rm h}}{s^\nu}\circ\widetilde{\bm\psi}\Big)\derpar{s^\nu}{x^\mu}\right)=
-\derpar{({\rm h}\circ\widetilde{\bm\psi})}{x^\mu}\, ,      
$$
which are equations \eqref{balanceq}.

Looking for integrable multivector fields solutions to \eqref{hameq2},
could reduce the number of arbitrary functions
to less than $n(m^2-1)+m^2-1 = (n+1)(m^2-1)$, or, 
if they do not exist, we would apply an integrability algorithm to find a
submanifold of $\widetilde{\cal P}$ (if it exists),
where there exist tangent integrable multivector fields.
\end{proof}

\begin{corollary}
\label{tangency}
If ${\bf X}_{\alpha}\in\vf^m(\widetilde{\cal P})$ is a solution
to equations \eqref{hameq2}, then
it is tangent to the submanifold $\mathcal{S}\subset\widetilde{\mathcal{P}}$ defined by the constraint ${\rm H}=p+{\rm h}=0$
(and to all the level sets of ${\rm H}$).    
\end{corollary}
\begin{proof}
Locally, ${\bf X}_{\alpha}=X_1\wedge\dotsb\wedge X_m$, 
and it is tangent to $\mathcal{S}$ if, and only if, $X_\mu$ are also tangent, for $\mu=1,\ldots,m$;
and this holds because
\begin{align}
\Lie_{X_\mu}{\rm H}&=\innp{X_\mu}\d{\rm H}=\innp{X_\mu}\alpha=(-1)^{m-1}\innp{X_\mu}\innp{{\bf X}_\alpha}\bd\widetilde\Theta 
\nonumber \\
&=(-1)^{m-1}\innp{X_\mu}\innp{X_1\wedge\dotsb\wedge X_m}\bd\widetilde\Theta=0 \, . 
\label{tangent}
\end{align}
In coordinates, this result also follows from equations  \eqref{eqconst}, which are the coordinate expression of 
$\Lie_{X_\mu}(p+{\rm h})=0$;
which means that equations \eqref{balanceq} are just this condition of tangency.
These equations do not have a counterpart in the restricted formalism and appear in the extended formalism as a consequence of the geometry.
\end{proof}

In addition to equation \eqref{hameq2}, the equations of an extended Hamiltonian multicontact system can be expressed in other equivalent alternative ways; as shown in the following:

\begin{definition}
\label{hameqbis2}
Let $(\widetilde{\mathcal{P}},\widetilde\Theta;\mathcal{H})$ be
an extended Hamiltonian multicontact system.
The field equations can be written in the following equivalent ways:
\begin{enumerate}[{\rm(i)}]\itemsep=-1pt
\item 
The \dfn{extended multicontact Herglotz--Hamilton--de Donder--Weyl equations for sections} $\widetilde{\bm\psi}\colon M\to\widetilde{\cal P}$ are 
\begin{equation}
\label{sect1H}
\widetilde{\bm\psi}^*(\widetilde\Theta+\mathcal{H})=0 \,,\quad
\widetilde{\bm\psi}^*\innp{Y}\bd\widetilde\Theta=- {(\widetilde{\bm\psi}^*\innp{Y}\alpha)\,\omega}
\,, \ \text{for every } Y\in\mathfrak{X}(\widetilde{\cal P}) \,, 
\end{equation}
or equivalently
\begin{equation}
\label{sect2H}
\innp{\widetilde{\bm\psi}^{(m)}}(\widetilde\Theta+\mathcal{H})\circ\widetilde{\bm\psi}=0 \,,\quad
\innp{\widetilde{\bm\psi}^{(m)}}(\bd\widetilde\Theta\circ\widetilde{\bm\psi})=(-1)^{m-1}{\alpha}\circ\widetilde{\bm\psi} \,. 
\end{equation}
where $\widetilde{\bm\psi}^{(m)}$ denotes the canonical lift of the section
$\widetilde{\bm\psi}$ to $\ds\bigwedge\nolimits^m\Tan\widetilde{\mathcal{P}}$ (see Appendix \ref{mvfc}).
\item 
The \dfn{extended multicontact Herglotz--Hamilton--de Donder--Weyl equations for $\widetilde\tau$-transverse locally decomposable multivector fields} ${\bf X}\in\mathfrak{X}^m(\widetilde{\cal P})$ are
\begin{equation}
\label{vfH}
\innp{{\bf X}_{\alpha}}(\widetilde\Theta+\mathcal{H})=0 \,,\quad
\innp{{\bf X}_{\alpha}}\bd\widetilde\Theta=(-1)^{m-1}\alpha \,, 
\end{equation}
where $\widetilde\tau$-transversality can be imposed by demanding $\innp{{\bf X}_{\alpha}}\omega=1$.
\item 
The \dfn{extended multicontact Herglotz--Hamilton--de Donder--Weyl equations for Ehresmann connections} $\nabla$ on $\widetilde{\cal P}\to M$ are
\begin{equation}
\label{EcH}
\innp{\nabla}(\widetilde\Theta+\mathcal{H})=(m-1)(\widetilde\Theta+\mathcal{H}) \, ,\quad
\innp{\nabla}\bd\widetilde\Theta=(m-1)\,\bd\widetilde\Theta+(-1)^{m-1}\omega\wedge\alpha\,. 
\end{equation}
\end{enumerate}
\end{definition}

The relation among these kinds of equations is summarised in the following result:

\begin{theorem}
\label{equivteor1}
Let $(\widetilde{\mathcal{P}},\widetilde\Theta;\mathcal{H})$ be
an extended Hamiltonian multicontact system.
\begin{enumerate}[\rm(i)]\itemsep=-1pt
\item
If ${\bf X}_{\alpha}\in\mathfrak{X}^m(\widetilde{\mathcal{P}})$ is a $\widetilde\tau$-transverse and integrable solution to \eqref{vfH}, then its integral sections satisfy \eqref{sect1H} (equivalently, \eqref{sect2H}).
\item
Conversely, if $\widetilde{\bm\psi}\colon M\to\widetilde{\mathcal{P}}$ satisfies \eqref{sect1H} or \eqref{sect2H}, then there exist a tubular neighbourhood 
$U\subset{\widetilde{\mathcal{P}}}$ of $\mathrm{Im}\,\widetilde{\bm\psi}$ and a $\widetilde\tau$-transverse integrable multivector field ${\bf X}_{\alpha}\in\mathfrak{X}^m(U)$ such that  
(i) $\widetilde{\bm\psi}$ is an integral section of ${\bf X}_{\alpha}$, and  
(ii) ${\bf X}_{\alpha}$ satisfies \eqref{vfH} on $\mathrm{Im}\,\widetilde{\bm\psi}$.
\item
Integrable Ehresmann connections in $\widetilde{\mathcal{P}}$ solving \eqref{EcH} are locally associated with equivalence classes of integrable, $\widetilde\tau$-transverse, locally decomposable multivector fields $\{{\bf X}_{\alpha}\}\subset\mathfrak{X}^m(\widetilde{\mathcal{P}})$ solving
\eqref{vfH}, and conversely.
\end{enumerate}
\end{theorem}
\begin{proof}
(i) \
The coordinate expressions of equations \eqref{sect1H} and \eqref{sect2H} are \eqref{HHDWeq} and  \eqref{balanceq}.
Then, this result has been proved in Theorem \ref{holsecreg3}.

\noindent (ii) \
If $\widetilde{\bm\psi}\colon W\subset M\to\widetilde{\mathcal{P}}$ is a local section solution to \eqref{sect1H} or \eqref{sect2H}, the map $\bigwedge\nolimits^m(\widetilde{\bm\psi}\vert_{W})$ defines 
a locally decomposable multivector field 
in $\widetilde{\bm\psi}(W)\subset\widetilde{\mathcal{P}}$ that is tangent to $\widetilde{\bm\psi}(W)$.
Using partitions of unity, this construction can be extended to the whole $M$,
obtaining a multivector field on $\mathrm{Im}\,\widetilde{\bm\psi}$
and such that  $\widetilde{\bm\psi}$ is an integral section of ${\bf X}_{\alpha}$, by construction.
Finally, using the local flow of $\widetilde\tau$-vertical vector fields,
this multivector field can be extended to a tubular neighbourhood $U$ of $\mathrm{Im}\,\widetilde{\bm\psi}$.
Then, since ${\bf X}_{\alpha}\circ\widetilde{\bm\psi}=\widetilde{\bm\psi}^{(m)}$, if equations \eqref{sect1H} or \eqref{sect2H} hold for $\widetilde{\bm\psi}$, 
then \eqref{vfH} holds for ${\bf X}_{\alpha}$ on $\mathrm{Im}\,\widetilde{\bm\psi}$.
(See \cite{LGMRR_23} for more details).

\noindent (iii) \
The local expression of a $\widetilde\tau$-transverse, locally decomposable multivector field ${\bf X}_{\alpha}\in\mathfrak{X}^m(\widetilde{\mathcal{P}})$ is \eqref{locmvf3}.
Then, the coordinate expression of the Ehresmann connections locally associated with 
the equivalence class $\{{\bf X}_{\alpha}\}$ is
$$
\nabla=\d x^\nu\otimes\left(\derpar{}{x^\nu}+\widetilde F_\nu^A\derpar{}{y^A}+\widetilde G^\mu_{A\nu}\derpar{}{p^\mu_A}+\widetilde G_\nu\derpar{}{p}+\widetilde g^\mu_\nu\derpar{}{s^\mu}\right) \, .
$$
Therefore,  \eqref{EcH} lead to the same 
 \eqref{eqsG0}, \eqref{eqsG1}, \eqref{eqsG2}, and \eqref{eqconst}
as for ${\bf X}_{\alpha}$.
\end{proof}

\begin{remark}
The tangency conditions \eqref{tangent} can be stated equivalently
\begin{itemize}\itemsep=-1pt
\item for Ehreshman connections solution to equations \eqref{EcH}, as $\innp{\nabla}\alpha=0$.
\item for sections solution to equations
\eqref{sect1H} or \eqref{sect2H}, as
$\widetilde{\bm\psi}^*\alpha=0$ or, equivalently,
$\d({\rm H}\circ\widetilde{\bm\psi})=0$.
\end{itemize}
\end{remark}

\subsection{Relation between extended and restricted Hamiltonian multicontact systems}
\protect\label{rerhs}

Starting from extended Hamiltonian multicontact systems, we can recover restricted Hamiltonian multicontact systems. In fact:

\begin{theorem}
\label{related}
Let $(\widetilde{\mathcal{P}},\widetilde\Theta;\mathcal{H})$ be an extended Hamiltonian multicontact system and $\alpha\in\df^1(\widetilde{\mathcal{P}})$ the associated Hamiltonian form.
Then, there exists a restricted Hamiltonian multicontact system $({\cal P}^*,\Theta_{\bf h},\omega)$
defined by a Hamiltonian section ${\bf h}\colon\mathcal{P}^*\to\widetilde{\mathcal{P}}$, 
such that:
\begin{enumerate}[{\rm(i)}]\itemsep=-1pt
\item
The relation between $\alpha$ and ${\bf h}$ is ${\bf h}^*\alpha=0$.
\item 
The dissipation forms $\sigma_\alpha$
associated with the multicontact structure $\widetilde\Theta$ and the form $\alpha$,
and $\sigma_{\Theta_{\bf h}}$
associated with the special multicontact structure $(\Theta_{\bf h},\omega)$, are related by
$\sigma_\alpha=\widetilde{\mathfrak{p}}^*\sigma_{\Theta_{\bf h}}$.
\item 
For every $\beta\in\df^k(\widetilde{\cal P})$,
the following relation holds:
$\bd\,{\bf h}^*\beta={\bf h}^*\bd\beta$.

As a consequence,
${\bf h}^*\bd\widetilde\Theta=\bd\Theta_{\bf h}$.
\item 
If ${\bf X}_{\alpha}\in\vf^m(\widetilde{\cal P})$ 
is a solution to  \eqref{vfH},
then there exists ${\bf X}_{\bf h}\in\vf^m(\mathcal{P}^*)$ that
is ${\bf h}$-related to ${\bf X}_{\alpha}$
and is a solution to  \eqref{vfHr}.

In particular, if ${\bf X}_{\alpha}$ is integrable, then
${\bf X}_{\bf h}$ is also integrable, and 
if $\widetilde{\bm\psi}\colon M\to\widetilde{\cal P}$ is an integral
section of ${\bf X}_{\alpha}$, then 
$\bm\psi=\widetilde{\mathfrak{p}}\circ\widetilde{\bm\psi}\colon M\to\mathcal{P}^*$
is an integral section of ${\bf X}_{\bf h}$.
\end{enumerate}
\end{theorem}
\begin{proof}
For $U\subset\widetilde{\mathcal{P}}$,
by \eqref{Hcoor} and \eqref{alphacoor}, we have 
${\rm H}\vert_U=p+{\rm h}$ and
$\alpha\vert_U=\d{\rm H}=\d p+\d{\rm h}$, respectively; where ${\rm h}=\widetilde{\mathfrak{p}}^*h\in\Cinfty(U)$ and $h\in\Cinfty(\widetilde{\mathfrak{p}}(U))$.
Then, we define a Hamiltonian section that is locally given by the function ${\rm h}$.
This means that the submanifold $\mathcal{S}={\rm Im}\,{\bf h}$ is locally defined by the constraint
$$
{\rm H}=p+{\rm h}(x^\nu,y^A,p^\nu_A, s^\nu)=0 \, .
$$
If $\jmath_{\mathcal{S}}\colon\mathcal{S}\hookrightarrow\widetilde{\mathcal{P}}$ is the natural embedding, then $\mathfrak{p}_{\mathcal{S}}=\widetilde{\mathfrak{p}}\circ\jmath_{\mathcal{S}}$
(the restriction of the projection $\widetilde{\mathfrak{p}}$ to $\mathcal{S}$)
is a diffeomorphism, whose inverse ${\bf h}_{\mathcal{S}}$ is 
defined by ${\bf h}=\jmath_{\mathcal{S}}\circ{\bf h}_{\mathcal{S}}$.
So, we have the diagram
\begin{equation}
\label{diaghamsec0}
\begin{tikzcd}[row sep=large, column sep=huge]
\mathcal{S} & \widetilde{\mathcal{P}} \\
& \mathcal{P}^*
\arrow["\jmath_{\mathcal{S}}", from=1-1, to=1-2]
\arrow["\mathfrak{p}_{\mathcal{S}}", from=1-1, to=2-2]
\arrow["\widetilde{\mathfrak{p}}"', from=1-2, to=2-2]
\arrow["\mathbf{h}"', bend right=30, from=2-2, to=1-2]
\arrow["\mathbf{h}_{\mathcal{S}}", bend left=30, from=2-2, to=1-1]
\end{tikzcd}
\end{equation}

\noindent(i) \ As ${\rm H}$ locally defines $\mathcal{S}$, we have $\jmath_{\mathcal{S}}^*{\rm H}=0$,
and hence $\jmath_{\mathcal{S}}^*\alpha=0$; therefore,
$$
{\bf h}^*\alpha={\bf h}^*\d{\rm H}=\d\,{\bf h}^*{\rm H}=\d({\bf h}_{\mathcal{S}}^*\jmath_{\mathcal{S}}^*{\rm H})=0 \, .
$$

\noindent(ii) \ 
The local expression \eqref{sigmalpha}
of the dissipation form $\sigma_\alpha$ for $\Theta$ and $\alpha$
is the same as that of the dissipation form $\sigma_{\Theta_{\bf h}}$
obtained in \eqref{Reebsigma} for $(\Theta_{\bf h},\omega)$.
This proves the claim.

\noindent (iii) \ 
As a consequence of (ii), the Lichnerowicz differential operators defined by $\sigma_\alpha$ and $\sigma_{\Theta_{\bf h}}$ also have the same local expressions and then
$$
\bd\,{\bf h}^*\beta=\d\,{\bf h}^*\beta+\sigma_\alpha\wedge{\bf h}^*\beta={\bf h}^*\d\beta+{\bf h}^*\sigma_\alpha\wedge{\bf h}^*\beta=
{\bf h}^*(\d\beta+\sigma_\alpha\wedge\beta)=
{\bf h}^*\bd\beta\, .
$$

\noindent(iv) \ If ${\bf X}_{\alpha}\in\vf^m(\widetilde{\cal P})$ is a solution
to equations \eqref{vfH}, by Corollary \ref{tangency}, it is tangent to $\mathcal{S}$ and then there exists ${\bf  X}_{\mathcal{S}}\in\vf^m(\mathcal{S})$
such that $(\jmath_{\mathcal{S}})_*{\bf  X}_{\mathcal{S}}={\bf X}_{\alpha}\vert_\mathcal{S}$.
Let ${\bf X}_{\bf h}\in\vf^m(\cal P^*)$ be defined by
${\bf X}_{\bf h}=(\mathfrak{p}_{\mathcal{S}})_*{\bf  X}_{\mathcal{S}}$. 
Therefore, since the Lichnerowicz differential operators $\bd$ defined by the dissipation forms $\sigma_\alpha$ and $\sigma_{\Theta_{\bf h}}$ are the same,
from the equation
$\innp{{\bf X}_{\alpha}}\overline{\mathrm{d}}\widetilde\Theta=(-1)^{m-1}{\alpha}$
and the condition $\jmath_{\mathcal{S}}^*{\alpha}=0$, 
we obtain:
\begin{align*}
0 &={\bf h}^*\big(\innp{{\bf X}_{\alpha}}\bd\widetilde\Theta-(-1)^{m-1}{\alpha}\big) = (\jmath_{\mathcal{S}}\circ{\bf h}_{\mathcal{S}})^*[\innp{{\bf X}_{\alpha}}\overline{\mathrm{d}}\widetilde\Theta-(-1)^{m-1}{\alpha}] \\ &=
{\bf h}_{\mathcal{S}}^*\big(\innp{{\bf  X}_{\mathcal{S}}}\jmath_{\mathcal{S}}^*\bd\widetilde\Theta-(-1)^{m-1}\jmath_{\mathcal{S}}^*{\alpha}\big)= 
\innp{{\bf X}_{\bf h}}({\bf h}_{\mathcal{S}}^*\circ\jmath_{\mathcal{S}}^*)\bd\widetilde\Theta \\ &= \innp{{\bf X}_{\bf h}}{\bf h}^*\bd\widetilde\Theta=\innp{{\bf X}_{\bf h}} \bd\,{\bf h}^*\widetilde\Theta=\innp{{\bf X}_{\bf h}}\bd\Theta_{\bf h}  \, .  
\end{align*}
Analogously, from the local expressions \eqref{thetaHcoor} and \eqref{ThetaH},
we conclude that 
${\bf h}^*(\widetilde\Theta+\mathcal{H})=\Theta_{\bf h}$,
and we obtain $\innp{{\bf X}_{\bf h}}\Theta_{\bf h}=0$.
Furthermore, as $\widetilde{\mathfrak{p}}\circ {\bf h}={\rm Id}_{\cal P^*}$,
we also have
$$
\innp{{\bf X}_{\bf h}}\omega={\bf h}^*\big(\innp{{\bf X}_{\alpha}}\omega\big)=1 \, .
$$
Finally, if ${\bf X}_{\alpha}$ is integrable,
as it is tangent to $\mathcal{S}$, the integral sections 
of ${\bf X}_{\alpha}$ passing
through any point of $\mathcal{S}$ remain in $\mathcal{S}$, and hence
they are the integral sections of ${\bf  X}_{\mathcal{S}}$,
so ${\bf X}_{\mathcal{S}}$ is integrable and,
as a consequence, ${\bf X}_{\bf h}$ is also integrable.    
\end{proof}

Conversely, extended Hamiltonian multicontact systems are obtained starting from
restricted Hamiltonian ones:

\begin{theorem}
\label{existuniq}
Let $({\cal P}^*,\Theta_{\bf h},\omega)$ be a
restricted Hamiltonian multicontact system defined by a Hamiltonian section ${\bf h}\colon\mathcal{P}^*\to\widetilde{\mathcal{P}}$. Then, there exists a unique form
${\alpha}\in\df^1(\widetilde{\cal P})$ such that:
\begin{enumerate}[{\rm(i)}] \itemsep = -1pt
\item The form $\alpha$ is exact, $\alpha = \d \mathrm{H}$, where $\mathrm{H}$ is the function defined by $-\widetilde \Theta +\widetilde{\mathfrak{p}}^\ast \Theta_{\bf h} = \mathrm{H}\,\omega$, and satisfies the properties of Proposition \ref{prop:existence_sigma}.
\item ${\bf h}^*{\alpha}=0$.
\item $\innp{Y}{\alpha}\neq0$, for every non-vanishing 
$Y\in\vf^{{\rm V}(\widetilde{\mathfrak{p}})}(\widetilde{\cal P})$ and, in particular, \(\textstyle\innp{\derpar{}{p}}{\alpha}=1\).
\end{enumerate}
Thus, $(\widetilde{\mathcal{P}},\widetilde\Theta;\mathcal{H})$, where $\mathcal{H}={\rm H}\,\omega=- \widetilde \Theta + \widetilde{\mathfrak{p}}^\ast \Theta_{\bf h}$, is an extended Hamiltonian multicontact system.
\begin{enumerate}[{\rm(iv)}] 
\item 
Furthermore, if ${\bf X}_{\bf h}\in\vf^m(\mathcal{P}^*)$ is an (integrable) solution to  \eqref{vfHr}, then
${\bf X}_{\alpha}={\bf h}_*{\bf X}_{\bf h}\in\vf^m(\widetilde{\cal P})$ 
is an (integrable) solution to  \eqref{vfH} along the image of the section ${\bf h}$.
\end{enumerate}
\end{theorem}
\begin{proof}
Suppose that there exist ${\alpha},{\alpha}'$ satisfying conditions (i), (ii), and (iii).
Taking into account Corollary \ref{orthog},
we have that, in $U\subset\widetilde{\cal P}$,
${\alpha}=\d p+\beta$ and ${\alpha}'=\d p+\beta'$,
where $\beta=\widetilde{\mathfrak{p}}^*\bar\beta$, $\beta'=\widetilde{\mathfrak{p}}^*\bar\beta'$,
being $\bar\beta,\bar\beta'\in\df^1(\widetilde{\mathfrak{p}}(U))$ exact $1$-forms.
From condition (ii) we have
$\jmath_{\mathcal{S}}^*{\alpha}=\jmath_{\mathcal{S}}^*{\alpha}'$; hence
$$
0=\jmath_{\mathcal{S}}^*({\alpha}-{\alpha}') \Longleftrightarrow
0=\jmath_{\mathcal{S}}^*(\beta-\beta')=(\widetilde{\mathfrak{p}}\circ\jmath_{\mathcal{S}})^*(\bar\beta-\bar\beta')\ \Longrightarrow
0=(\widetilde{\mathfrak{p}}\circ\jmath_{\mathcal{S}}\circ{\bf h}_{\mathcal{S}})^*(\bar\beta-\bar\beta')=\bar\beta-\bar\beta' \, , 
$$
since $\widetilde{\mathfrak{p}}\circ\jmath_{\mathcal{S}}\circ{\bf h}_{\mathcal{S}}=\widetilde{\mathfrak{p}}\circ{\bf h}={\rm Id}_{\widetilde{\mathfrak{p}}(U)}$;
then $\bar\beta=\bar\beta'$
and therefore $\alpha=\alpha'$.
This proves uniqueness.

The existence is trivial, since every Hamiltonian section ${\bf h}$
is locally given by a function $h\in\Cinfty(\widetilde{\mathfrak{p}}(U))$ such that
$p=-h(x^\nu,y^A,p_A^\nu, s^\nu)$. Hence
\begin{equation}
\label{alphaloc}
{\alpha}\vert_U=\d p+\d{\rm h}=\d{\rm H}\, .    
\end{equation}
Then, there exists a function ${\rm H}\in\Cinfty(\widetilde{\cal P})$,
whose local expression in $U$ is
${\rm H}=p+{\rm h}$, with ${\rm h}=\widetilde{\mathfrak{p}}^*h$, and $h\in\Cinfty(\widetilde{\mathfrak{p}}(U))$.
So $\alpha$ is an exact form and, taking into account \eqref{alphaloc} and that ${\cal S}=\mathrm{Im}\,(\bf h)$
is defined by the constraint $p+{\rm h}=0$, we obtain
$$
{\bf h}^*\alpha={\bf h}^*\d(p+{\rm h})=
\d{\bf h}^*(p+{\rm h})=
\d\big({\bf h_{\cal S}}^*{\jmath_{\cal S}}^*(p+{\rm h})\big)=0 \ .
$$ 
From \eqref{alphaloc}, property (iii) is immediate.
Furthermore, $\alpha$  trivially satisfies the properties of Proposition \ref{prop:existence_sigma}.

Finally, (iv) is proved as in the last part of the proof of Theorem \ref{related}.
\end{proof}

\subsection{Generic extended Hamiltonian multicontact systems (external sources)}
\label{gehms}

As we have seen in Definition \ref{hameqbis}, the field equations of an
extended Hamiltonian multicontact system are stated by giving an exact differential $1$-form $\alpha$ associated with the Hamiltonian density $\mathcal{H}$.
In this section, we study the case where this $1$-form is neither exact, nor closed;
which leads to considering another type of extended Hamiltonian multicontact systems.
In Hamiltonian mechanics, this corresponds to introducing external forces in the system \cite{God_69},
In Hamiltonian field theories, the non-closed $1$-form $\alpha$ would represent some kind of external source.

\begin{definition}
\label{hameqbis22}
Consider the multicontact manifold $(\widetilde{\mathcal{P}},\widetilde\Theta)$.
We say that the tuple $(\widetilde{\mathcal{P}},\widetilde\Theta;\mathcal{H},\widetilde{\bm\alpha})$ is
a \dfn{generic extended Hamiltonian multicontact system} if:
\begin{enumerate}[{\rm (i)}]\itemsep=-1pt
\item
$\mathcal{H}={\rm H}\,\omega\in\df^m(\widetilde{\mathcal{P}})$
is a \dfn{Hamiltonian density}, where  the \dfn{extended Hamiltonian function} ${\mathrm{H}}\in\Cinfty(\widetilde{\mathcal{P}})$ is locally expressed in $U\subset\widetilde{\mathcal{P}}$ as
${\rm H}=p+{\rm h}$, with $h\in\Cinfty(\widetilde{\mathfrak{p}}(U))$.
\item 
$\widetilde{\bm\alpha}\in\df^1(\widetilde{\mathcal{P}})$
is a non-closed form that satisfies the properties of Proposition \ref{prop:existence_sigma}
It is called a \dfn{generic Hamiltonian form}.
\end{enumerate}
The form $\widetilde{\bm\alpha}$ defines a dissipation form $\sigma_{\bm\alpha}$ which is used to construct the corresponding Lichnerowicz differential $\bd$
(Def. \ref{bard}).
Then, the field equations of the system are
\begin{equation}
\label{hameq2b}
\innp{{\bf X}_{\widetilde{\bm\alpha}}}(\widetilde\Theta+ \mathcal{H})= 0\, , \quad
\innp{{\bf X}_{\widetilde{\bm\alpha}}}\overline{\mathrm{d}}\widetilde\Theta=(-1)^{m-1}\widetilde{\bm\alpha}  \, ;
\end{equation}
for $\widetilde\tau$-transverse, locally decomposable multivector fields ${\bf X}_{\widetilde{\bm\alpha}}\in \vf^m(\widetilde{\mathcal{P}})$, 
where $\widetilde\tau$-transversality can be imposed by demanding $\innp{{\bf X}_{\widetilde{\bm\alpha}}}\omega=1$.
\end{definition}

If a non-closed form $\widetilde{\bm\alpha}$ satisfies the properties of Proposition \ref{prop:existence_sigma}, 
following the same reasoning as in Proposition \ref{Hprop} and Corollary \ref{orthog}, we conclude that $\widetilde{\bm\alpha}$ necessarily has the local expression
\begin{equation}
\label{alphagen}
\widetilde{\bm\alpha}=\d p+\beta_\mu\d x^\mu+\beta_A\d y^A+\beta^A_\mu\d p^\mu_A+\gamma_\mu\d s^\mu \, ,     
\end{equation} 
for some $\widetilde{\mathfrak{p}}$-basic local functions $\beta_\mu,\beta_A,\beta_\mu^A,\gamma_\mu\in\Cinfty(\widetilde{\mathcal{P}})$
(that is, they depend only on the coordinates $(x^\mu,y^A,p^\mu_A,s^\mu)$).
By relabeling the functions in this expression, we can write it as
\begin{align}
\widetilde{\bm\alpha}&=\d(p+{\rm h})+\Big(\beta_\mu-\derpar{{\rm h}}{x^\mu}\Big)\d x^\mu+\Big(\beta_A-\derpar{{\rm h}}{y^A}\Big)\d y^A+\Big(\beta^A_\mu-\derpar{{\rm h}}{p^\mu_A}\Big)\d p^\mu_A+\Big(\gamma_\mu-\derpar{{\rm h}}{s^\mu}\Big)\d s^\mu
\\ &=
\d{\rm H}+\widetilde\beta_\mu\d x^\mu+\widetilde\beta_A\d y^A+\widetilde\beta^A_\mu\d p^\mu_A+\widetilde\gamma_\mu\d s^\mu  \\
&=
\alpha+\widetilde\beta_\mu\d x^\mu+\widetilde\beta_A\d y^A+\widetilde\beta^A_\mu\d p^\mu_A+\widetilde\gamma_\mu\d s^\mu= \alpha +\widetilde\beta\,, \label{genericalpha}
\end{align}
where $\alpha=\d{\rm H}$ is the Hamiltonian form associated with the Hamiltonian density $\mathcal{H}$.

The local expression of the dissipation form associated with $\widetilde\Theta$ and $\widetilde{\bm\alpha}$ is 
$\sigma_{\widetilde{\bm\alpha}}=\gamma_\mu\,\d x^\mu$.

Revisiting the computations of Theorem \ref{holsecreg3} and using \eqref{alphagen}, we easily obtain the following:

\begin{theorem}
\label{holsecreg3b}
Let $(\widetilde{\mathcal{P}}, \widetilde\Theta;\mathcal{H},\widetilde{\bm\alpha})$ 
be a generic extended Hamiltonian multicontact system.
Then:
\begin{enumerate}[\rm(i)]\itemsep=-1pt
\item 
There are locally decomposable multivector fields
${\bf X}_{\widetilde{\bm\alpha}}\in\vf^m(\widetilde{\cal P})$
satisfying equations \eqref{hameq2b}.

In a system of natural coordinates they depend on $n(m^2-1)+m^2-1$ arbitrary functions.
\item
If ${\bf X}_{\widetilde{\bm\alpha}}$ are integrable, their integral sections $\widetilde{\bm\psi}(x) =(x^\mu,y^A(x^\mu),p^\nu_A(x^\mu),p(x^\mu), s^\nu(x^\mu))$
are the solutions to the \dfn{generalised extended Herglotz--Hamilton--de Donder--Weyl equations} 
\begin{equation}
\label{HHDWeqb}
\frac{\partial s^\mu}{\partial x^\mu}=\Big(p_A^\mu\beta^A_\mu-{\rm h}\Big)\circ\widetilde{\bm\psi}\,, \quad
\derpar{y^A}{x^\mu}= \beta_\mu^A\circ\widetilde{\bm\psi} \, ,\quad
\derpar{p_A^\mu}{x^\mu}=-(\beta_A+\gamma_\nu p^\nu_A)\circ\widetilde{\bm\psi} \, .
\end{equation}
\begin{equation}
\label{HHDWeqpb}
\frac{\partial p}{\partial x^\nu}=-\Big(\big(\beta_\nu\circ\widetilde{\bm\psi}\big)+\big(\beta_A\circ\widetilde{\bm\psi}\big)\big(\beta_\nu^A\circ\widetilde{\bm\psi}\big)+
\big(\beta_\mu^A\circ\widetilde{\bm\psi}\big)\derpar{p^\mu_A}{x^\nu}+\big(\gamma_\mu\circ\widetilde{\bm\psi}\big)\derpar{s^\mu}{x^\nu}\Big)\, .  
\end{equation} 
\end{enumerate}
\end{theorem}

\begin{remark}
For generic extended  Hamiltonian multicontact systems, the field equations can also be expressed in alternative equivalent ways as stated in Definition \ref{hameqbis2}.
\end{remark}

The restricted Hamiltonian formalism for generic extended Hamiltonian systems is as follows:
If $\alpha=\d{\rm H}$ is the Hamiltonian form introduced in the expression \eqref{genericalpha},
which is associated with the Hamiltonian density $\mathcal{H}$,
we can define a Hamiltonian section ${\bf h}\colon\mathcal{P}^*\to\widetilde{\mathcal{P}}$
such that conditions (i), (ii) and (iii) of Theorems \ref{related} and \ref{existuniq} hold. 
Then we define the form
$$
\bm\alpha={\bf h}^*\widetilde{\bm\alpha}={\bf h}^*\widetilde\beta\in\df^1(\mathcal{P}^*).
$$
(This form can also be obtained taking into account that,
as the expression \eqref{genericalpha} shows, $\widetilde\beta$ is $\widetilde{\mathfrak{p}}$-projectable, and then there exists $\bm\alpha\in\df^1(\mathcal{P}^*)$ such that
$\widetilde\beta=\widetilde{\mathfrak{p}}^*\bm\alpha$).
Therefore:

\begin{definition}
The tuple $(\mathcal{P}^*,\Theta_{\bf h};\bm\alpha)$ is said to be a \dfn{generic restricted Hamiltonian multicontact system}.

\noindent Taking the dissipation form $\sigma_{\bm\alpha}={\bf h}^*\sigma_{\widetilde{\bm\alpha}}=\gamma_\mu\d x^\mu$,
we can construct the corresponding Lichnerowicz differential operator $\bd$.
Then the field equations for this system are
\begin{equation}
\label{otraeq}
\innp{{\bf X}_{\bf h}}\Theta_{\bf h}=0,\quad
\innp{{\bf X}_{\bf h}}\bd\Theta_{\bf h}=(-1)^{m-1}\bm\alpha  \,,
\end{equation}
for $\overline\tau$-transverse locally decomposable multivector fields ${\bf X}_{\bf h}\in\mathfrak{X}^m({\cal P}^*)$,
where $\overline\tau$-transversality can be imposed by demanding $\innp{{\bf X}_{\bf h}}\omega=1$
\end{definition}

It is immediate to prove that the multivector fields ${\bf X}_{\widetilde{\bm\alpha}}$ and ${\bf X}_{\bf h}$,
solutions to equations \eqref{hameq2b} and \eqref{otraeq}, respectively, 
are ${\bf h}$-related (on $\Im\,{\bf h}$).

In coordinates, the integral sections ${\bm\psi}_{\bf h}(x) =(x^\mu,y^A(x^\nu),p^\mu_A(x^\nu),s^\mu(x^\nu))$ of integrable multivector fields solutions to equations
\eqref{otraeq} are the solutions to the \dfn{generalised restricted Herglotz--Hamilton--de Donder--Weyl equations}
\begin{equation}
\frac{\partial s^\mu}{\partial x^\mu}=\big(p_A^\mu\beta_\mu^A-h\big)\circ\widetilde{\bm\psi}_{\bf h}\,, \quad
\derpar{y^A}{x^\mu}=\beta_\mu^A\circ\widetilde{\bm\psi}_{\bf h} \, ,\quad
\derpar{p_A^\mu}{x^\mu}=-\big(\beta_A+\gamma_\nu p^\nu_A\big)\circ\widetilde{\bm\psi}_{\bf h} \,.
\end{equation}

\section{Hamiltonian premulticontact systems}
\protect\label{arhs}

In the usual (restricted) multicontact formulation of Hamiltonian field theories,
there are many cases  where the Hamiltonian field equations are not stated in $\mathcal{P}^*$, but rather in submanifolds of it.
For instance, this happens when considering the Hamiltonian formalism associated with an almost-regular Lagrangian, 
where the Legendre transformation is a surjective submersion and the Hamiltonian phase space is its image, which is a submanifold of $\mathcal{P}^*$ (see \cite{LGMRR_23}).
In the extended Hamiltonian multicontact formalism, this situation corresponds to taking 
a submanifold of $\widetilde{\mathcal{P}}$.

\subsection{Restricted Hamiltonian premulticontact systems}
\label{rHps}

(See \cite{LGMRR_23} for other details).

Consider the bundle $\kappa^1\colon J^{1*}\pi\to E$ (see Diagram \eqref{firstdiag})
and let $P_\circ\hookrightarrow J^{1*}\pi$ be a $\kappa^1$-transverse submanifold; so we have the subbundle $\kappa_\circ^1\colon P_\circ\to E$.
Consider the induced submanifold
$$ \jmath_\circ\colon \mathcal{P}_\circ^*= P_\circ\times_M \bigwedge\nolimits^{m-1}\Tan^*M
\longhookrightarrow
{\cal P}^*= J^{1*}\pi\times_M \bigwedge\nolimits^{m-1}\Tan^*M \,. $$
It is a $\tau$-transverse submanifold and then $\mathcal{P}_\circ^*\to \mathscr{E}\to M$ is a subbundle of $\mathcal{P}^*\to \mathscr{E}\to M$. 
Natural adapted coordinates in $\mathcal{P}_\circ^*$ are denoted by $(x^\mu,y^A,p_i^\lambda,s^\mu)$
(as some original multimomentum coordinates have been expressed as functions of the remaining multimomentum coordinates and $(x^\mu,y^A)$).

\begin{definition}
A \dfn{Hamiltonian section} is a map
${\bf h}_\circ\colon \mathcal{P}_\circ^*\to \widetilde{\mathcal{P}}$
such that the map ${\bf h}\colon \mathcal{P}^*\to \widetilde{\mathcal{P}}$
satisfying ${\bf h}_\circ={\bf h}\circ\jmath_\circ$
is a section of the projection $\widetilde{\mathfrak{p}}$.
Locally, ${\bf h}_\circ$ is determined by a function
$h_\circ\in\Cinfty(U)$, where $U\subset\mathcal{P}_\circ^*$, given by
\[
{\bf h}_\circ(x^\mu,y^A,p_i^\lambda,s^\mu)=\bigl(x^\mu,y^A,p_i^\lambda,p=- h_\circ(x^\nu,y^B,p_j^\eta,s^\nu),s^\mu\bigr)\,,
\]
which is called a \dfn{local Hamiltonian function} associated with ${\bf h}_\circ$.

\noindent The \dfn{Hamilton--Cartan form} associated with ${\bf h}_\circ$ is
\begin{equation}
\label{thetaHcoor0}
\Theta_{{\bf h}_\circ} \coloneqq {\bf h}_\circ^{\,*}\widetilde\Theta=\jmath_\circ^{\,*}\Theta_{\bf h}\in \df^m(\mathcal{P}_\circ^*) \,.
\end{equation}
\end{definition} 

We have the diagram
\begin{equation}
\label{diagbig}
\begin{tikzcd}
{J^{1*}\pi} & {\mathcal{P}^* = J^{1*}\pi\times_M\bigwedge^{m-1}\Tan^*M} & {\widetilde{\mathcal{P}} = \mathcal{M}\pi\times_M\bigwedge^{m-1}\Tan^*M} \\
{P_\circ} & {\mathcal{P}_\circ^* = P_\circ\times_M \bigwedge^{m-1}\Tan^*M} \\
&& {\bigwedge^{m-1}\Tan^*M} \\
E \\
& {M}
\arrow["\varrho", from=1-2, to=1-1]
\arrow["{\mathbf{h}}", bend left = 20, from=1-2, to=1-3]
\arrow["{\widetilde{\mathfrak{p}}}", from=1-3, to=1-2]
\arrow["{\jmath_\circ}"', hook, from=2-2, to=1-2]
\arrow["{\mathbf{h}_\circ}"', bend right = 20, from=2-2, to=1-3]
\arrow["{\varrho_\circ}", from=2-2, to=2-1]
\arrow["\upsilon"', from=2-2, to=3-3]
\arrow["{\overline{\tau}_\circ}", from=2-2, to=5-2]
\arrow["{j_{P_\circ}}", hook, from=2-1, to=1-1]
\arrow["{\kappa_\circ^1}"', from=2-1, to=4-1]
\arrow["{\overline{\kappa}_\circ^1}", from=2-1, to=5-2]
\arrow["{\tau_\Lambda}", from=3-3, to=5-2]
\arrow["\pi"', shift right=2, from=4-1, to=5-2]
\end{tikzcd}
\end{equation}

On $\mathcal{P}^*_\circ$ we use adapted coordinates $(x^\mu,y^A,p^\lambda_i,s^\mu)$, where
$p^\lambda_i$ are the free multimomenta; for the remaining pairs $(\eta,j)\in\Lambda$, the
multimomenta are determined by the constraints $p^\eta_j=g^\eta_j(x^\nu,y^B,p^\lambda_i)$.
We write $h_\circ$ for the local Hamiltonian function and use $P^\mu_A$ to indicate all the multimomentum coordinates, both the constrained and free ones. Then $\omega=\d^mx$ and
\begin{equation}
\label{eq:ThetaP0}
\Theta_{{\bf h}_\circ} = -P^\mu_A\,\d y^A\wedge\d^{m-1}x_\mu+h_\circ\,\d^mx+\d s^\mu\wedge\d^{m-1}x_\mu \, ,
\end{equation}
\begin{equation}
\label{eq:ThetaP02}
\d\Theta_{{\bf h}_\circ}=-\d P^\mu_A\wedge\d y^A\wedge\d^{m-1}x_\mu+\d h_\circ\wedge\d^mx \, .
\end{equation}

There are now three possible scenarios:
\begin{enumerate}\itemsep=-1pt
\item 
The pair $(\mathcal{P}_\circ^*,\Theta_{{\bf h}_\circ})$ is a multicontact
manifold.
Then, the formalism is the same as described in Section \ref{rhs}
but on $(\mathcal{P}^*_\circ,\Theta_{{\bf h}_\circ})$
instead of $(\mathcal{P}^*,\Theta_{\bf h})$.
(In Section \ref{ex:burguer} we present an example of this situation).
\item
The pair $(\mathcal{P}_\circ^*,\Theta_{{\bf h}_\circ})$ is neither a multicontact manifold nor a premulticontact manifold.
This case is not of interest.
\item 
The pair $(\mathcal{P}_\circ^*,\Theta_{{\bf h}_\circ})$ is a premulticontact
manifold.
This is the case studied below.
\end{enumerate}

To ensure that the pair $(\mathcal{P}_\circ^*,\Theta_{{\bf h}_\circ},\omega)$ is a premulticontact
manifold, we have to impose some additional conditions, which we study next.

Let $V(\overline\kappa^1_\circ)\subset\Tan P_\circ$ be the distribution of vectors tangent to
$P_\circ$ that are vertical over $M$, and let $\varrho_\circ\colon\mathcal{P}^*_\circ\to P_\circ$
be the natural projection. For every $q\in\mathcal{P}^*_\circ$ and every
$X\in V_{\varrho_\circ(q)}(\overline\kappa^1_\circ)$, there is a unique lift
$\widehat{X}_q \in V_q(\overline{\tau}_\circ)$ such that
\[
\Tan_q\varrho_\circ(\widehat{X}_q)=X ,\quad \iota(\widehat{X}_q)\Theta_{{\bf h}_\circ}=0 \,.
\]
In coordinates, taking $\ds X=X^A\derpar{}{y^A}\Big|_{\varrho_\circ(q)}+X^\lambda_i\derpar{}{p^\lambda_i}\Big|_{\varrho_\circ(q)}$, then
$\ds\widehat{X}_q=X^A\derpar{}{y^A}\Big|_q+X^\lambda_i\derpar{}{p^\lambda_i}\Big|_q+ P^\mu_A(q)\,X^A\derpar{}{s^\mu}\Big|_q$.
When there is no confusion, we write $\widehat X$ instead of $\widehat X_q$.
For every $q\in\mathcal{P}^*_\circ$, let us define
\begin{equation}\label{Kdef}
K_q\coloneqq\{X\in V_{\varrho_\circ(q)}(\overline\kappa^1_\circ)\, \mid\, \iota(\widehat{Y}_q)\iota(\widehat{X}_q)\d\Theta_{{\bf h}_\circ}=0\,,
\ \text{for every } Y\in V_{\varrho_\circ(q)}(\overline\kappa^1_\circ)\}\,.
\end{equation}
As shown in the proof of Proposition \ref{prop:restricted-ranks}, $K_q$ depends only on $\varrho_\circ(q)$.

\begin{proposition}\label{prop:restricted-ranks}
Let $m\geq 2$ and $q \in \mathcal{P}^*_\circ$. Then:
\begin{enumerate}[{\rm (i)}]\itemsep=-1pt
\item $\rank\mathfrak{R}_q=\rank K_q+m$.
\item For every $X\in K_q$ there is a unique $f_q(X)\in\mathbb{R}$ such that
$\iota(\widehat X)\d\Theta_{{\bf h}_\circ}=f_q(X)\,\omega$, and $f_q\colon K_q\to\mathbb{R}$ is linear. In coordinates,
\[ f_q(X)=\widehat X(h_\circ)+\sum_{(\mu,A)\in\Lambda}X^A\derpar{g^\mu_A}{x^\mu} \,. \]
\item 
$\rank(\ker\Theta_{{\bf h}_\circ} \cap \ker\d\Theta_{{\bf h}_\circ})_q=\rank K_q-\rank f_q$. 
\item $\Theta_{{\bf h}_\circ}$ is nowhere vanishing, and $\Theta_{{\bf h}_\circ}\notin\mathrm{Im}\,\flat_{\d\Theta_{{\bf h}_\circ}}$ at every point.
\end{enumerate}
\end{proposition}
\begin{proof}
Let $\ds Z=Z^\nu\derpar{}{x^\nu}\Big|_q+Z^A\derpar{}{y^A}\Big|_q+Z^\lambda_i\derpar{}{p^\lambda_i}\Big|_q+Z_s^\mu\derpar{}{s^\mu}\Big|_q \in \Tan_q\mathcal{P}^*_\circ$ be an arbitrary tangent vector, where
we call $\ds Z_\circ=Z^A\derpar{}{y^A}\Big|_{\varrho_\circ(q)}+Z^\lambda_i\derpar{}{p^\lambda_i}\Big|_{\varrho_\circ(q)}\in V_{\varrho_\circ(q)}(\overline\kappa^1_\circ)$. Note that if $Z\in\ker\omega_q$, then $\Tan_q\varrho_\circ(Z)=Z_\circ$.

Using the expression in coordinates in \eqref{eq:ThetaP0}, we get
\[
\innp{Z}\Theta_{{\bf h}_\circ}=\big(Z_s^\mu-P^\mu_AZ^A+h_\circ Z^\mu\big)\d^{m-1}x_\mu
+Z^\nu\big(P^\mu_A\,\d y^A-\d s^\mu\big)\wedge\d^{m-2}x_{\mu\nu}\,.
\]
Hence $\innp{Z}\Theta_{{\bf h}_\circ}=0$ if, and only if, $Z^\nu=0$
and $Z_s^\mu=P^\mu_AZ^A$, that is, there exists $X \in V_{\varrho_\circ(q)}(\overline{\kappa}^1_\circ)$ such that its lift $\widehat X_q$ coincides with $Z$. Therefore
\begin{equation}
\label{eq:kerTheta0}
(\ker\Theta_{{\bf h}_\circ})_q=\{\widehat X_q\mid X\in V_{\varrho_\circ(q)}(\overline\kappa^1_\circ)\}\subset\ker\omega_q\,.
\end{equation}
On the other hand, contracting $Z\in\ker\omega_q$ with \eqref{eq:ThetaP02} gives 
\begin{equation}\label{eq:iZdTheta0v}
\innp{Z}\d\Theta_{{\bf h}_\circ}=\varphi_\mu(Z_\circ)\wedge\d^{m-1}x_\mu+\big(Z(h_\circ)+F_AZ^A\big)\,\d^mx\,,
\end{equation}
where the operator $\ds\d^{\rm v}$ denotes the differential along the vertical directions over $M$ (in particular, 
since the functions $P^\mu_A$ do not depend on $s^\nu$, $\d^{\rm v}P^\mu_A$ contains only $\d y^B$ and $\d p^\lambda_i$),
\[
F_A\coloneqq\sum_{\mu=1}^m\derpar{P^\mu_A}{x^\mu}=\sum_{\mu\,:\,(\mu,A)\in\Lambda}\derpar{g^\mu_A}{x^\mu}\,,
\qquad
\varphi_\mu(Z_\circ)\coloneqq Z^A\,\d^{\rm v}P^\mu_A-Z_\circ(P^\mu_A)\,\d y^A\,.
\]

Finally, for
$X,Y\in V_{\varrho_\circ(q)}(\overline\kappa^1_\circ)$, applying \eqref{eq:iZdTheta0v} to $Z=\widehat X_q$ and contracting with $\widehat Y_q$, we get $\iota(\widehat Y_q)\iota(\widehat X_q)\d\Theta_{{\bf h}_\circ}=\varphi_\mu(X)(Y)\,\d^{m-1}x_\mu$. Since $\varphi_\mu(X)$ only contains $\d y^B$ and
$\d p^\lambda_i$, we conclude that
\begin{equation}\label{eq:Kphi}
K_q=\{X\in V_{\varrho_\circ(q)}(\overline\kappa^1_\circ)\mid \varphi_\mu(X)=0\,,\ \mu=1,\dots,m\}\,.
\end{equation}
As the functions $P^\mu_A$ do not depend on $s^\nu$, this condition depends only on $\varrho_\circ(q)$, and so does $K_q$.

\smallskip
\noindent(i) \ 
By definition, $\mathfrak{R}\subset\ker\omega$. Let $Z,Y\in\ker\omega_q$, equation \eqref{eq:iZdTheta0v} gives
\[
\innp{Y}\innp{Z}\d\Theta_{{\bf h}_\circ}=\varphi_\mu(Z_\circ)(Y_\circ)\,\d^{m-1}x_\mu\,,\quad
Y_\circ=\Tan_q\varrho_\circ(Y)\,.
\]
Hence, by
\eqref{eq:Kphi}, $Z\in\mathfrak{R}_q$ if, and only if, $Z_\circ\in K_q$, therefore, we obtain
\[
\mathfrak{R}_q=\{\widehat X_q\mid X\in K_q\}\oplus\Big\langle\derpar{}{s^1}\Big|_q,\dots,\derpar{}{s^m}\Big|_q\Big\rangle\,,
\]
and thus $\rank\mathfrak{R}_q=\dim K_q+m$.

\smallskip
\noindent(ii) \
Let $X\in K_q$. Then $\varphi_\mu(X)=0$ by \eqref{eq:Kphi}, and $\widehat X\in\ker\omega$, so
\eqref{eq:iZdTheta0v} gives
\[
\iota(\widehat X)\d\Theta_{{\bf h}_\circ}=\big(\widehat X(h_\circ)+F_AX^A\big)\,\d^mx \eqqcolon f_q(X)\,\omega\,.
\]
The function $f_q(X)$ is unique because $\omega_q\neq 0$. It is linear in $X$ because the lift is
linear and so is the contraction.

\smallskip
\noindent(iii) \
Let $Z\in(\ker\Theta_{{\bf h}_\circ}\cap\ker\d\Theta_{{\bf h}_\circ})_q$. By \eqref{eq:kerTheta0}, $Z=\widehat X$ with
$X\in V_{\varrho_\circ(q)}(\overline\kappa^1_\circ)$, and $Z\in\ker\omega$. In \eqref{eq:iZdTheta0v}, the forms
$\varphi_\mu(X)\wedge\d^{m-1}x_\mu$ and $\d^mx$ are linearly independent, so
$\innp{Z}\d\Theta_{{\bf h}_\circ}=0$ implies $\varphi_\mu(X)=0 $ for every $\mu$, (that is, $X\in K_q$), and then
$f_q(X)=0$ by (ii). Conversely, if $X\in K_q$ and $f_q(X)=0$, then $\widehat X\in\ker\Theta_{{\bf h}_\circ}$ and
$\innp{\widehat X}\d\Theta_{{\bf h}_\circ}=0$ by (ii). Since the lift is linear and injective,
\[
\rank(\ker\Theta_{{\bf h}_\circ}\cap\ker\d\Theta_{{\bf h}_\circ})_q=\dim\ker f_q=\dim K_q-\rank f_q\,,
\]
and $\rank f_q\in\{0,1\}$ because $f_q$ is a linear form.

\smallskip
\noindent(iv) \
At every point, $\ds\iota\Big(\derpar{}{ s^\nu}\Big)\Theta_{{\bf h}_\circ}=\d^{m-1}x_\nu\neq 0$, so $\Theta_{{\bf h}_\circ}$ is nowhere
vanishing. Suppose now that $\innp{Z}\d\Theta_{{\bf h}_\circ}=\Theta_{{\bf h}_\circ}$ at some point $q$, for some
$Z\in\Tan_q\mathcal{P}^*_\circ$. Since the functions $P^\mu_A$ do not depend on $s^\nu$,
\eqref{eq:ThetaP02} gives $\ds\iota\Big(\derpar{}{ s^\nu}\Big)\d\Theta_{{\bf h}_\circ}=\derpar{h_\circ}{s^\nu}\,\d^mx$.
Hence, for $\nu=1,\dots,m$,
\[
\d^{m-1}x_\nu=\iota\Big(\derpar{}{ s^\nu}\Big)\Theta_{{\bf h}_\circ}
=\iota\Big(\derpar{}{ s^\nu}\Big)\innp{Z}\d\Theta_{{\bf h}_\circ}
=-\innp{Z}\iota\Big(\derpar{}{ s^\nu}\Big)\d\Theta_{{\bf h}_\circ}
=-\derpar{h_\circ}{s^\nu}\,\innp{Z}\d^mx\,.
\]
Thus, the $m$ linearly independent forms $\d^{m-1}x_\nu$ would all be multiples of the single
form $\innp{Z}\d^mx$, which is impossible because $m\geq 2$.
\end{proof}

It follows from the Proposition \ref{prop:restricted-ranks} and Definition \ref{genpremulticontact} that the pair $(\mathcal{P}_\circ^*,\Theta_{{\bf h}_\circ})$ defines a premulticontact manifold if both $\dim K_q$  and $\rank f_q$ are constant.

\begin{remark}
We assume the manifold $(\mathcal{P}^*_\circ, \Theta_{{\bf h}_\circ})$ to be premulticontact, and we let $X_\circ$ be a section of K and $\widehat{X}_\circ$ its lift. Then, using Cartan's calculus we get 
\[ \Lie_{\widehat{X}_\circ}\Theta_{{\bf h}_\circ} =\d\,\iota(\widehat{X}_\circ)\Theta_{{\bf h}_\circ}+\iota(\widehat{X}_\circ)\d\Theta_{{\bf h}_\circ} =\iota(\widehat{X}_\circ)\d\Theta_{{\bf h}_\circ} =f_{X_\circ}\,\omega \, .\]
Hence the function $f_{X_\circ}$ measures the failure of $\widehat{X}_\circ$ to be an infinitesimal symmetry of $\Theta_{{\bf h}_\circ}$.
\end{remark}
Then, to make the pair $(\mathcal{P}^*_\circ, \Theta_{{\bf h}_\circ})$ special premulticontact, we need to fix $f = 0$. Thus, from Remark \ref{alpha_especial_and_alpha_pre}, we can define the corresponding Lichnerowicz differential operator $\bd$, and
\[
\bd \Theta_{{\bf h}_\circ}=\d \Theta_{{\bf h}_\circ}+\sigma_{\Theta_{{\bf h}_\circ}}\wedge \Theta_{{\bf h}_\circ}\in \df^{m+1}(\mathcal{P}_\circ^*) \,.
\]

\begin{definition}
The triple $(\mathcal{P}_\circ^*,\Theta_{{\bf h}_\circ},\omega)$
is called a \dfn{restricted Hamiltonian premulticontact system}.
\end{definition}

\begin{remark}
For a restricted Hamiltonian premulticontact system
$({\cal P}_\circ^*,\Theta_{{\bf h}_\circ},\omega)$,
the field equations for sections
$\bm{\psi}_\circ\colon M\to \mathcal{P}_\circ^*$,
$\overline\tau_\circ$-transverse multivector fields
${\bf X}_{{\bf h}_\circ}\in \vf^m(\mathcal{P}_\circ^*)$,
and Ehresmann connections $\nabla_{{\bf h}_\circ}$ on
$\mathcal{P}_\circ^*$
are analogous to \eqref{sect1Hr}--\eqref{EcHr}, replacing
$\Theta_{\bf h}$ with $\Theta_{{\bf h}_\circ}$:
\begin{align}
\label{sect1Hb0}
& \bm{\psi}_\circ^*\Theta_{{\bf h}_\circ}=0 \,,
&&
\bm{\psi}_\circ^*\innp{Y_\circ}\,\bd \Theta_{{\bf h}_\circ}=0,
\ \text{ for every $Y_\circ\in\mathfrak{X}({\cal P}_\circ^*)$} \, ,
\\
\label{sect2H0}
& \innp{\bm{\psi}_\circ^{(m)}}
(\Theta_{{\bf h}_\circ}\circ \bm{\psi}_\circ)=0 \, ,
&&
\innp{\bm{\psi}_\circ^{(m)}}
(\bd\Theta_{{\bf h}_\circ}\circ \bm{\psi}_\circ)=0 \, ,
\\
\label{vfHb0}
& \innp{\bfX_{{\bf h}_\circ}}\,\Theta_{{\bf h}_\circ}=0 \, ,
&&
\innp{\bfX_{{\bf h}_\circ}}\,\bd \Theta_{{\bf h}_\circ}=0 \, ,
\\
\label{EcH00}
& \innp{\nabla_{{\bf h}_\circ}}\,\Theta_{{\bf h}_\circ}
=(m-1)\Theta_{{\bf h}_\circ} \, ,
&&
\innp{\nabla_{{\bf h}_\circ}}\,\bd\Theta_{{\bf h}_\circ}
=(m-1)\,\bd\Theta_{{\bf h}_\circ} \, .
\end{align}
In general, these equations are not compatible in $\mathcal{P}_\circ^*$.
Hence, a constraint algorithm is required to determine a final
constraint submanifold
${\cal P}_f^*\subseteq{\cal P}_\circ^*$,
if it exists, where there are (integrable) multivector fields or connections solutions to the field equations, such that they are tangent to
${\cal P}_f^*$.
(See \cite{LMMMR_05} for the premultisymplectic case).
\end{remark}

\subsection{Extended Hamiltonian premulticontact systems}

Let $ \jmath_\circ\colon \mathcal{P}_\circ^*= P_\circ\times_M \bigwedge\nolimits^{m-1}\Tan^*M
\hookrightarrow
{\cal P}^*$
be the submanifold introduced in the previous Section \ref{rHps}.
Then, we define the submanifold $\widetilde\jmath_\circ\colon \widetilde{\mathcal{P}}_\circ
\hookrightarrow
\widetilde{\cal P}=\mathcal{M}\pi\times_M \bigwedge^{m-1}\Tan^*M$ by
$$
\widetilde{\mathcal{P}}_\circ\coloneqq\widetilde{\mathfrak{p}}^{-1}(\mathcal{P}^*_\circ)
=\widetilde{P}_\circ\times_M \bigwedge\nolimits^{m-1}\Tan^*M \,,
$$
where $\widetilde P_\circ\coloneqq\mathfrak{p}^{-1}(P_\circ)\hookrightarrow \mathcal{M}\pi$,
which is a $\widetilde{\kappa}^1$-transverse submanifold of the bundle
$\widetilde{\kappa}^1\colon\mathcal{M}\pi\to E$ (see Diagram \eqref{firstdiag}),
since $\widetilde\kappa^1=\kappa^1\circ\mathfrak{p}$, and $P_\circ$ is $\kappa^1$-transverse
(the coordinate $p$ is not constrained on $\widetilde{\mathcal{P}}_\circ$).
Recalling that $\widetilde\tau=\overline\tau\circ\widetilde{\mathfrak{p}}\colon\widetilde{\mathcal{P}}\to M$ and $\widetilde\kappa=\tau\circ\widetilde{\mathfrak{p}}\colon\widetilde{\mathcal{P}}\to\mathscr{E}$, 
the maps $\kappa_\circ=\widetilde\kappa\circ\widetilde\jmath_\circ\colon\widetilde{\mathcal{P}}_\circ\to\mathscr{E}$ and $\widetilde\tau_\circ=\widetilde\tau\circ\widetilde\jmath_\circ=\tau_M\circ\kappa_\circ\colon\widetilde{\mathcal{P}}_\circ\to M$ are surjective submersions, and we have the diagram
\begin{equation}
\label{diagsub}
\begin{tikzcd}[row sep=large, column sep=huge]
&\mathcal{P}_\circ^* & & \mathcal{P}^* &  
\\
& \widetilde{\mathcal{P}}_\circ & & \widetilde{\mathcal{P}} \\
& & \mathscr{E} \\
	& & M
\arrow["\tau_\circ"{pos=0.2}, bend left=15, from=1-2, to=3-3]
\arrow["\overline{\tau}_\circ"', bend right=60, from=1-2, to=4-3]
\arrow["\jmath_\circ", from=1-2, to=1-4]
\arrow["\overline{\tau}", bend left=60, from=1-4, to=4-3]
\arrow["\tau"{pos=0.3}, bend right=15, from=1-4, to=3-3]
\arrow["\widetilde{\mathfrak{p}}_\circ"', from=2-2, to=1-2]
\arrow["\overline{\mathfrak{p}}_\circ", crossing over, from=2-2, to=1-4]
\arrow["\widetilde\jmath_\circ", crossing over, from=2-2, to=2-4]
\arrow["\kappa_\circ"', from=2-2, to=3-3]
\arrow["\widetilde{\tau}_\circ"', bend right=20, from=2-2, to=4-3]
\arrow["\widetilde{\mathfrak{p}}", from=2-4, to=1-4]
\arrow["\widetilde\kappa", from=2-4, to=3-3]
\arrow["\widetilde{\tau}", bend left=20, from=2-4, to=4-3]
\arrow["\tau_M", from=3-3, to=4-3]
\end{tikzcd}
\end{equation}
where $\widetilde{\mathfrak{p}}_\circ\colon\widetilde{\mathcal{P}}_\circ\to\mathcal{P}^*_\circ$
is the restriction of the projection $\overline{\mathfrak{p}}_\circ=\widetilde{\mathfrak{p}}\circ\widetilde\jmath_\circ
\colon\widetilde{\mathcal{P}}_\circ\to\mathcal{P}^*$ to the image.

The natural adapted coordinates in $\widetilde{\mathcal{P}}_\circ$ are denoted by $(x^\mu,y^A,p_i^\lambda,p,s^\mu)$. 

Notice that if $\zeta^j\in\Cinfty(\mathcal{P}^*)$ are constraints defining locally $\mathcal{P}_\circ^*$ in $\mathcal{P}^*$,
then the constraints $\widetilde{\mathfrak{p}}^*\zeta^j\in\Cinfty(\widetilde{\mathcal{P}})$
define locally $\widetilde{\mathcal{P}}_\circ$ in $\widetilde{\mathcal{P}}$.

\begin{definition}
\label{canform0}
The \dfn{canonical form} in $\widetilde{\P}_\circ$ is $\widetilde\Theta_\circ \coloneqq \widetilde\jmath_\circ^{\,*}\widetilde\Theta\in\df^m(\widetilde{\mathcal{P}}_\circ)$.
\end{definition}

As was the case in the previous section, we have the following three scenarios:
\begin{enumerate}\itemsep=-1pt
\item 
The pair $(\widetilde{\mathcal{P}}_\circ,\widetilde\Theta_\circ)$ is a multicontact
manifold.
Then, the formalism is the same as described in Section \ref{subsec:extended-Hamiltonian}
but on $(\widetilde{\mathcal{P}}_\circ,\widetilde\Theta_\circ)$
instead of $(\widetilde{\mathcal{P}},\widetilde\Theta)$.
(See the example in Section \ref{ex:burguer}).
\item
The pair $(\widetilde{\mathcal{P}}_\circ,\widetilde\Theta_\circ)$ is neither a multicontact manifold nor a premulticontact manifold.
This case is not of interest.
\item 
The pair $(\widetilde{\mathcal{P}}_\circ,\widetilde\Theta_\circ)$ is a premulticontact manifold,
which is the case we study next.
\end{enumerate}

\begin{remark}
Following similar steps as in the proof of Proposition \ref{prop:restricted-ranks},
one can show that the pair $(\widetilde{\mathcal{P}}_\circ,\widetilde\Theta_\circ)$ is a premulticontact
manifold if, and only if, the subspaces $K_q$ defined in \eqref{Kdef} have constant dimension.
In particular, if both $\dim K_q$ and $\rank f_q$ are constant, so that
$(\mathcal{P}^*_\circ,\Theta_{{\bf h}_\circ})$ is a premulticontact manifold, then
$(\widetilde{\mathcal{P}}_\circ,\widetilde\Theta_\circ)$ is a premulticontact manifold too.
\end{remark}

\begin{definition}
\label{hameqbis3}
Consider a premulticontact manifold $(\widetilde{\mathcal{P}}_\circ,\widetilde\Theta_\circ)$ and
a \dfn{Hamiltonian density}
$\mathcal{H}_\circ\in\df^m(\widetilde{\mathcal{P}}_\circ)$ which is a $\widetilde\tau_\circ$-semibasic form
$\mathcal{H}_\circ={\mathrm{H}}_\circ\,\omega$, where ${\mathrm{H}}_\circ\in\Cinfty(\widetilde{\mathcal{P}}_\circ)$ is the associated \dfn{Hamiltonian function}.

\noindent We say that $(\widetilde{\mathcal{P}}_\circ,\widetilde\Theta_\circ;\mathcal{H}_\circ)$ is
an \dfn{extended Hamiltonian premulticontact system} if the exact form $\alpha_\circ:=\d{\rm H}_\circ\in\df^1(\widetilde{\mathcal{P}}_\circ)$ satisfies the properties of Proposition \ref{prop:existence_sigma}.
It is called the \dfn{Hamiltonian form} associated with $\mathcal{H}_\circ$
and, obviously, $\d\mathcal{H}_\circ=\alpha_\circ\wedge\omega$.

\noindent The form $\alpha_\circ$ defines a dissipation form $\sigma_{\alpha_\circ}$ which is used to construct the corresponding Lichnerowicz differential $\bd$
(Def. \ref{bard}).
Then, the field equations of the system are
\begin{equation}
\label{hameq20}
\innp{{\bf X}_{\alpha_\circ}}(\widetilde\Theta_\circ + \mathcal{H}_\circ) = 0\, , \quad
\innp{{\bf X}_{\alpha_\circ}}\overline{\mathrm{d}}\widetilde\Theta_\circ=(-1)^{m-1}\alpha_\circ  \, ;
\end{equation}
where ${\bf X}_{\alpha_\circ}\in \vf^m(\widetilde{\mathcal{P}}_\circ)$ are $\widetilde{\tau}_\circ$-transverse, locally decomposable multivector fields,
and $\widetilde{\tau}_\circ$-transversality can be imposed by demanding $\innp{{\bf X}_{\alpha_\circ}}\omega=1$.
\end{definition}

\begin{remark}
Notice that the condition on $\alpha_\circ$ in Definition \ref{hameqbis3} holds if, and only if,
the linear form $f$ of Proposition \ref{prop:restricted-ranks}~(ii), associated with
${\rm h}_\circ={\rm H}_\circ-p$, vanishes identically.
Indeed, the hypothesis of Proposition \ref{prop:existence_sigma} reads
\[
\derpar{{\rm H}_\circ}{p}=1 \, , \quad
\widehat X({\rm H}_\circ)+F_AX^A=0 \, , \ \text{for every } X\in K \, .
\]
Since $\widehat X(p)=0$, the second condition is $\widehat X({\rm h}_\circ)+F_AX^A=f(X)=0$.
\end{remark}

\begin{remark}
As for extended Hamiltonian multicontact systems, in general, multivector fields solution to  \eqref{hameq20}
do not necessarily have to be integrable, and this condition must be imposed additionally.
\end{remark}

As in Proposition \ref{Hprop} and Corollary \ref{orthog}, we have the following characterisation of Hamiltonian densities and Hamiltonian forms defining extended Hamiltonian premulticontact systems:

\begin{proposition}
\label{Hprop2}
Let $(\widetilde{\mathcal{P}}_\circ, \widetilde\Theta_\circ;\mathcal{H}_\circ)$ be an extended Hamiltonian premulticontact system
and $\alpha_\circ$ the associated Hamiltonian form. Then,
for every
$\widetilde{\mathfrak{p}}_\circ$-vertical vector field $Y_\circ\in\vf^{{\rm V}(\widetilde{\mathfrak{p}}_\circ)}(\widetilde{\cal P}_\circ)$, we have 
$$
\innp{Y_\circ}\d\widetilde\Theta_\circ=-Y_\circ({\rm H}_\circ)\,\omega \, .
$$
As a consequence, in a chart of natural coordinates $U_\circ\subset\widetilde{\mathcal{P}}_\circ$, 
adapted to the bundle structure (with $\omega=\d^mx$), we find
$\ds\derpar{{\rm H}_\circ}{p}=1$ and hence, there exists
${\rm h}_\circ=\widetilde{\mathfrak{p}}_\circ^*h_\circ\in\Cinfty(U_\circ)$, with $h_\circ\in\Cinfty(\widetilde{\mathfrak{p}}_\circ(U_\circ))$, such that
\begin{equation}
{\rm H}_\circ\underset{U_\circ}{=}p+{\rm h}_\circ \, .
\label{Hcoor2}
\end{equation}
\end{proposition}
\begin{proof}
In natural coordinates,
$$\widetilde\Theta_\circ=-P^\mu_A\,\d y^A\wedge\d^{m-1}x_\mu-p\,\d^mx+\d s^\mu\wedge\d^{m-1}x_\mu\, ,$$
where the functions $P^\mu_A$ do not depend on $p$. Hence
$$\ds\innp{\derpar{}{p}}\widetilde\Theta_\circ=0 \ \text{ and } \ \ds\innp{\derpar{}{p}}\d\widetilde\Theta_\circ=-\omega \, ,$$
so $\partial/\partial p\in\ker\flat_{\widetilde\Theta_\circ}$.
Every $\widetilde{\mathfrak{p}}_\circ$-vertical vector field is $\ds Y_\circ=g\,\derpar{}{p}$, for some function $g$, and then
the hypothesis of Proposition \ref{prop:existence_sigma} gives
\[
\innp{Y_\circ}\d\widetilde\Theta_\circ=-\innp{Y_\circ}(\alpha_\circ\wedge\omega)
=-\alpha_\circ(Y_\circ)\,\omega=-Y_\circ({\rm H}_\circ)\,\omega \, .
\]
Taking $g=1$ we obtain $\partial{\rm H}_\circ/\partial p=1$, and therefore ${\rm H}_\circ-p$ does not depend on $p$
on $U_\circ$, which gives \eqref{Hcoor2}.
\end{proof}

As an obvious corollary, from the previous Proposition \ref{Hprop2}, we get: 

\begin{corollary}
\label{orthog2}
For the Hamiltonian form $\alpha_\circ$ the following relation holds, for every
$\widetilde{\mathfrak{p}}_\circ$-vertical vector field $Y_\circ\in\vf^{{\rm V}(\widetilde{\mathfrak{p}}_\circ)}(\widetilde{\cal P}_\circ)$,
\begin{equation}
\label{ortogonal2}
{\alpha_\circ}(Y_\circ)\omega = -\innp{Y_\circ}\d\widetilde\Theta_\circ \, ; 
\end{equation}
hence, in a chart of natural coordinates, we obtain \
$\ds\innp{\derpar{}{p}}{\alpha_\circ}=1$.
Therefore, there exists an exact and $\widetilde{\mathfrak{p}}_\circ$-basic local $1$-form $\beta_\circ\in\df^1(U_\circ)$
such that
\begin{equation}
\alpha_\circ \underset{U_\circ}{=}\d p+\beta_\circ
\equiv\d p+\d{\rm h}_\circ \, .
\label{alphacoor2}
\end{equation}    
\end{corollary}

As in the multicontact case, the field equations for extended Hamiltonian premulticontact systems
can be written as follows:

\begin{definition}
\label{hameqbis20}
Let $(\widetilde{\mathcal{P}}_\circ,\widetilde\Theta_\circ;\mathcal{H}_\circ)$ be
an extended Hamiltonian premulticontact system.
The field equations can be written in the following equivalent ways:
\begin{enumerate}[{\rm(i)}]\itemsep=-1pt
\item 
The \dfn{extended premulticontact Herglotz--Hamilton--de Donder--Weyl equations for sections} $\widetilde{\bm\psi}_\circ\colon M\to\widetilde{\cal P}_\circ$ are
\begin{equation}
\label{sect1H0}
\widetilde{\bm\psi}_\circ^*(\widetilde\Theta_\circ+\mathcal{H}_\circ)=0 \,,\qquad
\widetilde{\bm\psi}_\circ^*\innp{Y}\bd\widetilde\Theta_\circ=- {(\widetilde{\bm\psi}_\circ^*\innp{Y}\alpha_\circ)\,\omega} \,,\quad
\text{for every }Y\in\mathfrak{X}(\widetilde{\cal P}_\circ) \, , 
\end{equation}
or equivalently
\begin{equation}
\label{sect2H0b}
\innp{\widetilde{\bm\psi}_\circ^{(m)}}(\widetilde\Theta_\circ+\mathcal{H}_\circ)\circ\widetilde{\bm\psi}_\circ=0 \,,\qquad
\innp{\widetilde{\bm\psi}_\circ^{(m)}}(\bd\widetilde\Theta_\circ\circ\widetilde{\bm\psi}_\circ)=(-1)^{m-1}\alpha_\circ \circ\widetilde{\bm\psi}_\circ \,. 
\end{equation}
\item 
The \dfn{extended premulticontact Herglotz--Hamilton--de Donder--Weyl equations for $\widetilde{\tau}_\circ$-transverse locally decomposable multivector fields} ${\bf X}_{\alpha_\circ}\in\mathfrak{X}^m(\widetilde{\cal P}_\circ)$ are
\begin{equation}
\label{vfH0}
\innp{{\bf X}_{\alpha_\circ}}(\widetilde\Theta_\circ+\mathcal{H}_\circ)=0 \,,\qquad
\innp{{\bf X}_{\alpha_\circ}}\bd\widetilde\Theta_\circ=(-1)^{m-1}{\alpha_\circ} \,, 
\end{equation}
where $\widetilde{\tau}_\circ$-transversality can be imposed by demanding $\innp{{\bf X}_{\alpha_\circ}}\omega=1$.
\item 
The \dfn{extended premulticontact Herglotz--Hamilton--de Donder--Weyl equations for Ehresmann connections} $\nabla_\circ$ on $\widetilde{\cal P}_\circ\to M$ are
\begin{equation}
\label{EcH0}
\innp{\nabla_\circ}(\widetilde\Theta_\circ+\mathcal{H}_\circ)=(m-1)(\widetilde\Theta_\circ+\mathcal{H}_\circ)\,,\qquad
\innp{\nabla_\circ}\bd\widetilde\Theta_\circ=(m-1)\bd\widetilde\Theta_\circ+(-1)^{m-1}\omega\wedge\alpha_\circ\,. 
\end{equation}
\end{enumerate}
\end{definition}

\begin{remark}
As the geometric structure of these kinds of systems is premulticontact,
in general, the field equations are not compatible on $\widetilde{\mathcal{P}}_\circ$,
and a constraint algorithm is required to determine a final
constraint submanifold $\widetilde{\cal P}_f\subseteq\widetilde{\mathcal{P}}_\circ$,
if it exists, where there are multivector fields solutions which are tangent to $\widetilde{\cal P}_f$.
\end{remark}

Then, as for extended multicontact systems, we have the following result:

\begin{theorem}
\label{holsecreg4}
Let $(\widetilde{\mathcal{P}}_\circ,\widetilde\Theta_\circ;\mathcal{H}_\circ)$ be
an extended Hamiltonian premulticontact system.
Then, if there are $\widetilde{\tau}_\circ$-transverse, locally decomposable multivector fields
${\bf X}_{\alpha_\circ}\in\vf^m(\widetilde{\cal P}_\circ)$
and connections
solution to equations \eqref{hameq20} and \eqref{EcH0}, respectively
(maybe on a submanifold $\widetilde{\mathcal{P}}_f\subseteq\widetilde{\mathcal{P}}_\circ$), then their integral sections $\widetilde{\bm\psi}_\circ(x) =(x^\mu,y^A(x^\mu),p^\lambda_i(x^\mu),p(x^\mu), s^\nu(x^\mu))$
are the solutions to the equations \eqref{sect1H0} or \eqref{sect2H0b}
(on $\widetilde{\mathcal{P}}_f$).
\end{theorem}

Finally, following the same pattern as for Corollary \ref{tangency},
we can prove:

\begin{corollary}
\label{tangency2}
If ${\bf X}_{\alpha_\circ}\in\vf^m(\widetilde{\cal P}_\circ)$ is a solution
to equations \eqref{vfH0}, then
it is tangent to the submanifold $\mathcal{S}_\circ\subset\widetilde{\mathcal{P}}_\circ$ defined by the constraint ${\rm H}_\circ=p+{\rm h}_\circ=0$
(and to all the level sets of ${\rm H}_\circ$).    
\end{corollary}

\begin{remark}
Generic extended Hamiltonian premulticontact systems can be defined in an analogous way as in Section \ref{gehms}.
\end{remark}

\subsection{Relation between extended and 
restricted Hamiltonian premulticontact systems}
\protect\label{rearrhs}

Finally, the relation between extended and 
restricted Hamiltonian premulticontact systems is established 
as in Section \ref{rerhs}. 

First, observe that every Hamiltonian section ${\bf h}_\circ\colon\mathcal{P}^*_\circ\to\widetilde{\mathcal{P}}$
defines a section $\overline{\bf h}_\circ\colon\mathcal{P}^*_\circ\to\widetilde{\mathcal{P}}_\circ$
of the projection $\widetilde{\mathfrak{p}}_\circ$,
such that ${\bf h}_\circ=\widetilde{\jmath}_\circ\circ\overline{\bf h}_\circ$
(see Diagrams \eqref{diagbig} and \eqref{diagsub}).
Then, following the same patterns as in the proofs of Theorems \ref{related} and \ref{existuniq}, we can prove the following:

\begin{theorem}
\label{related20}
Let $(\widetilde{\mathcal{P}}_\circ,\widetilde\Theta_\circ;\mathcal{H}_\circ)$ 
be an extended Hamiltonian premulticontact system.
Then, there exists a restricted Hamiltonian premulticontact system 
$({\cal P}_\circ^*,\Theta_{{\bf h}_\circ},\omega)$ defined by a Hamiltonian section $\overline{\bf h}_\circ\colon\mathcal{P}^*_\circ\to\widetilde{\mathcal{P}}_\circ$
such that $\overline{\bf h}_\circ^*\alpha_\circ=0$.

\noindent Furthermore, if ${\bf X}_{\alpha_\circ}\in\vf^m(\widetilde{\cal P}_\circ)$ 
is a solution to  \eqref{vfH0},
then there exists ${\bf X}_{{\bf h}_\circ}\in\vf^m(\mathcal{P}_\circ^*)$ which
is $\overline{\bf h}_\circ$-related to ${\bf X}_{\alpha_\circ}$
and is a solution to  \eqref{vfHb0}.

\noindent In particular, if ${\bf X}_{\alpha_\circ}$ is integrable, then
${\bf X}_{{\bf h}_\circ}$ is also integrable, and
if $\widetilde{\bm\psi}_\circ\colon M\to\widetilde{\cal P}_\circ$ is an integral
section of ${\bf X}_{\alpha_\circ}$, then 
$\overline{\bm\psi}_\circ=\widetilde{\mathfrak{p}}_\circ\circ\widetilde{\bm\psi}_\circ\colon M\to\mathcal{P}_\circ^*$
is an integral section of ${\bf X}_{{\bf h}_\circ}$.
\end{theorem}

\begin{proposition}
\label{existuniq0}
Let $({\cal P}_\circ^*,\Theta_{{\bf h}_\circ},\omega)$ be a
restricted Hamiltonian premulticontact system defined by a Hamiltonian section ${\bf h}_\circ\colon\mathcal{P}_\circ^*\to\widetilde{\mathcal{P}}$,
and let $\mathcal{S}_\circ={\rm Im}\,{\overline{\bf h}}_\circ\hookrightarrow\widetilde{\cal P}_\circ$. Assume also that the linear form $f$ of Proposition \ref{prop:restricted-ranks}~(ii) vanishes identically.
Then, there exists a unique form
$\alpha_\circ\in\df^1(\widetilde{\cal P}_\circ)$ such that:
\begin{enumerate}[{\rm(i)}] \itemsep = -1pt
\item The form $\alpha_\circ$ is {exact}, $\alpha_\circ = \d {\mathrm{H}_\circ}$, where ${\mathrm{H}_\circ}$ is defined by $- \widetilde{\Theta}_\circ + \widetilde{\mathfrak{p}}_\circ^\ast \Theta_{{\bf h}_\circ} = {\mathrm{H}_\circ}\,\omega$.
\item $\overline{\bf h}_\circ^{\, \ast}\alpha_\circ=0$.
\item $\innp{Y_\circ}{\alpha_\circ}\neq0$, for every non-vanishing 
$Y_\circ\in\vf^{{\rm V}(\widetilde{\mathfrak{p}}_\circ)}(\widetilde{\cal P}_\circ)$ and, in particular, \(\textstyle\innp{\parder{}{p}}{\alpha_\circ}=1\).
\end{enumerate}
Thus, $(\widetilde{\mathcal{P}}_\circ,\widetilde\Theta_\circ;\mathcal{H}_\circ)$, where $\mathcal{H}_\circ={\mathrm{H}_\circ}\,\omega=- \widetilde{ \Theta}_\circ + \widetilde{\mathfrak{p}}_\circ^\ast \Theta_{{\bf h}_\circ}$, 
is an extended Hamiltonian premulticontact system.
\begin{enumerate}[{\rm(iv)}] 
\item 
Furthermore, if ${\bf X}_{{\bf h}_\circ}\in\vf^m(\mathcal{P}_\circ^*)$ is an (integrable) solution to  \eqref{vfHb0}, then
${\bf X}_{\alpha_\circ}=(\overline{\bf h}_\circ)_*{\bf X}_{{\bf h}_\circ}\in\vf^m(\widetilde{\cal P}_\circ)$ 
is an (integrable) solution to  \eqref{vfH0}.
\end{enumerate}
\end{proposition}

\section{Examples}
\protect\label{exs}

\subsection{Klein--Gordon equation and the telegrapher equation}

The {\sl Klein--Gordon equation}, 
\begin{equation}\label{klein-gordon}
(\Box + {\rm m}^2) y
\equiv \parderr{y}{x_\mu}{x^\mu}+{\rm m}^2 y = 0
\,,
\end{equation}
is one of the most important equations in classical and quantum field theory \cite{IZ}.
In it, $y(x^\mu)$ is a scalar field defined in space-time $\R^4$ endowed with the Minkowski metric $g_{\mu\nu}$
(with signature $+---$),
${\rm m}\geq 0$ is a constant parameter, $\ds\Box\equiv\parderr{}{x_\mu}{x^\mu}$ denotes the d'Alembert operator in $\R^4$, 
where $x_\mu=g_{\mu\nu}x^\nu$.

The multisymplectic formulation of this theory \cite{GGR-2023}
is developed using space-time coordinates $(x^\mu)$, $\mu=0,\ldots,3$, and
denoting by $y$ the field variable and by $y_\mu$ the corresponding multivelocities.
Then $(x^\mu,y,y_\mu)$ are the natural local coordinates in the multivelocity phase space $J^1\pi$.
The standard Lagrangian function for the Klein--Gordon equation is
\begin{equation}
\label{klein-gordon-lagrangian}
L_0(x^\mu,y,y_\mu)=
\frac{1}{2}\,y_\mu\, y^\mu-\frac{1}{2} {\rm m}^2 y^2 \,,
\end{equation}
where $y^\mu=g^{\mu\nu}y_\nu$.
This Lagrangian function can be slightly modified to include a more generic potential,
$\displaystyle L_0=\frac{1}{2}\,y_\mu\,y^\mu-V(y)$; however, we restrict ourselves to the particular case where $V=\frac{1}{2}{\rm m}^2 y^2$.

\subsubsection{Restricted Hamiltonian formalism}

An action-dependent version of this theory has been studied in \cite{LGMRR_25},
where 
the following modification of the Lagrangian $L_0$ is proposed:
$$
L = L_0(x^\mu,y,y_\mu)+\gamma_\mu s^\mu = 
\frac{1}{2}\,y_\mu y^\mu-\frac{1}{2} {\rm m}^2 y^2+\gamma_\mu s^\mu \, .
$$
where $\gamma_\mu\in\R$.
Now, we develop the restricted Hamilton--de Donder--Weyl formalism for this Lagrangian.

The adapted coordinates of the restricted and extended action-dependent multicontact bundles, $\overline{\tau}\colon{\mathcal{P}^*}\to\R^4$
and $\widetilde{\tau}\colon\widetilde{\P}\to\R^4$,
are denoted $(x^\mu,y,p^\mu,s^\mu)$ and
$(x^\mu,y,p^\mu,p,s^\mu)$, respectively.
Denoting 
$\mathcal{P} \colon J^1 \pi \times_{M}\bigwedge\nolimits^{m - 1} \Tan^*M$,
the {\sl restricted} and {\sl extended Legendre maps} associated with the Lagrangian $L$, respectively $\F L\colon\P\to\mathcal{P}^*$ and $\widetilde{\F L}\colon\mathcal{P}\to\widetilde{\mathcal{P}}$ (see \cite{LGMRR_23, LGMRR_25}), are given by
\begin{align} 
\F L(x^\mu,y,y_\mu,s^\mu)&=
\Big(x^\mu,y,p^\mu=\parder{L}{y_\mu},s^\mu\Big)=
(x^\mu,y,y^\mu,s^\mu)\,, \\
\widetilde{\F L}(x^\mu,y,y_\mu,s^\mu)&=\Big(x^\mu,y,p^\mu=\parder{L}{y_\mu},p=L-\parder{L}{y_\mu}y_\mu,s^\mu\Big) \\ &=
\Big(x^\mu,y,y^\mu,-\frac{1}{2} y^\mu y_\mu-\frac{1}{2} {\rm m}^2 y^2+\gamma_\mu s^\mu,s^\mu\Big)\,.
\end{align}  
Note that $\F L$ is a diffeomorphism and, hence, the Lagrangian $L$ is hyperregular.
We can construct the {\sl Hamiltonian section} ${\bf h}=\widetilde{\F L}\circ\F L^{-1}\colon{\cal P}^*\to\widetilde{\cal P}$,
with associated local Hamiltonian function 
\begin{equation}
\label{KGham}
h = \frac{1}{2} p^\mu p_\mu+\frac{1}{2} {\rm m}^2 y^2-\gamma_\mu s^\mu\,.
\end{equation}
The multicontact Hamiltonian $m$-form in $\mathcal{P}^*$ associated with ${\bf h}$ \eqref{thetaHcoor} is
$$
\Theta_{\bf h} = -p^\mu\d y\wedge\d^3 x_\mu +h\,\d^4 x + \d s^\mu\wedge \d^3 x_\mu \,,
$$
and $\sigma_{\Theta_{\bf h}}=-\gamma_\mu \d x^\mu$. The multicontact Herglotz--Hamilton--de Donder--Weyl equations \eqref{vfHr} for $\overline\tau$-transverse $4$-multivector fields $\bfX_{\mathbf{h}}=\bigwedge_{\mu=0}^3 X_\mu\in\vf^4(\mathcal{P}^\ast)$ with local expression
\begin{equation} X_\mu =\parder{}{x^\mu}+f_\mu\parder{}{y}+F_\mu^\nu\parder{}{p^\nu}+g_\mu^\nu\parder{}{s^\nu}\,, 
\end{equation}
lead to
\begin{equation}
g_\mu^\mu = \frac{1}{2}p^\mu p_\mu - \frac{1}{2}{\rm m}^2y^2 + \gamma_\mu s^\mu\,, \quad f_\mu = p_\mu \,,\quad F_\mu^\mu = - {\rm m}^2 y + \gamma_\mu p^\mu \,. 
\end{equation}
The restricted Herglotz--Hamilton--de Donder--Weyl equations \eqref{sect1Hr} or \eqref{sect2Hr} for the integral sections $\bm{\psi}_{\bf h}(x^\nu)=(x^\mu,y(x^\nu),p^\mu(x^\nu),s^\mu(x^\nu))$ 
of $\bfX_{\mathbf{h}}$ 
read as follows:
\begin{equation}
\parder{s^\mu}{x^\mu} = \frac{1}{2}p^\mu p_\mu - \frac{1}{2}{\rm m}^2y^2 + \gamma_\mu s^\mu\,, \quad
\parder{y}{x^\mu} = p_\mu \,,\quad
\parder{p^\mu}{x^\mu} = - {\rm m}^2 y + \gamma_\mu p^\mu \,,
\end{equation}
and, combining the second and third equations above, we obtain
\begin{equation}
\label{eq:KG-dampings-2}
\parderr{y}{x_\mu}{x^\mu} + {\rm m}^2 y = \gamma^\mu \parder{y}{x^\mu}\,,
\end{equation}
which is the \dfn{action-dependent Klein--Gordon equation}
(or {\sl Klein--Gordon equation with damping}).

By taking $\gamma_\mu = (-\gamma, 0, 0, 0)$, we recover the \dfn{telegrapher equation} \cite{chen2004} with mass term in $3+1$ dimensions
\begin{equation}
\label{eq:telegraph-equation}
\parderr{y}{x_\mu}{x^\mu} + \gamma\parder{y}{t} + {\rm m}^2 y = 0\,,
\end{equation} 
as a particular case of the Klein--Gordon equation with damping.

\subsubsection{Extended Hamiltonian formalism}

The Hamiltonian density is
$$
\mathcal{H}={\rm H}\,\d^4x=(p+{\rm h})\d^4x=\Big(p+\frac{1}{2} p^\mu p_\mu+\frac{1}{2} {\rm m}^2 y^2-\gamma_\mu s^\mu\Big)\d^4x\, ,
$$
and, from Theorem \ref{existuniq}, we can write locally,
$${\alpha} = \d p+\d{\rm h} = \d p+p_\mu \d p^\mu + {\rm m}^2 y \, \d y-\gamma_\mu \d s^\mu \, .$$
The canonical multicontact $4$-form is given by
\begin{equation*}
\widetilde{\Theta} =\, -p^\mu\d y\wedge\d^3 x_\mu -p\,\d^4 x + \d s^\mu\wedge \d^3 x_\mu\, .
\end{equation*}
The condition of Proposition \ref{prop:existence_sigma} holds for the form $\alpha$, since $\ds\ker\flat_{\widetilde\Theta}=\Big\langle\derpar{}{p}\Big\rangle$ and, hence,
$$
\innp{\derpar{}{p}}\d\widetilde\Theta+\innp{\derpar{}{p}}(\alpha\wedge\omega)=-\omega+\omega=0 \, .
$$
Finally, the dissipation form is $\sigma_\alpha=-\gamma_\mu \d x^\mu$, and the corresponding  Lichnerowicz differential operator $\bd$ is well-defined. Therefore
$$
\bd\widetilde{\Theta}=-\d p^\mu \wedge\d y\wedge\d^3 x_\mu-\d p\wedge\d^4x-\gamma_\mu p^\mu\,\d y\wedge\d^4x+\gamma_\mu\d s^\mu\wedge\d^4x \, .
$$

In a natural coordinate chart of $\widetilde{\mathcal{P}}$,
the expression of $\widetilde{\tau}$-transverse and locally decomposable $m$-multivector fields
satisfying $\innp{\bfX_\alpha}\omega=1$ is
\begin{align*}
{\bf X}_\alpha=\bigwedge_{\mu=0}^3 \Big(\parder{}{x^\mu}+f_\mu\parder{}{y}+F_\mu^\nu\parder{}{p^\nu}+G_\mu \parder{}{p} + g_\mu^\nu\parder{}{s^\nu}\Big) \, .
\end{align*}
For such multivector fields, the field equations \eqref{vfH} lead to
\begin{gather*}
g_\mu^\mu = \frac{1}{2}p^\mu p_\mu - \frac{1}{2}{\rm m}^2y^2 + \gamma_\mu s^\mu\,, \quad f_\mu = p_\mu \,,\quad F_\mu^\mu = - {\rm m}^2 y + \gamma_\mu p^\mu \,, \\
G_\mu = -{\rm m}^2y\,p_\mu-p_\nu F^\nu_\mu+\gamma_\nu g^\nu_\mu \, .
\end{gather*}
For the integral sections $\widetilde{\bm{\psi}}(x^\nu)=(x^\mu,y(x^\nu),p^\mu(x^\nu),p(x^\nu),s^\mu(x^\nu))$,
the extended Herglotz--Hamilton--de Donder--Weyl equations \eqref{HHDWeq}  are
\begin{gather}
\parder{s^\mu}{x^\mu} = \frac{1}{2}p^\mu p_\mu - \frac{1}{2}{\rm m}^2y^2 + \gamma_\mu s^\mu\,, \quad
\parder{y}{x^\mu} = p_\mu \,,\quad
\parder{p^\mu}{x^\mu} = - {\rm m}^2 y + \gamma_\mu p^\mu \,, \\
\parder{p}{x^\mu}=-{\rm m}^2y\,p_\mu-p_\nu\parder{p^\nu}{x^\mu}+\gamma_\nu \parder{s^\nu}{x^\mu}\, .
\end{gather}
As in the restricted formalism,
combining the second and third equations above, we obtain the action-dependent Klein--Gordon equation 
\eqref{eq:KG-dampings-2};
and, taking $\gamma_\mu = (-\gamma, 0, 0, 0)$, we recover the telegrapher equation 
\eqref{eq:telegraph-equation}.

\subsection{The heat equation and Burgers' equation}
\label{ex:burguer}

Burgers' equation \cite{Bat_15} is a convection-diffusion nonlinear partial differential equation \cite{Sal_15} 
that appears in fluid mechanics, nonlinear acoustics, gas dynamics, traffic flow, and in other fields of applied mathematics.
It reads as follows
\begin{equation}
\label{eq:burgers}
   \derpar{u}{t}+ u\derpar{u}{x} = k\parder{^2u}{x^2}\,,
\end{equation}
where $k\in\R^+$ is a diffusion coefficient. It is closely related to the heat equation,
\begin{equation}
\label{eq:heat}
  \derpar{u}{t} = k\parder{^2u}{x^2}\,.
\end{equation}
In fact, Burgers' equation can be obtained as a contactification of the heat equation \cite{Bur_48,Hopf_50}, as we will see in this example (see also \cite{GGMRR_20}).

\subsubsection{Restricted Hamiltonian formalism}

Before contactifying the heat equation, we need to state its Hamiltonian formulation. It is well known that the heat equation is not variational. However, it can be made variational by introducing an additional dependent variable $v$
\cite{Bat_38,MF_53}.

Then, for the action-dependent version of the heat equation,
the configuration bundle of the theory is $\mathscr{E}\to\R^2$, where the coordinates in $\R^2$ and $\mathscr{E}$ are $(t,x)$ and $(t,x,u,v,s^x,s^t)$, respectively, and the volume form in $\R^2$ is $\omega_{_M}=\d t\wedge\d x$.
The adapted coordinates of the  multicontact bundles $\tau\colon\mathcal{P}\to\R^2$,
$\overline{\tau}\colon{\mathcal{P}^*}\to\R^2$, and $\widetilde{\tau}\colon\widetilde{\mathcal{P}}\to\R^2$
are denoted $(t,x,u,v,u_t,u_x,v_t,v_x,s^t,s^x)$, $(t,x,u,v,p^t,p^x,q^t,q^x,s^t,s^x)$, and $(t,x,u,v,p^t,p^x,q^t,q^x,p,s^t,s^x)$, respectively.

The restricted Hamiltonian formalism for this theory arises from the Lagrangian 
(see \cite{LGMRR_25} for details)
\begin{equation}
\label{lagbur}
L=-ku_xv_x-\frac{1}{2}(v u_t - uv_t)- \gamma us^x\in\Cinfty(\mathcal{P})\,,
\end{equation}
{where $ \gamma\in\R\setminus\{ 0\}$}.
The restricted Legendre map $\F L\colon\mathcal{P}\to\mathcal{P}^*$ associated with this Lagrangian is
\begin{equation} \F L(t,x;u,v,u_t,u_x,v_t,v_x,s^t,s^x) = (t,x; u, v, p^t, p^x, q^t, q^x,s^t,s^x)\,, \end{equation}
where
\begin{equation}
p^t = \parder{L}{u_t} = -\frac12v\,,\quad p^x = \parder{L}{u_x} = -kv_x\,,\quad q^t = \parder{L}{v_t} = \frac12u\,,\quad q^x = \parder{L}{v_x} = -ku_x\,.
\end{equation}
Hence, the restricted Hamiltonian formalism takes place in the image of the Legendre map $\mathcal{P}_\circ^\ast=\F L(\mathcal{P})$,
which is a submanifold of $\mathcal{P}^*$ defined by the two constraints
\begin{equation} 
\label{consheat}
p^t + \frac{1}{2}v = 0\,,\quad q^t - \frac{1}{2}u = 0\,, 
\end{equation}
and the Lagrangian function $L$ is almost-regular.
We use the coordinates $(t,x;u,v,p^x,q^x,s^t,s^x)$ on $\mathcal{P}_\circ^\ast$. 
Therefore, there is a Hamiltonian section ${\bf h}_\circ\colon\mathcal{P}_\circ^\ast\to\widetilde{\mathcal{P}}$ associated with $L$,
which is defined by the Hamiltonian function
(see Diagram \eqref{diagbig})
\begin{equation}
\label{ham_circ}
h_\circ=-\frac{1}{k}p^xq^x+\gamma us^x\,. 
\end{equation}
The Hamiltonian $2$-form \eqref{thetaHcoor0} is
\begin{align}
\Theta_{{\bf h}_\circ} &= p^x\d u\wedge\d t+q^x\d v\wedge\d t+\frac{1}{2}v\,\d u\wedge\d x-\frac12 u\,\d v\wedge\d x \nonumber \\
&\quad -\Big(\frac1kp^xq^x-\gamma us^x\Big)\d t\wedge\d x+\d s^t\wedge\d x-\d s^x\wedge\d t   \, .
\end{align}
Let us check that $\Theta_{{\bf h}_\circ}\notin{\rm Im}\,\flat_{\d\Theta_{{\bf h}_\circ}}$,
as stated in Proposition \ref{prop:restricted-ranks}~(iv).
In this example,
\[
\d h_\circ=-\frac1k\big(q^x\,\d p^x+p^x\,\d q^x\big)+\gamma\,s^x\,\d u+\gamma\,u\,\d s^x \, ,
\]
and
\begin{equation}
\label{eq:dTheta-heat}
\d\Theta_{{\bf h}_\circ}=\d p^x\wedge\d u\wedge\d t+\d q^x\wedge\d v\wedge\d t
-\d u\wedge\d v\wedge\d x+\d h_\circ\wedge\d t\wedge\d x \, .
\end{equation}
Since $h_\circ$ does not depend on $s^t$, the form \eqref{eq:dTheta-heat}
does not contain $\d s^t$; so
\[
\innp{\derpar{}{s^t}}\d\Theta_{{\bf h}_\circ}=0 \, ,
\quad
\innp{\derpar{}{s^t}}\Theta_{{\bf h}_\circ}=\d x \, .
\]
Suppose that there exists a vector field $Z\in\mathfrak{X}(\mathcal{P}^*_\circ)$ such that
$\Theta_{{\bf h}_\circ}=\innp{Z}\d\Theta_{{\bf h}_\circ}$. Then
\[
\d x=\innp{\derpar{}{s^t}}\Theta_{{\bf h}_\circ}
=\innp{\derpar{}{s^t}}\innp{Z}\d\Theta_{{\bf h}_\circ}
=-\innp{Z}\innp{\derpar{}{s^t}}\d\Theta_{{\bf h}_\circ}=0 \, ,
\]
which is a contradiction. Therefore, $\Theta_{{\bf h}_\circ}\notin{\rm Im}\,\flat_{\d\Theta_{{\bf h}_\circ}}$
at all points of $\mathcal{P}^*_\circ$.
The same holds for the Lichnerowicz differential. As $\sigma_{\Theta_{{\bf h}_\circ}} = \gamma u\,\d x$, we have
$\bd\Theta_{{\bf h}_\circ}=\d\Theta_{{\bf h}_\circ}+\gamma\,u\,\d x\wedge\Theta_{{\bf h}_\circ}$, and
\[
\innp{\derpar{}{s^t}}\bd\Theta_{{\bf h}_\circ}
=-\gamma\,u\,\d x\wedge\innp{\derpar{}{s^t}}\Theta_{{\bf h}_\circ}
=-\gamma\,u\,\d x\wedge\d x=0 \, ;
\]
hence, the argument above shows that there is no vector field $Z$ such that
$\Theta_{{\bf h}_\circ}=\innp{Z}\bd\Theta_{{\bf h}_\circ}$ either.

Thus, the pair $(\Theta_{{\bf h}_\circ},\omega)$ defines a variational multicontact structure in $\mathcal{P}_\circ^*$.
So, being the Lagrangian \eqref{lagbur}, it defines a premulticontact system in the Lagrangian formalism of the theory (see \cite{LGMRR_25}), but a multicontact system in the Hamiltonian formalism.

For a 2-multivector field ${\bf X}_{\mathbf{h}_\circ} =X_t\wedge X_x\in\vf^2({\cal P}^*_\circ)$, 
such that $\innp{{\bf X}_{\mathbf{h}_\circ}}\omega=1$, with
\begin{align}
X_t&=\parder{}{t}+f_t^u\parder{}{u}+f_t^v\parder{}{v}+F_t^x\parder{}{p^x}+G_t^x\parder{}{q^x}+g_t^t\parder{}{s^t}+g_t^x\parder{}{s^x}\, , \\
X_x&=\parder{}{x}+f_x^u\parder{}{u}+f_x^v\parder{}{v}+F_x^x\parder{}{p^x}+G_x^x\parder{}{q^x}+g_x^t\parder{}{s^t}+g_x^x\parder{}{s^x}\, ,
\end{align}
the Hamiltonian field equations \eqref{vfHb0} are
\begin{gather}
f^v_t = F^x_x + \gamma(s^x + up^x)\,,\quad f^u_t   = - G^x_x -\gamma uq^x\,,\quad f^u_x = -\frac1k q^x\,,\quad f^v_x = -\frac1k p^x\,, \\
g^t_t+g_x^x=p^xf^u_x+q^xf_x^v-\frac12 vf_t^u+\frac12 uf_t^v+\frac1k p^xq^x-\gamma u s^x  \,.
\end{gather}
For the integral sections of those multivector fields which are integrable, 
$\bm{\psi}_{\bf h}(t,x)=(t,x,u(t,x),v(t,x),p^x(t,x),q^x(t,x),s^t(t,x),s^x(t,x))$,
the restricted Herglotz--Hamilton--de Donder--Weyl equations obtained from the preceding equations are
\begin{gather}
\parder{v}{t} = \parder{p^x}{x} + \gamma(s^x + up^x)\,,\quad \parder{u}{t}  = - \parder{q^x}{x} -\gamma uq^x\,,\quad \parder{u}{x} = -\frac1k q^x\,,\quad \parder{v}{x} = -\frac1k p^x\,,
\\
\parder{s^t}{t} + \parder{s^x}{x} = p^x \parder{u}{x} + q^x\parder{v}{x} -\frac12 v\parder{u}{t} + \frac12 u\parder{v}{t} + \frac1k p^xq^x - \gamma u s^x  \,, \label{eq:burgers-ham-3}
\end{gather}
and there are no compatibility constraints.
Combining the first four equations, we obtain
\begin{equation}
\parder{u}{t} - \gamma ku\parder{u}{x} = k\parder{^2u}{x^2}\,, \quad
\parder{v}{t} + \gamma ku\parder{v}{x} = -k\parder{^2v}{x^2} + \gamma s^x
\label{eq:burgers-ham-2}\,.
\end{equation}
If we set the value of $\gamma$ to be $\gamma = -\dfrac{1}{k}$, the first equation \eqref{eq:burgers-ham-2} becomes Burgers' equation \eqref{eq:burgers}.

Taking $v=p^x=s^t=s^x=0$ and $q^x=-k\dparder{u}{x}$
solves the whole system for any solution $u$ to Burgers' equation \eqref{eq:burgers}.

\subsubsection{Extended Hamiltonian formalism}

For the extended formalism of this theory,
consider the submanifold $\widetilde{\mathcal{P}}_\circ\hookrightarrow\widetilde{\mathcal{P}}$
defined in $\widetilde{\mathcal{P}}$ by the constraints \eqref{consheat}. 
The natural coordinates in $\widetilde{\mathcal{P}}_\circ$ are $(t,x;u,v,p^x,q^x,p,s^t,s^x)$. 

the Hamiltonian density is
$$
\mathcal{H}_\circ={\rm H}_\circ\,\d t\wedge\d x=(p+{\rm h}_\circ)\d t\wedge\d x=\Big(p-\frac{1}{k}p^xq^x+\gamma us^x\Big)\d t\wedge\d x\, .
$$
and the Hamiltonian 1-form is now given by 
$$
{\alpha}_\circ = \d p+\d{\rm h}_\circ = \d p -\frac{1}{k} p^x\d q^x-\frac{1}{k} q^x\d p^x + \gamma s^x\d u + \gamma u \d s^x \, ,
$$
The canonical $2$-form of Definition \ref{canform0} is given by
\begin{align*}
\widetilde{\Theta}_\circ =&\, p^x\d u\wedge\d t+q^x\d v\wedge\d t+\frac{1}{2}v\,\d u\wedge\d x-\frac{1}{2}u\,\d v\wedge\d x -p\,\d t\wedge\d x+\d s^t\wedge\d x-\d s^x\wedge\d t \, ,
\end{align*}
which is a multicontact form in $\widetilde{\mathcal{P}}_\circ$.
The condition of Proposition \ref{prop:existence_sigma} holds, since $\ds\ker\flat_{\widetilde\Theta_\circ}=\Big\langle\derpar{}{p}\Big\rangle$ and then,
$$
\innp{\derpar{}{p}}\d\widetilde\Theta_\circ+\innp{\derpar{}{p}}(\alpha_\circ\wedge\omega)=-\omega+\omega=0 \, .
$$ 
Hence, the existence of a dissipation form is assured and, in coordinates, 
reads $\sigma_{\alpha_\circ}=\gamma u\,\d x$. So, we can define the corresponding $\bd$ operator, and
\begin{align*}
\bd\widetilde{\Theta}_\circ = & \, \d p^x \wedge \d u \wedge\d t+\d q^x \wedge \d v\wedge\d t-\d u \wedge \d v \wedge \d x -\d p\wedge\d t\wedge\d x\\
&+\gamma u\, (p^x\d x\wedge\d u\wedge\d t + \,q^x\d x\wedge \d v\wedge\d t-\,\d x\wedge \d s^x\wedge\d t) \, .
\end{align*}

The coordinate expression in $\widetilde{\mathcal{P}}_\circ$ of a $\widetilde{\tau}_\circ$-transverse and locally decomposable $2$-multivector fields
satisfying $\innp{{\bf X}_{\alpha_\circ}}\omega=1$ is
\begin{align*}
{\bf X}_{\alpha_\circ} =& \left(\derpar{}{t}+f_t^u\derpar{}{u}+f_t^v\derpar{}{v} + F^x_t\derpar{}{p^x} + G^x_t \derpar{}{q^x} +G_t\derpar{}{p} + g^t_t\derpar{}{s^t}+ g^x_t\derpar{}{s^x}\right)\\
&\wedge  \left(\derpar{}{x}+f_x^u\derpar{}{u}+f_x^v\derpar{}{v} + F^x_x\derpar{}{p^x} + G^x_x \derpar{}{q^x} +G_x\derpar{}{p} + g^t_x \derpar{}{s^t}+ g^x_x\derpar{}{s^x}\right) \, .
\end{align*}
For such multivector fields, the field equations \eqref{vfH0} lead to
\begin{align}
f^v_t = F^x_x & + \gamma(s^x + up^x)\,,\quad f^u_t =  - G^x_x -\gamma uq^x\,,\quad f^u_x = 
- \frac{1}{k} q^x\,,\quad f^v_x =  -\frac{1}{k} p^x\,, 
\label{one}    \\
G_x &=  f^u_t f^v_x - f^v_t f^u_x + \gamma u\, (p^x f^u_x + q^x f^v_x - g^x_x) \\ 
&=\frac{1}{k}(q^x F^x_x+p^x G^x_x+\gamma\,q^xs^x)-\gamma\,u\,g^x_x \, ,
\label{two} \\
G_t &= - F^x_tf^u_x + F^x_x f^u_t - G^x_tf^v_x + G^x_xf^v_t + \gamma u\, (p^x f^u_t + q^x f^v_t - g_t^x) \\ 
&=\frac{1}{k}(q^x F^x_t+p^x G^x_t)
-\gamma s^x f^u_t-\gamma u g^x_t \, , \label{three} \\
g^t_t+g_x^x &= p^xf^u_x+q^xf_x^v-\frac12 vf_t^u+\frac{1}{2} uf_t^v+\frac{1}{k} p^xq^x - \gamma u s^x \, , 
\end{align}
where the second equalities of the equations \eqref{two} and \eqref{three} follow using some equations \eqref{one}.

For the integral sections $\widetilde{\bm{\psi}}_\circ(t,x)=(t,x,u(t,x),v(t,x),p^x(t,x),q^x(t,x),p(t,x),s^t(t,x),s^x(t,x))$ of those multivector fields ${\bf X}_{\alpha_\circ}$ which are integrable,
the extended Herglotz--Hamilton--de Donder--Weyl equations are
\begin{align}
\parder{v}{t}=&\parder{p^x}{x} + \gamma(s^x+up^x)\,,\quad \parder{u}{t}=-\parder{q^x}{x}-\gamma uq^x\,,\quad \parder{u}{x} = -\frac{1}{k} q^x\,,\quad \parder{v}{x} = -\frac{1}{k} p^x\,,
\label{unas} \\
\parder{p}{x} &=\parder{u}{t} \parder{v}{x}-\parder{v}{t}\parder{u}{x}+\gamma u\Big(p^x\parder{u}{x}+q^x \parder{v}{x}-\parder{s^x}{x}\Big) \, , 
\\ 
&=\frac{1}{k}\Big(q^x\derpar{p^x}{x}+p^x\derpar{q^x}{x}+\gamma\,q^xs^x\Big)-\gamma\,u\,\derpar{s^x}{x}=-\dv{({\rm h}_\circ\circ\widetilde{\bm\psi}_\circ)}{x} \, , 
\label{eq:burgers-ham-3b} \\
\pdv{p}{t} &=-\pdv{p^x}{t}\pdv{u}{x} + \pdv{p^x}{x} \pdv{u}{t} -  \pdv{q^x}{t} \pdv{v}{x} + \pdv{q^x}{x} \pdv{v}{t} + \gamma u \Big(p^x \pdv{u}{t} + q^x \pdv{v}{t}-\pdv{s^x}{t}\Big) \\
&=\frac{1}{k}\Big(q^x\parder{p^x}{t}+p^x\parder{q^x}{t}\Big)
-\gamma s^x\parder{u}{t}-\gamma u \parder{s^x}{t}=-\dv{({\rm h}_\circ\circ\widetilde{\bm\psi}_\circ)}{t} \, ,
\label{eq:burgers-ham-3c} \\
& \qquad\parder{s^t}{t} + \parder{s^x}{x} = p^x \parder{u}{x} + q^x\parder{v}{x} -\frac{1}{2} v\parder{u}{t} + \frac{1}{2} u\parder{v}{t} +\frac{1}{k} p^xq^x - \gamma u s^x  \,, 
\label{otras}
\end{align}
Combining the first four equations \eqref{unas}, we obtain the following:
\begin{equation}
\parder{u}{t} - \gamma ku\parder{u}{x} = k\parder{^2u}{x^2}\,, \quad
\parder{v}{t} + \gamma ku\parder{v}{x} = -k\parder{^2v}{x^2} + \gamma s^x
\label{eq:burgers-ham}\, .    
\end{equation}
Setting the constant $\gamma$ to $\gamma = -\dfrac{1}{k}$, 
the first equation \eqref{eq:burgers-ham} becomes Burgers' equation \eqref{eq:burgers}.

\section{Conclusions and outlook}
\protect\label{conclusions}

In this paper, we have developed the extended Hamiltonian formalism for
action-dependent (Herglotz-type) classical field theories within the framework
of multicontact geometry. The field theory is described by the canonical multicontact structure together with the Hamiltonian density $\mathcal{H} = \mathrm{H} \dd^m x$, which can be defined by a Hamiltonian section. This generalises to action-dependent
field theories both the extended multisymplectic Hamiltonian formalism of
conservative field theories and Kuwabara's extended formalism for time-dependent
mechanics.

The main conclusions of this work can be summarised as follows:
\begin{enumerate}[{\rm (i)}] \itemsep = -1 pt
\item The canonical form $\widetilde\Theta$ on $\widetilde{\mathcal{P}}$ does not define a
special multicontact structure in the previously available sense, since its Reeb
distribution has rank $m+1$ and $\flat_{\widetilde\Theta}$ has a non-trivial
kernel spanned by $\tparder{}{p}$. This has led us to adopt a more general
notion of (pre)multicontact form, defined through its canonical
(pre)multisymplectisation, which includes the previous definitions and covers
both the extended and the restricted structures.
\item In the extended setting, the dissipation form is no longer canonical: it
depends on an auxiliary $1$-form, and it is precisely the Hamiltonian section that
fixes it. Hence, an extended Hamiltonian multicontact system is determined
by a Hamiltonian density $\mathcal{H}$, which allows us to define a Hamiltonian $1$-form $\alpha\in\Omega^1(\widetilde{\mathcal{P}})$, which then determines $\sigma_\alpha$. This formulation allows to treat non-closed $\alpha$, which give a description of external sources.
\item We have proved the local existence of solutions to the extended field
equations, given by $\widetilde\tau$-transverse, locally decomposable multivector
fields tangent to the constraint submanifold (Corollary~\ref{tangency}), and the
equivalence of the formulations of the Herglotz--Hamilton--de Donder--Weyl
equations in terms of sections, multivector fields and Ehresmann connections
(Theorem~\ref{equivteor1}). These equations contain, in addition, an equation
for the variable $p$, equation  \eqref{balanceq}, which has no counterpart in the restricted
formalism and provides geometric property of tangency.
\item This tangency condition can be employed then to establish the precise relation between the extended and the
restricted formalisms: solutions on $\widetilde{\mathcal{P}}$ project onto solutions on
$\mathcal{P}^*$ (Theorem~\ref{related}) and, conversely, every restricted
Hamiltonian multicontact system determines a unique exact $1$-form
$\alpha$ whose solutions are obtained from the restricted ones
(Theorem~\ref{existuniq}). Hence, both descriptions are equivalent.
\item These results have been extended to the almost-regular case
(Section~\ref{arhs}), when both formalisms are premulticontact
(Theorem~\ref{related20} and Proposition~\ref{existuniq0}). In the extended
framework, the primary constraints arising from the Legendre map and the
Hamiltonian constraint are treated on the same footing on a submanifold
$\widetilde{\mathcal{P}}_\circ\subset\widetilde{\mathcal{P}}$.
\end{enumerate}

The formalism has been illustrated with two examples.
The first one is the action-dependent
Klein--Gordon equation, which arises from a hyperregular Lagrangian and
contains the telegrapher equation as a particular case.
The second one is the heat equation,
whose action-dependent version is associated with an almost-regular Lagrangian but leads to a multicontact Hamiltonian formalisms,
and, for a suitable choice of the dissipation parameter, yields Burgers'
equation. 
The latter shows how the extended (pre)multicontact formalism handles
the contactification of non-variational equations.

Several lines of research remain open. First, as pointed out in the
Introduction, the extended framework seems to be the natural setting for the
study of the algebraic structures of action-dependent field theories. 
We expect that the
canonical structure of $\widetilde{\mathcal{P}}$ provides the appropriate setting to relate the brackets of
Hamiltonian forms on multicontact manifolds introduced in~\cite{LIR_25} with
the multisymplectic Poisson brackets of conservative theories. Other natural problems are the
study of symmetries and dissipated quantities, the reduction of extended
Hamiltonian multicontact systems with symmetry, the Hamilton--Jacobi theory in
this framework, and the design of geometric numerical integrators that preserve
the extended multicontact structure. We plan to address some of these questions
in forthcoming works.

\appendix
\section{Multivector fields and connections}
\protect\label{mvfc}

The basic notions on multivector fields and connections are reviewed in this section. For further details, see \cite{BCGGG_91,CCI_91,EMR_98,KMS_93,Mar_97}.

Let ${\cal M}$ be a differentiable manifold.
\dfn{Multivector fields of degree $m$}
or \dfn{$m$-multivector fields} on ${\cal M}$ (with $m\leq\dim\mathcal{M}$) are contravariant skew-symmetric tensor fields of order $m$. 
The set of $m$-multivector fields on ${\cal M}$ is denoted by $\vf^m({\cal M})$.

A multivector field $\bfX\in \vf^m({\cal M})$ is \dfn{locally decomposable} if, for every ${\rm p}\in{\cal M}$, there exists an open neighbourhood $U_{\rm p} \subset{\cal M}$ such that 
\begin{equation}
\restr{\bfX}{U_{\rm p}} = X_1\wedge\dotsb\wedge X_m \,,\quad \text{for some }X_1,\dotsc,X_m \in\vf(U_{\rm p}) \,.
\end{equation}
Locally decomposable $m$-multivector fields are locally associated with $m$-dimensional distributions on $\mathcal{M}$.
This induces a partition of $\vf^m(\mathcal{M})$ into equivalence classes $\{ {\bf X}\}\subset\vf^m(\mathcal{M})$ consisting of locally decomposable multivector fields associated with the same distribution.
If ${\bf X},{\bf X}'\in\{ {\bf X}\}$, then for every $U\subset \mathcal{M}$ there exists a non-vanishing function $f\in \Cinfty(U)$ such that 
${\bf X}'\big\vert_U=f{\bf X}$.

An \dfn{integrable multivector field} is a locally decomposable multivector field whose associated distribution is integrable, that is, involutive.

The \dfn{contraction} of a locally decomposable multivector field $\bfX\in \vf^m({\cal M})$ with a differential form $\Omega\in\df^k({\cal M})$
is the natural pairing between contravariant and covariant tensor fields:
\begin{equation}
\restr{\innp{\bfX}\Omega}{U_p} = \begin{dcases}
\innp{X_m}\dotsb\innp{X_1}\Omega\,, & k \geq m\,,\\
0\,, & k < m\,.
\end{dcases}
\end{equation}
The \dfn{Lie derivative} of a form  $\alpha\in\df^k(\mathcal{M})$ with respect to $\bfX $ is the graded bracket (of degree $m-1$)
$$
\Lie_\bfX \alpha\coloneqq[\d , \innp{\bfX}]\alpha=
(\d\innp{\bfX} - (-1)^m\innp{\bfX}\d)\alpha \,.
$$
If $\bfX\in\vf^m(\mathcal{M})$ and $\bfY\in\vf^n(\mathcal{M})$,
the \dfn{Schouten--Nijenhuis bracket} of $\bfX ,{\bf Y}$ is
the graded commutator of $\Lie_\bfX$ and $\Lie_\bfY$; that is,
the multivector field $[\bfX,\bfY]\in\vf^{m+n-1}(\mathcal{M})$, such that,
\begin{equation}
\label{liebrac}
\Lie_{[\bfX,\bfY]}=[\Lie_\bfX,\Lie_\bfY]=\Lie_\bfX\Lie_\bfY- (-1)^{(m -1)(n-1)}\Lie_\bfY\Lie_\bfX \,.
\end{equation}
For (locally) decomposable multivector fields, it is given as follows: 
given $\bfX=X_1\wedge\cdots\wedge X_m$ and $\bfY=Y_1\wedge\cdots\wedge Y_n$, their Schouten--Nijenhuis bracket is
\begin{equation}
[\bfX,\bfY]=\sum_{i,j}(-1)^{i+j}[X_i,Y_j]\wedge 
X_1\wedge\cdots\wedge X_{i-1}\wedge X_{i+1}\wedge\cdots\wedge X_m\wedge Y_1\wedge\cdots\wedge Y_{j-1}\wedge Y_{j+1}\wedge\cdots\wedge Y_n \,,    
\end{equation}
and it extends linearly to any multivector fields in general.

Let $\varkappa\colon {\cal M}\rightarrow M$ be a fibre bundle
with $\dim M=m$ and $\dim{\cal M}=N+m$.

A multivector field $\bfX \in \vf^m({\cal M})$ is said to be $\varkappa$-\textsl{transverse} if
$\innp{\bfX}(\varkappa^*\beta)\vert_{\rm p} \neq 0$
for every ${\rm p} \in{\cal M}$ and every non-vanishing form $\beta \in \df^m(M)$.

Let $(x^\mu,z^i)$ be local coordinates on ${\cal M}$, where $x^\mu$ are coordinates on the base $M$ and $z^i$ are fibre coordinates ($1\leq\mu\leq m$, $1\leq i\leq\dim\mathcal{M}-m$).
A representative of a class $\{\bfX\}\subset\vf^m({\cal M})$
of locally decomposable $\varkappa$-transverse multivector fields can be locally written as 
\begin{equation}
\bfX= \bigwedge^m_{\mu=1} \bfX_\mu= 
\bigwedge^m_{\mu=1} \left( \frac{\partial}{\partial x^\mu} + F_\mu^i\frac{\partial}{\partial z^i}\right)\,, \quad F^i_\mu \in \Cinfty({\cal M}) \,.
\label{mvf}
\end{equation}
If $M$ is orientable with volume form $\omega_{_M}\in\df^m(M)$, then
$\bfX$ is $\varkappa$-transverse if, and only if,
$\innp{\bfX}(\varkappa^*\omega_{_M})\neq 0$.
This condition can be normalised by imposing
$\innp{\bfX}(\varkappa^*\omega_{_M})=1$,
which fixes the representative in \eqref{mvf}.

If $\bfX \in \vf^m({\cal M})$ is locally decomposable, $\varkappa$-transverse, and integrable,
then its integral submanifolds are locally the images of sections 
$\psi\colon U\subset M \rightarrow{\cal M}$ of $\varkappa$, called \dfn{integral sections} of $\bfX$.
Thus, for every ${\rm p}\in{\cal M}$, there exist $x\in M$ and an integral section $\psi\colon U_x\subset M\to {\cal M}$ such that ${\rm p} = \psi(x)$.  
In coordinates, if $\psi(x^\mu)=(x^\mu,z^i(x^\nu))$, then
\[
\frac{\partial z^i}{\partial x^\mu}=F_\mu^i \,.
\]

Furthermore, the \dfn{canonical prolongation} of a section 
$\psi\colon U\subset M\to \mathcal{M}$ to 
$\ds\bigwedge\nolimits^m\Tan{\cal M}$ is the section 
$\ds\psi^{(m)}\colon U\subset M\to\bigwedge\nolimits^m\Tan\mathcal{M}$ 
defined by
\[
\psi^{(m)}\coloneqq\bigwedge\nolimits^m\Tan\psi\circ{\bf Y}_{\omega_{_M}}\,,
\]
where 
$\ds\bigwedge\nolimits^m\Tan\psi$ is the natural extension of $\psi$ 
to the corresponding multitangent bundles,
and ${\bf Y}_{\omega_{_M}}\in\vf^m(M)$ is the unique multivector field such that $\iota({\bf Y}_{\omega_{_M}})\omega_{_M}=1$.
Then $\psi$ is an integral section of ${\bf X}\in\vf^m({\cal M})$ if, and only if,
${\bf X}\circ\psi=\psi^{(m)}$.

An \dfn{Ehresmann connection} on $\varkappa\colon \mathcal{M}\to M$
is a $\varkappa$-semibasic $1$-form $\nabla$ on $\mathcal{M}$ with values in $\Tan \mathcal{M}$, that is, a $(1,1)$-tensor field $\nabla\in\mathcal T^1_1(\mathcal{M})$,
such that $\innp{\nabla}\alpha =\alpha$ for every $\varkappa$-semibasic form $\alpha\in\df^1(\mathcal{M})$
(here, $\innp{\nabla}\alpha$ denotes the natural contraction between tensor fields).

An Ehresmann connection defines a splitting of $\Tan\mathcal{M}$ into vertical and horizontal distributions, and thus can be regarded as the horizontal projector.

The connection is \dfn{integrable} if its horizontal distribution is integrable (equivalently, if the connection is flat). In this case, its integral submanifolds are local sections $\psi\colon U\subset M\to \mathcal{M}$ of the projection $\varkappa$.

Finally, equivalence classes of locally decomposable and $\varkappa$-transverse multivector fields $\{{\bf X}\} \subseteq \vf^m (\mathcal{M})$
are in one-to-one correspondence with orientable Ehresmann connections on $\mathcal{M}\to M$. This correspondence is characterised by the fact that the horizontal distribution of the connection coincides with the distribution associated with $\{ {\bf X}\}$.
Under this correspondence, integrable classes correspond to flat orientable Ehresmann connections.

In natural coordinates, a connection reads
\[
\nabla= dx^\mu\otimes\left( \frac{\partial}{\partial x^\mu} + F_\mu^i\frac{\partial}{\partial z^i}\right) \,.
\]

\subsection*{Acknowledgements}

M. de Le\'on and R. Izquierdo acknowledge the financial support of the 
{\sl Ministerio de Ciencia, Innovaci\'on y Universidades} (Spain), grant PID2022-137909NB-C21, of AEI.
M. de Le\'on also acknowledges financial support from the Severo Ochoa Program for Centers of Excellence in R\&D, CEX-2023-001347-S.
N. Rom\'an-Roy and A. Villanova acknowledge the financial support of the 
{\sl Generalitat de Catalunya, Departament de Recerca i Universitats}, through AGAUR grant 2025 SGR 00878 (GEOMVAP).

\bibliographystyle{abbrv}
{\small
\bibliography{references}

@article{AB_51,
    author = {J. L. Anderson and P. G. Bergmann},
    title = {{Constraints in covariant field theories}},
    journal = {Phys. Rev.},
    volume = {{\bf 83}},
    number = {5},
    pages = {1018--1025},
    year = {1951},
    note = {\href{https://doi.org/10.1103/PhysRev.83.1018}{10.1103/PhysRev.83.1018}}
}

@article{Bat_15,
    author = {Harry Bateman},
    title = {{Some Recent Researches on the Motion of Fluids}},
    journal = {Mon. Weather Rev.},
    volume = {{\bf 43}},
    pages = {163--170},
    year = {1915},
    note = {\href{https://doi.org/10.1175/1520-0493(1915)43%3C163:SRROTM%3E2.0.CO;2}{10.1175/1520-0493(1915)43\textless 163:SRROTM\textgreater 2.0.CO;2}}
}

@article{Bat_38,
    author = {Harry Bateman},
    title = {{On dissipative systems and related variational principles}},
    journal = {Phys. Rev.},
    volume = {{\bf 38}},
    pages = {815},
    year = {1938},
    note = {\href{https://doi.org/10.1103/PHYSREV.38.815}{10.1103/PHYSREV.38.815}}
}

@article{Bra_18,
    author = {Alessandro Bravetti},
    title = {{Contact geometry and thermodynamics}},
    journal = {Int. J. Geom. Methods Mod. Phys.},
    volume = {{\bf 16}},
    number = {supp01},
    pages = {1940003},
    year = {2018},
    note = {\href{https://doi.org/10.1142/S0219887819400036}{10.1142/S0219887819400036}}
}

@article{Bur_48,
author = {J.M. Burgers},
title = {A Mathematical Model Illustrating the Theory of Turbulence},
journal = {Adv. Appl. Mechs.},
volume = {{\bf 1}},
pages = {171-199},
year = {1948},
note = {\href{https://doi.org/10.1016/S0065-2156(08)70100-5}{10.1016/S0065-2156(08)70100-5}}
}

@article{GGB_03,
    author = {B. Georgieva and R. Guenther and T. Bodurov},
    title = {Generalized variational principle of Herglotz for several independent variables. First Noether-type theorem},
    journal = {J. Math. Phys.},
    volume = {44},
    number = {9},
    pages = {3911-3927},
    year = {2003},
    note = {\href{https://doi.org/10.1063/1.1597419}{10.1063/1.1597419}}
    }

@article{Hopf_50,
author = {E. Hopf},
title = {The partial differential equation $u_t + uu_x = \mu u_{xx}$},
journal = {Comm. Pure App. Math.},
volume = {3},
number = {3},
pages = {201-230},
year = {1950},
note = {\href{https://doi.org/10.1002/cpa.3160030302}{10.1002/cpa.3160030302}}
}

@book{chen2004,
  title={Home Networking Basis: Transmission Environments and Wired/wireless Protocols},
  author={W.Y. Chen},
  publisher={Prentice Hall PTR},
  series={Communications engineering},
  year={2004}
}

@book{MF_53,
    author = {P.M. Morse and H. Feshbach},
    title = {{Methods of Theoretical Physics, Part I}},
    publisher = {McGraw-Hill},
    edition = {},
    year = {1953}
}

@book{BCGGG_91,
    author = {Robert L. Bryant and S. S. Chern and Robert B. Gardner and Hubert L. Goldschmidt and P. A. Griffiths},
    title = {{Exterior Differential Systems}},
    publisher = {Springer-Verlag},
    year = {1991},
    address = {New York},
    series = {Mathematical Sciences Research Institute Publications},
    volume = {18},
    edition = {1st},
    note = {\href{https://doi.org/10.1007/978-1-4613-9714-4}{10.1007/978-1-4613-9714-4}}
}

@article{BCT_17,
    author = {Alessandro Bravetti and Hans Cruz and Diego Tapias},
    title = {{Contact Hamiltonian mechanics}},
    journal = {Ann. Phys.},
    volume = {{\bf 376}},
    pages = {17--39},
    year = {2017},
    note = {\href{https://doi.org/10.1016/j.aop.2016.11.003}{10.1016/j.aop.2016.11.003}}
}

@article{BGG_17,
    author = {A. J. Bruce and K. Grabowska and J. Grabowski},
    title = {{Remarks on Contact and Jacobi Geometry}},
    journal = {Symmetry Integr. Geom.: Methods Appl. (SIGMA)},
    volume = {{\bf 13}},
    year = {2017},
    pages = {059},
    note = {\href{https://doi.org/10.3842/SIGMA.2017.059}{10.3842/SIGMA.2017.059}}
}

@article{CBG_21,
    author = {Hans Cruz-Prado and Alessandro Bravetti and A. Garcia-Chung},
    title = {{From geometry to coherent dissipative dynamics in quantum mechanics}},
    journal = {Quantum Rep.},
    volume = {{\bf 3}},
    number = {4},
    pages = {664--683},
    year = {2021},
    note = {\href{https://doi.org/10.3390/quantum3040042}{10.3390/quantum3040042}}
}

@article{CCI_91,
    author = {Jose F. Cariñena and Michael Crampin and Luis A. Ibort},
    journal = {Diff. Geom. Appl.},
    number = {4},
    pages = {345--374},
    title = {{On the multisymplectic formalism for first order field theories}},
    volume = {{\bf 1}},
    year = {1991},
    note = {\href{https://doi.org/10.1016/0926-2245(91)90013-Y}{10.1016/0926-2245(91)90013-Y}}
}

@article{CCM_18,
    author = {Florio M. Ciaglia and Hans Cruz and Giuseppe Marmo},
    title = {{Contact manifolds and dissipation, classical and quantum}},
    journal = {Ann. Phys.},
    volume = {{\bf 398}},
    pages = {159--179},
    year = {2018},
    note = {\href{https://doi.org/10.1016/j.aop.2018.09.012}{10.1016/j.aop.2018.09.012}}
}

@article{Dir_50,
    author = {P. A. M. Dirac},
    title = {{Generalized Hamiltonian Dynamics}},
    journal = {Can. J. Math.},
    volume = {{\bf 2}},
    year = {1950},
    pages = {129--148},
    note = {\href{https://doi.org/10.4153/CJM-1950-012-1}{10.4153/CJM-1950-012-1}}
}

@article{LEMR_2007,
    author = {Arturo Echeverría-Enríquez and Manuel de Le\'on and Miguel Carlos Muñoz-Lecanda and Narciso Román-Roy},
    title = {{Extended Hamiltonian systems in multisymplectic field theories}},
    year = {2007},
    journal = {J. Math. Phys.},
    volume = {{\bf 48}},
    number = {11},
    pages = {112901},
    note = {\href{https://doi.org/10.1063/1.2801875}{10.1063/1.2801875}}
}

@article{EMR_98,
    author = {Arturo Echeverría-Enríquez and Miguel Carlos Muñoz-Lecanda and Narciso Román-Roy},
    title = {{Multivector fields and connections: Setting Lagrangian equations in field theories}},
    journal = {J. Math. Phys.},
    year = {1998},
    volume = {{\bf 39}},
    number = {9},
    pages = {4578--4603},
    note = {\href{https://doi.org/10.1063/1.532525}{10.1063/1.532525}}
}

@article{PR_2002,
title = {Geometry of Hamiltonian $n$-vector fields in multisymplectic field theory},
journal = {J. Geom. Phys.},
volume = {{\bf 44}},
number = {1},
pages = {52--69},
year = {2002},
note = {\href{https://doi.org/10.1016/S0393-0440(02)00031-1}{10.1016/S0393-0440(02)00031-1}},
author = {Cornelius Paufler and Hartmann Romer},
}

@book{God_69,
  title = {G{\'e}om{\'e}trie Diff{\'e}rentielle et M{\'e}canique Analytique},
  author = {Godbillon, Claude},
  year = {1969},
  publisher = {Hermann},
  address = {Paris},
  pages = {184}
}

@article{GG_22,
    author = {Katarzyna Grabowska and Janusz Grabowski},
    title = {{A geometric approach to contact Hamiltonians and contact Hamilton--Jacobi theory}},
    journal = {J. Phys. A: Math. Theor.},
    volume = {{\bf 55}},
    number = {43},
    pages = {435204},
    year = {2022},
    note = {\href{https://doi.org/10.1088/1751-8121/ac9adb}{10.1088/1751-8121/ac9adb}}
}

@article{GGMRR_20,
    author = {Jordi Gaset and Xavier Gràcia and Miguel C. Muñoz-Lecanda and Xavier Rivas and Narciso Román-Roy},
    title = {{A contact geometry framework for field theories with dissipation}},
    journal = {Ann. Phys.},
    volume = {{\bf 414}},
    pages = {168092},
    year = {2020},
    note = {\href{https://doi.org/10.1016/j.aop.2020.168092}{10.1016/j.aop.2020.168092}}
}

@article{GRR_26,
    author = {Jordi Gaset and Xavier Rivas and Narciso Román-Roy},
    title = {{A survey on geometric frameworks for action-dependent classical field theories and their relationship}},
    journal = {Geom. Mech.},
    volume = {{\bf 3}},
    number = {2},
    pages = {171--220},
    year = {2026},
    note = {\href{https://doi.org/10.1142/S2972458926400010}{10.1142/S2972458926400010}}
}

@article{GGMRR_20a,
    author = {Jordi Gaset and Xavier Gràcia and Miguel C. Muñoz-Lecanda and Xavier Rivas and Narciso Román-Roy},
    title = {{New contributions to the Hamiltonian and Lagrangian contact formalisms for dissipative mechanical systems and their symmetries}},
    journal = {Int. J. Geom. Methods Mod. Phys.},
    volume = {{\bf 17}},
    number = {6},
    pages = {2050090},
    year = {2020},
    note = {\href{https://doi.org/10.1142/S0219887820500905}{10.1142/S0219887820500905}}
}

@article{GGMRR_21,
    author = {Jordi Gaset and Xavier Gràcia and Miguel C. Muñoz-Lecanda and Xavier Rivas and Narciso Román-Roy},
    title = {{A $k$-contact Lagrangian formulation for nonconservative field theories}},
    journal = {Rep. Math. Phys.},
    volume = {{\bf 87}},
    number = {3},
    pages = {347--368},
    year = {2021},
    note = {\href{https://doi.org/10.1016/S0034-4877(21)00041-0}{10.1016/S0034-4877(21)00041-0}}
}

@unpublished{GIMM_98,
      title = {{Momentum Maps and Classical Relativistic Fields. Part I: Covariant Field Theory}}, 
      author = {Mark J. Gotay and James Isenberg and Jerrold E. Marsden and Richard Montgomery},
      year = {1998},
      note = {\href{https://arxiv.org/abs/physics/9801019}{arxiv:physics/9801019}} 
}

@article{GLMR_24,
    author = {Jordi Gaset and Manuel Lainz and Arnau Mas and Xavier Rivas},
    title = {{The Herglotz variational principle for dissipative field theories}},
    journal = {Geom. Mech.},
    volume = {{\bf 1}},
    number = {2},
    pages = {153--178},
    year = {2024},
    note = {\href{https://doi.org/10.1142/S2972458924500060}{10.1142/S2972458924500060}}
}

@article{GNH_78,
    title = {{Presymplectic manifolds and the Dirac–Bergmann theory of constraints}},
    author = {Mark J. Gotay and James M. Nester and G. Hinds},
    journal = {J. Math. Phys.},
    volume = {{\bf 19}},
    number = {11},
    pages = {2388--2399},
    year = {1978},
    note = {\href{https://doi.org/10.1063/1.523597}{10.1063/1.523597}}
    }

@article{GGR-2023,
    author = { Joaquim Gomis and Arnoldo Guerra and Narciso Román-Roy},
    title = {{Multisymplectic constraint analysis of scalar field theories, Chern-Simons gravity, and bosonic string theory}},
    year = {2023},
    journal = {Nuc. Phys. B},
    volume = {{\bf 987}},
    pages = {116069},
    note = {\href{https://doi.org/10.1016/j.nuclphysb.2022.116069}{10.1016/j.nuclphysb.2022.116069}}
}

@book{KMS_93,
    author = {Ivan Kolář and Peter W. Michor and Jan Slovák},
    title = {{Natural operations in differential geometry}},
    year = {1993},
    publisher = {Springer Berlin, Heidelberg},
    note = {\href{https://doi.org/10.1007/978-3-662-02950-3}{10.1007/978-3-662-02950-3}}
}

@article{Ku-84,
    title = {{Time-dependent mechanical symmetries and extended Hamiltonian
systems}},
    author = {R. Kuwabara},
    year = {1984},
    journal = {Rep. Math. Phys.},
    volume = {{\bf 19}},
    number = {1},
    pages = {27--38},
    note = {\href{https://doi.org/10.1016/0034-4877(84)90023-5}{10.1016/0034-4877(84)90023-5}}
}

@article{LPAF2018,
    author = {M.J. Lazo and J. Paiva and J.T.S. Amaral and G.S.F. Frederico},
    title = {{An action principle for action-dependent Lagrangians: Toward an action principle to non-conservative systems}},
    journal = {J. Math. Phys.},
    volume = {{\bf 59}},
    number = {3},
    pages = {032902},
    year = {2018},
    note = {\href{https://doi.org/10.1063/1.5019936}{10.1063/1.5019936}}
}

@article{LGGMR_23,
    author = {Manuel de León and Jordi Gaset and Xavier Gràcia and Miguel C. Muñoz-Lecanda and Xavier Rivas},
    title = {{Time-dependent contact mechanics}},
    journal = {Monatsh. Math.},
    volume = {{\bf 201}},
    pages = {1149--1183},
    year = {2023},
    note = {\href{https://doi.org/10.1007/s00605-022-01767-1}{10.1007/s00605-022-01767-1}}
}

@article{LGMRR_23,
    author = {Manuel de León and Jordi Gaset and Miguel Carlos Muñoz-Lecanda and Xavier Rivas and Narciso Román-Roy},
    title = {{Multicontact formulation for non-conservative field theories}},
    journal = {J. Phys. A: Math. Theor.},
    volume = {{\bf 56}},
    number = {2},
    pages = {025201},
    year = {2023},
    note = {\href{https://doi.org/10.1088/1751-8121/acb575}{10.1088/1751-8121/acb575}}
}

@article{LGMRR_25,
    author = {Manuel de León and Jordi Gaset and Miguel Carlos Muñoz-Lecanda and Xavier Rivas and Narciso Román-Roy},
    title = {{Practical introduction to action-dependent field theories}},
    journal = {Fortschr. Phys.},
    volume = {73},
    number = {5},
    pages = {e70000},
    year = {2025},
    note = {\href{https://doi.org/10.1002/prop.70000}{10.1002/prop.70000}}
}

@article{LIR_25,
    author = {Manuel de León and Rubén Izquierdo-López and Xavier Rivas},
    title = {{Brackets in multicontact geometry and multisymplectization}},
    journal = {Mediterr. J. Math.},
    volume = {{\bf 23}},
    number = {8},
    pages = {1--44},
    year = {2026},
    note = {\href{https://doi.org/10.1007/s00009-026-03077-4}{10.1007/s00009-026-03077-4}}
}

@article{LL_19,
    author = {Manuel de León and Manuel Laínz-Valcázar},
    title = {{Contact Hamiltonian systems}},
    journal = {J. Math. Phys.},
    volume = {{\bf 60}},
    number = {10},
    pages = {102902},
    year = {2019},
    note = {\href{https://doi.org/10.1063/1.5096475}{10.1063/1.5096475}}
}

@article{LLMP_03,
    author = {Manuel de León and Belén López and Juan C. Marrero and Edith Padrón},
    title = {{On the computation of the Lichnerowicz--Jacobi cohomology}},
    journal = {J. Geom. Phys.},
    year = {2003},
    volume = {{\bf 44}},
    number = {4},
    pages = {507--522},
    note = {\href{https://doi.org/10.1016/S0393-0440(02)00056-6}{10.1016/S0393-0440(02)00056-6}}
}

@article{LMMMR_05,
    author = {Manuel de León and Jesús Marín-Solano and Juan Carlos Marrero and Miguel Carlos Muñoz-Lecanda and Narciso Román-Roy},
    title = {{Premultisymplectic constraint algorithm for field theories}},
    journal = {Int. J. Geom. Meth. Mod. Phys.},
    volume = {2},
    number = {5},
    pages = {839--871},
    year = {2005},
    note = {\href{https://doi.org/10.1142/S0219887805000880}
    {10.1142/S0219887805000880}},
}

@incollection{LMS_03,
    author = {Manuel de León and David {Martín de Diego} and Aitor Santamaría-Merino},
    title = {{Tulczyjew triples and Lagrangian submanifolds in classical field theories}},
    year = {2003},
    booktitle = {Applied Differential Geometry and Mechanic},
    pages = {21--47},
    publisher = {Univ. of Gent, Academia Press},
    address = {Gent},
    note = {\href{https://arxiv.org/abs/math-ph/0302026}{arxiv:math-ph/0302026}}
}

@article{Mar_97,
    author = {Charles-Michel Marle},
    title = {{The Schouten--Nijenhuis bracket and interior products}},
    journal = {J. Geom. Phys.},
    volume = {{\bf 23}},
    number = {3--4},
    pages = {350--359},
    year = {1997},
    note = {\href{https://doi.org/10.1016/S0393-0440(97)80009-5}{10.1016/S0393-0440(97)80009-5}}
}

@article{Mar_88,
    author = {Geoffrey Martin},
    title = {{A Darboux theorem for multi-symplectic manifolds}},
    year = {1988},
    pages = {133--138},
    volume = {{\bf 16}},
    journal = {Lett. Math. Phys.},
    note = {\href{https://doi.org/10.1007/BF00402020}{10.1007/BF00402020}}
}

@article{RRV_26,
    author = {Xavier Rivas and Narciso Román-Roy and Anna Villanova},
    title = {{Skinner--Rusk formalism of action-dependent multicontact field theories}},
    journal = {Int. J. Geom. Methods Mod. Phys.},
    volume = {{\bf 23}},
    number = {10},
    pages = {2050090},
    year = {2026},
    note = {\href{https://doi.org/10.1142/S0219887826502130}{10.1142/S0219887826502130}}
}

@article{Rom_09,
    author = {Narciso Román-Roy},
    title = {{Multisymplectic Lagrangian and Hamiltonian formalisms of classical field theories}},
    journal = {Symmetry Integr. Geom.: Methods Appl. (SIGMA)},
    volume = {{\bf 5}},
    number = {100},
    year = {2009},
    note = {\href{https://doi.org/10.3842/SIGMA.2009.100}{10.3842/SIGMA.2009.100}},
}

@article{Ryv_26,
    author = {Leonid Ryvkin},
    title = {{Darboux type theorems in multisymplectic geometry}},
    journal = {arXiv:2503.03672 [math.SG]},
    volume = {},
    pages = {},
    year = {2026},
    note = {\href{https://doi.org/10.48550/arXiv.2503.03672
}{10.48550/arXiv.2503.03672}} 
}

@book{Sal_15,
    author = {Sandro Salsa},
    title = {{Partial Differential Equations in Action: From Modelling to Theory}},
    series = {Unitext},
    volume = {86},
    edition = {2nd},
    publisher = {Springer International Publishing},
    year = {2015},
    note = {\href{https://doi.org/10.1007/978-3-319-15093-2}{10.1007/978-3-319-15093-2}}
}

@article{Vit_15,
    author = {Luca Vitagliano},
    title = {{$L_\infty$-algebras from multicontact geometry}},
    journal = {Diff. Geom. Appl.},
    year = {2015},
    volume = {{\bf 39}},
    pages = {147--165},
    note = {\href{https://doi.org/10.1016/j.difgeo.2015.01.006}{10.1016/j.difgeo.2015.01.006}}
}

@book{IZ,
  author    = {Claude Itzykson and Jean-Bernard Zuber},
  title     = {{Quantum Field Theory}},
  publisher = {McGraw-Hill},
  address   = {New York},
  year      = {1980},
  isbn      = {978-0486445687}
}

@article {Herglotz,
author={G.~Herglotz},
title={Ber\"uhrungstransformationen},
journal={Lectures at the University of G\"ottingen, G\"ottingen}, year={1930},
}
}

\end{document}